%% file: thesis.tex
\documentclass[12pt,a4paper,amscd,amssymb]{report}
\usepackage[centertags]{amsmath}
\usepackage{amsthm, amsfonts}
\usepackage{graphics,color}
\usepackage{amssymb,letterspace}
\usepackage{epsfig}
\usepackage{amscd, amssymb,latexsym}
\usepackage{setspace}
\usepackage{epsf}

\newcommand{\bea}{\begin{eqnarray}}
\newcommand{\eea}{\end{eqnarray}}
\newcommand{\be}{\begin{equation}}
\newcommand{\ee}{\end{equation}}

\usepackage{setspace,graphicx,axodraw,multirow} 
\usepackage[stable]{footmisc}
\usepackage{cite,bbm}

\begin{document}

\newpage
\pagestyle{empty}
    \setcounter{footnote}{0}
    \null\vfil
    \vskip 20pt
    \begin{center}
      \setlength{\parskip}{0pt}
      {\textbf{UNIVERSITY OF SOUTHAMPTON}\par}
      { FACULTY OF ENGINEERING, SCIENCE AND MATHEMATICS \par}
      { Department of Physics and Astronomy \par}
      \vfill
      \vskip 50pt
      { \bf HOLOGRAPHIC DESCRIPTIONS OF QCD \par}
      \vskip 10pt
      { by \par}
      \vskip 10pt
      { Andrew John Mark Tedder \par}
      \vfill
      \vskip 50pt
      {A thesis submitted in partial fulfillment for the \par}
      {degree of Doctor of Philosophy \par}
      \bigskip
      \bigskip
      { May 2008 \par}
      \bigskip
    \end{center}
    \par
    \vfil\null
\newpage
\pagestyle{plain}
\pagenumbering{roman}

\newpage
 {\small \begin{center}UNIVERSITY OF SOUTHAMPTON\\\vspace{0.35cm}
\underline{Abstract}\\\vspace{0.35cm}
FACULTY OF SCIENCE\\\vspace{0.35cm}
PHYSICS\\\vspace{0.35cm}
\underline{Doctor of Philosophy}\\\vspace{0.35cm}
HOLOGRAPHIC DESCRIPTIONS OF QCD\\\vspace{0.35cm}
by Andrew John Mark Tedder
\end{center}

The AdS/CFT correspondence has long been used as a tool for understanding non-perturbative phenomena in gauge theories because it is an example of a `strong-weak' duality: when one side of the duality is weakly coupled, the other is strongly coupled and vice-versa. Hence strongly coupled phenomena can be studied by looking at the weakly coupled side of the duality. In its original form the correspondence proposes a duality between type IIB superstring theory on $AdS_5 \times S^5$ and an $\mathcal{N}=4$ supersymmetric Yang-Mills theory in four dimensions. In this thesis we investigate proposed duals to QCD itself. Duals to QCD fall into two categories: `top-down' and `bottom-up'. We take inspiration from both by truncating a consistent solution to the type IIB supergravity equations of motion (top-down). This model demonstrates dynamical chiral symmetry breaking, has a running coupling and contains a holographic description of the vector meson sector. By artificially extending the existing U(1) symmetry to SU(2) (bottom-up) we then obtain a holographic description of the axial vector sector.  We show that this model reproduces the masses and decay constants of the lightest mesons to the 10\% level. By regulating the UV with a sharp cut-off we can reproduce the $\rho$ meson masses to within 2\%. Finally we demonstrate that this model can be used to reproduce a very good agreement with hadronization data for particle production over a range of four orders of magnitude.}

\newpage
\onehalfspacing
\tableofcontents
\listoffigures
\listoftables
\doublespacing

\chapter*{Declaration of Authorship}

The work described in this thesis was carried out in collaboration with Professor Nick Evans and Tom Waterson. The following list details our original work and gives references for the material.

\begin{itemize}
\item Chapter \ref{ch:IRimp}: Nick Evans, Andrew Tedder and Tom Waterson, JHEP v.01 2007, p.058, arXiv:hep-ph/0603249
\item Chapter \ref{ch:UVimp}: Nick Evans and Andrew Tedder, Phys. Lett. B642 2006, p.546-550, arXiv:hep-ph/0609112
\item Chapter \ref{ch:Hadronisation}: Nick Evans and Andrew Tedder, 2007, arXiv:0711.0300 (hep-ph), accepted for publication in Physics Review Letters
\end{itemize}

There is also some original work of ours in chapter \ref{ch:CM} (the quantitative glueball spectrum, and the $m_{\rho}:m_{\pi}^2$ graph) and chapter \ref{ch:AdS-QCD} (the possible effect of gluonic contributions to $g_5$). However no claims to originality are made for the rest of chapters \ref{ch:CM} and \ref{ch:AdS-QCD}, and all of chapters \ref{ch:intro} to \ref{ch:flavour}, whose content has been complied from a variety of sources.

\chapter*{Acknowledgements}

I would like to thank my supervisor, Professor Nick Evans, for coming up with such interesting research topics, guiding me towards publication and for humouring my more fanciful ideas. The students, post-doctorates and staff at Southampton have all been very friendly and willing to help me. I like to think I made them consider the simple questions that they thought were beneath them! In particular I single out Ed Threlfall, Andreas Juettner, James Ettle and Michael `one hour' Donnellan for their general help. \vspace{0.5cm}\\Jonathan Shock and Tom Waterson guided my education in AdS/CFT matters for which I am grateful. I have also had great fun doing outreach activities with Pearl John, who is always quick to remind me that not everyone knows what a quark is.\vspace{0.5cm}\\I am very grateful to my parents who have been very supportive throughout my life, and taught me the value of a balanced education from a young age.\vspace{0.5cm}\\Financially I thank the University of Southampton for funding me through my PhD studies.

\pagenumbering{arabic}

\input{./Files/intro}

\input{./Files/AdS-CFT}

\input{./Files/Flavour}

\input{./Files/CM}

\input{./Files/AdS-QCD}

\input{./Files/IRimproving}

\input{./Files/UVimproving}

\input{./Files/Hadronisation}

\input{./Files/conclusion}

\input{./Files/biblio}
\end{document}

%% file: Files/intro.tex
\chapter{Introduction} \label{ch:intro}

\section{Introduction Overview}

The advance of physics in the last few centuries has been relentless. One of the main themes of this advance has been the desire for unification, the ultimate aim being to have one theory that describes all the fundamental forces. A major step towards this goal was achieved in the nineteenth century, when separate descriptions of electricity and magnetism were unified into a single theory of electromagnetism. It wasn't until the 1930s that it was fully appreciated that electromagnetism was an example of a gauge theory. \\

A gauge theory, in its most general sense, is a model with an invariance under a local symmetry of some of the variables in the theory \cite{Abers:1973qs,Yang:1954ek}. In the case of QED, the phases on all the fields can be changed locally, and so long as we introduce a gauge field which connects the points of local relabelling, the physics remains unchanged. In QED, for example, the quanta of these gauge fields are called photons. \\

QED is an example of an Abelian gauge theory: the order in which two gauge transformations, $\mathcal{G}_1,\mathcal{G}_2$, are performed is irrelevant. It is possible to construct gauge theories that are non-Abelian, meaning that the order in which two (or more) gauge transformations are performed does matter. In fact it was discovered in the 1970s that this was exactly what was needed to describe both the weak force and the strong force, which make up two more of the four fundamental forces. The force which we have yet to come to is gravity.\\

Currently gravity is a bit of an enigma in the particle physics world: classically it is well described by the theory of general relativity \cite{schutz,wald}, and there are no experimental measurements which contradict its predictions. However it is hard to test gravity at a single particle level because it is so weak in comparison to the other three forces. However almost everyone believes that general relativity needs modification; as it stands at the moment it cannot be quantized, which would make it unique amongst all the other forces. How to quantize gravity has been the bane of quantum physicists for generations, and there are currently two prongs of attack: loop quantum gravity \cite{Rovelli:1997yv}, and string theory \cite{polchinski, Green:1987mn, Green:1987sp}, on which we will spend some further time in section \ref{sch:string}. However the theme of this thesis is QCD, so it is prudent to review some of the features of this non-Abelian gauge theory.

\section{QCD}

In the late 1940s and early 1950s, only a few `elementary particles' were known: the proton, neutron, electron, neutrino and photon. Almost the entire observable universe consists of just these particles. However, some puzzling unstable particles had been seen in cosmic rays. Furthermore, physicists were keen to learn about the nuclear force which was presumed to bind protons and neutrons together to form nuclei. This led to the construction of larger and larger particle colliders throughout the decades, the most recent being the Large Hadron Collider in Switzerland. These accelerators have revealed the existence of hundreds of new particles, and almost all of them can be classed as \textbf{hadrons}. The discovery of these hadrons was not greeted with glee: it seemed incredulous to believe that such a large number of particles could all be fundamental. Fortunately, the initial skepticism has been validated. A lot of work in the 1950s and 1960s has shown that all hadrons (which include the protons and neutrons) are composite particles: they themselves are made up of even smaller particles, called quarks. It is currently believed that there are six flavours of quark, each of which exist in one of three `colours': red, green or blue. Almost all known hadrons can be accounted for by combining a suitable combination of quarks in varying levels of excitation. The entire spectrum can be explained by allowing gluons (the quanta of the QCD gauge field) to form particles too.

Why quarks are not seen as free entities of their own is a very deep question, and is essentially the question of \textbf{colour confinement}. Physicists hypothesise that the force between two quarks does not diminish as they are separated (contrast this with QED and, say, a positron and electron). Therefore it would take an infinite amount of energy to separate two quarks. Hence quarks are forever bound into colourless combinations, such as red-green-blue or red-antired. These colourless combinations are the hadrons we see in particle detectors and cosmic rays. Confinement, although widely believed, is yet to be analytically proven.

When quarks were first discussed in theoretical papers it was not certain if they were just convenient mathematical fictions, or whether they were physical entities. An excellent paper \cite{Bjorken:1968dy} predicted what would be seen in deep inelastic scattering of electrons and protons if the quarks did physically exist. Essentially, the idea was to replicate the idea of Rutherford Scattering by firing electrons at the protons and neutrons in nuclei and analyzing the resulting scattering pattern. The experiment was performed at the Stanford Linear Accelerator Center (SLAC) in 1969. The results matched the theoretical predictions, and since then quarks have been embraced as physical entities. \cite{Perkins:1982xb} contains a good review of this experimental evidence.

\subsection{The QCD Lagrangian} The QCD Lagrangian is given by \cite{peskin}

{\bea \mathcal{L} &=& \bar{\psi}\left(i \partial_{\nu}\gamma^{\nu}+ g A^a_{\mu}\gamma^{\mu}t^a-m \right)\psi -\frac{1}{4}F^a_{\mu\nu}F^{\mu\nu a} \nonumber\\
F^a_{\mu\nu} &=& \partial_{\mu}A_{\nu}^a-\partial_{\nu}A_{\mu}^a+g f^{abc}A^b_{\mu}A^c_{\nu} \label{QCDlag} \eea}
Mathematically $\psi$ is a Dirac fermion in the fundamental representation of SU(3), $A$ is the gauge field, which is in the adjoint representation of SU(3), $f^{abc}$ are the structure constants of SU(3), $g$ is a number and $m$ is a mass. Physically, $\psi$ represents the quarks, with $m$ being their mass. $A$ represents the gluons, and $g$ is the QCD coupling constant. Greek letters label space-time indices, and Roman letters label gauge group indices.\\

The Lagrangian fully describes QCD. It is deceptively simple, and like a fractal that on close inspection is ever more complex, so too is QCD. For example, it is not immediately clear from (\ref{QCDlag}) how to calculate hadron masses, decay constants or scattering cross-sections. In what follows we choose to mention those aspects which will be of relevance in this thesis.

\subsection{Renormalization} \label{sch:renorm}

The tricky subject of renormalization is dealt with in many textbooks \cite{peskin, ryder} and is the result of efforts made in the 1940s when it was realised that simple calculations, even in QED, resulted in divergent integrals. It is now accepted that these divergences, if finite in number\footnote{A theory is, by definition, renormalizable, if there are only a finite number of divergences. If there are an infinite number, the theory is said to be non-renormalizable. For details on how we know whether a theory is renormalizable or not, see \cite{peskin}}, can be reabsorbed into the parameters of the Lagrangian. Hence quantities such as the quark mass, $m$, the QCD coupling constant, $g$, and the field strengths, $A_{\mu}$, as they appear in the Lagrangian in equation (\ref{QCDlag}) are infinite. These quantities are called the \textit{bare} values, and when we speak of the quark mass, QCD coupling constant etc. we are actually referring to the redefined renormalized quantities.\\

To properly construct a systematic method of renormalization, we have to define a renormalization scale which is the energy at which we define our renormalized quantities ($m,g,A_{\mu}$). We then see what happens to these quantities as we vary the energy scale at which we are working. It turns out that the quantities vary as we change the energy scale. What must be remembered at all times is that the physics is independent of our choice of renormalization scale: experimentally measurable quantities are completely blind to which renormalization scheme the theorist has chosen. The implication of this statement is explored in the following section, via the renormalization group equation.

\subsection{The Renormalization Group Equation} \label{sch:callan}

The first step towards calculating hadron masses, decay constants and cross-sections from (\ref{QCDlag}) is the calculation of correlation functions. In fact this is the most difficult step.

Let us consider the bare connected n-point correlation function, $\Gamma^{(n)}_B$ in $\phi^4$ theory\footnote{$\phi^4$ theory is one of the simplest examples of an interacting quantum field theory. Its Lagrangian is $\mathcal{L}=\frac{1}{2}(\partial_{\mu}\phi)^2-\frac{1}{2}m^2\phi^2-\frac{\lambda}{4!}\phi^4$. More details can be found in \cite{peskin}. However, its exact nature is unimportant here. We use it because its renormalization group equation is particularly simple.}. It is defined as{\be \Gamma^{(n)}_B(x_1,x_2,...,x_n)\equiv \langle 0|T \phi_0(x_1)\phi_0(x_2)...\phi_0(x_n)|0 \rangle \ee}where $\phi_0(x)$ refers to the bare fields, before renormalization. They are related to the renormalized fields by a rescaling: $\phi_0(x) = Z_{\phi}^{1/2}\phi(x)$. The bare quantity, $\Gamma^{(n)}_B$ is a function of the bare coupling constant, $g_0$, and some cutoff, $\Lambda$. The renormalized quantity, $\Gamma_R^{(n)}$ is a function of the renormalized coupling constant $g$ and the renormalization scale $M$.{\be \Gamma^{(n)} (k_i;g,M)=Z^{(n/2)}_{\phi}\Gamma^{(n)}_B(k_i;g_0,\Lambda) \ee}The bare theory is independent of the renormalization scale, $M$, so we can write{\be \left(M\frac{d}{d M} \right) \left(Z^{(-n/2)}_{\phi}\Gamma^{(n)}(k_i;g,M) \right) =0 \ee}It is easy to show that this equation can be rewritten as{\be \left[M\frac{\partial}{\partial M}+\beta \frac{\partial}{\partial g}-\frac{n}{2}\gamma \right] \Gamma^{(n)}(k_i;g,M)=0 \label{callan}\ee}with the definitions{\bea \beta &=& M \frac{\partial g}{\partial M} \label{beta}\\
 \gamma  &=& M \frac{\partial \ln Z_{\phi}}{\partial M} \label{gamma}\eea}A little thought will also convince us that $\beta$ and $\gamma$ can only be functions of $g$: by dimensional analysis they cannot depend upon $M$, and are clearly independent of either $n$ or $\{x_i\}$. Using the method of characteristics, we can solve (\ref{callan}):{\be \Gamma^{(n)}(k_i;g,M)=\Gamma^{(n)}(k_i;g(\rho);M) \exp \left(-\frac{n}{2}\displaystyle\int_1^\rho \gamma(g(\rho)) \frac{dx}{x} \right) \label{RGsoln}\ee}with the characteristic equation $\rho \frac{d g(\rho)}{d \rho} = \beta(g(\rho))$, and the initial condition $g(\rho=1)=g$.

Equivalent equations to (\ref{beta}) and (\ref{gamma}) can be derived for all quantum field theories, and the definitions of $\beta$ and $\gamma$ are identical to those definitions in (\ref{beta}) and (\ref{gamma}).

Next we turn to the significance of $\beta$ and $\gamma$.

\subsection{The beta function} \label{betasec}

We defined the beta function, $\beta(g)$ in (\ref{beta}), and from its definition we can see that the beta function tells us how the renormalised coupling constant changes as we vary the renormalisation scale. The strength of the coupling constant at any particular energy is an important quantity: it determines when perturbation theory is valid. Perturbation theory is the most successful tool we have for solving quantum field theories: it was used very successfully in QED, so it would be excellent if we could use the same technique for QCD.

At low energies QCD is not weakly coupled: the coupling constant is large and perturbation theory is not applicable. But what about at high energies - perhaps we could use perturbation theory then?

To see what happens, let us assume that at high energies the theory is weakly coupled, and check that this doesn't lead us to a reductio ad absurdum argument.

If $g$ is small, we can approximate the beta function at these energies by a Taylor series expansion:{\be \beta(g) = A + B\bar{g} + C \bar{g}^2 + D\bar{g}^3\ldots \label{betaexpan} \ee}
where $\bar{g}$ is the deviation of $g$ from its value at $p=M$. We define the value of $g$ at $p=M$ as $g_R$: $g_R \equiv g(p=M)$.

Any perturbative deviation from $g_R$ will involve some sort of interaction. With each interaction, at least one power of the coupling constant will enter. Hence we can set $A=0$.

$B, C, D \ldots$ can be calculated quantitatively\footnote{In QCD with 6 flavours of quark, $B=C=0$ and $D=-\frac{7}{16\pi^2}$\cite{peskin}}; the method is detailed in \cite{peskin}. Here we are only concerned with the qualitative effects of the the leading coefficient of (\ref{betaexpan}). The first non-zero contribution to (\ref{betaexpan}) is canonically called $\beta_1$. There are three possibilities:

\begin{enumerate}
\item $\beta_1>0$
\item $\beta_1=0$
\item $\beta_1<0$
\end{enumerate}

For $\beta_1(g)>0$, the coupling constant becomes small in the infrared region of the theory. This is the case with QED, and allows us to successfully use perturbation theory at every-day energies. However, in the high energy, ultraviolet regime, the coupling constant gets larger and larger, eventually reaching the point where perturbation theory is no longer valid. Fortunately, in QED, the energy level where this occurs is much much higher than any reachable today. Furthermore, new physics is expected to appear a long time before we reach such energies.

For $\beta_1(g)=0$, the coupling constant doesn't run, and so the bare coupling constant is equal to the renormalised coupling constant for all energies. This is the case for many supersymmetric theories. It is also the case with {\bf scale invariant field theories} and {\bf conformal field theories}: if the coupling constant doesn't run, and there is no other source of scale in the theory (e.g. a mass scale could be introduced by having massive quarks), then the theory is defined to be scale invariant. A conformal field, however, is invariant to metric rescalings ($ds^2 \rightarrow \lambda^2 ds^2$ for arbitrary, real $\lambda$). All conformal field theories are scale invariant (the metric can be used to measure distance, in a conformal field theory we can always rescale the metric, so it is impossible to unambiguously define a distance), however not all scale invariant theories are conformal. Known examples are rare though \cite{Belavin:1984vu,DiFrancesco:1997nk,Schottenloher:1997pw,Ginsparg:1988ui}.

For $\beta_1(g)<0$, the coupling constant gets smaller at large energies, and conversely gets large at small energies. This is the case for QCD (see next section). So although we cannot use perturbation theory at low energies for QCD, we can use it for high energies; typically those found at large particle accelerators in Switzerland, Germany and California.

\subsubsection{The beta function for QCD}

In 1973, the one-loop contribution to $\beta(g)$ for QCD was calculated to be{\be \beta_1(g_s) = \frac{g_s^3}{16\pi^2} \left(-\frac{11}{3}N_c + \frac{2n_f}{3} \right) \label{QCDbeta}\ee}with $n_f$ being the number of quark flavours, and $N_c$ deriving from the $SU(N_c)$ gauge group. There are widely believed to be six flavours of quark, so $\beta_1(g_s)<0$, meaning that at high energies the quarks interact weakly. They are said to be {\bf asymptotically free}.

\begin{figure}
  \hfill
    \begin{center}  
      \includegraphics{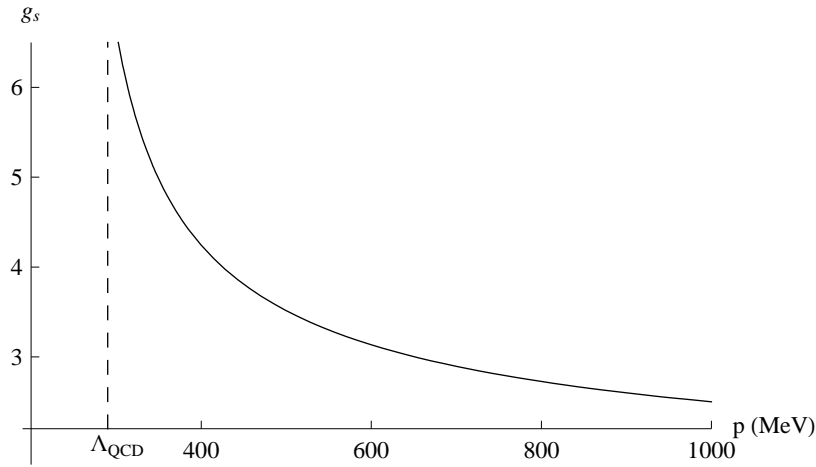}
    \end{center}
 \hfill 
\caption{Plot of the solution to the first loop beta function for QCD} \label{betaplot}
\end{figure}

We can go further, and solve (\ref{QCDbeta}):{\be g_s(p)^2 = \frac{g_R^2}{1+\frac{7 g_R^2}{16\pi^2}\log(p^2/M^2)} \label{solnbeta}\ee}whose solution is plotted in figure \ref{betaplot}. Note that $p$ enters the solution, for it is the only dimensionful parameter available to make the $\log$ term dimensionless. So in fact we can swap the $M$ derivative with a $p$ derivative in the definition of the $\beta$ function in (\ref{beta}):{\be \beta = M\frac{\partial g}{\partial M}= \frac{\partial g}{\partial \log M} \longleftrightarrow \beta = \frac{\partial g}{\partial \log p} \label{reparam} \ee}This reparameterization will come in useful in the next section.

Figure \ref{betaplot} brings to our attention two important features of QCD. Firstly, $g$ gets smaller and smaller as $p \rightarrow \infty$. So QCD is indeed asymptotically free, and we can use perturbation theory at high energies.

Secondly, figure \ref{betaplot} demonstrates the {\bf anomalous breaking of scale invariance}. If we were to look at the massless QCD Lagrangian, ie. (\ref{QCDlag}) with $m$ set to zero, we would see that there is no dimensionful parameter. We say that, classically, massless QCD is scale invariant. However, this would be a puzzle: phenomenologically there is clearly an inherent mass scale in QCD: the hadronic masses are not continuous, so any one could be used to define a scale. On studying the $\beta$ function we would be reassured. Figure \ref{betaplot} clearly introduces a scale into the theory, commonly called $\Lambda_{\mathrm{QCD}}$. It is normally defined to be the energy at which the coupling constant becomes infinite. In figure \ref{betaplot} this is approximately 246 MeV (although a more careful analysis, including higher order corrections to $\beta(g)$ sets $\Lambda_{\mathrm{QCD}}$ at $\sim215$ MeV). For all energies where $g_s \geq 1$, perturbation theory is no longer valid. Quantum effects have broken the classical symmetry: we say that the scale invariance has been broken anomalously.

\subsection{The 't Hooft Expansion}\label{sch:hooft}

Equation (\ref{QCDbeta}) provides an ideal opportunity to talk about the 't Hooft expansion, a particularly clever way of performing perturbation theory about the strong scale $\Lambda_{\mathrm{QCD}}$. We have just shown that at and around $\Lambda_{\mathrm{QCD}}$, the coupling constant is greater than one, which means a perturbative expansion such as in (\ref{betaexpan}) would be invalid. However in \cite{'tHooft:1974hx,'tHooft:1973jz} it was pointed out that there is a second dimensionless parameter in QCD: that of `$N_c$' in the gauge group $SU(N_c)$. 't Hooft showed that gauge theories simplify at large $N$ and that they have a perturbative expansion in terms of $1/N$. We summarise his analysis, and that of \cite{bible}, in this section.

We stated in (\ref{QCDbeta}) that the $\beta$-function for a pure $SU(N_c)$ gauge theory is given by{\be \beta(g_s) = -\frac{11}{3}N_c \frac{g_s^3}{16\pi^2} + \mathcal{O}(g_s^5) \ee}so the leading terms remain the same if we let $N\rightarrow \infty$, so long as we keep $\lambda \equiv g_s^2N$ fixed (one can show the higher order terms also stay the same in this limit). This limit is known as the \textbf{'t Hooft limit}.

The 't Hooft expansion was originally formulated in terms of a $U(N_c)$ gauge theory, with all matter in the adjoint representation (although this can be generalised to include fundamental matter). Let us assume, that as in QCD, all three-point terms are proportional to $g_s$ and all four-point terms are proportional to $g_s^2$, so the Lagrangian takes the schematic form{\be \mathcal{L} \sim \mathrm{Tr}(d\Phi_i d\Phi_i) + g_s
c^{ijk} \mathrm{Tr}(\Phi_i \Phi_j \Phi_k) + g_s^2 d^{ijkl} \mathrm{Tr}(\Phi_i
\Phi_j \Phi_k \Phi_l) \ee}for some constants $c^{ijk}$ and $d^{ijkl}$. We can rescale the fields by $\tilde{\Phi}_i \equiv g_s \Phi_i$, to reach{\be \mathcal{L} \sim \frac{1}{g_s^2} \left[ \mathrm{Tr}(d\tilde{\Phi}_i
d\tilde{\Phi}_i) + c^{ijk} \mathrm{Tr}(\tilde{\Phi}_i \tilde{\Phi}_j \tilde{\Phi}_k) + d^{ijkl} \mathrm{Tr}(\tilde{\Phi}_i \tilde{\Phi}_j
\tilde{\Phi}_k \tilde{\Phi}_l) \right] \label{hooftlag}\ee}where $1/ g_{YM}^2 = N/ \lambda$. Now we want to know what happens in the limit $N \rightarrow \infty, \lambda = \mathrm{fixed}$. Although $1/g_s^2$ tends to infinity, this is countered by the fact that the number of components in the fields also goes to infinity.

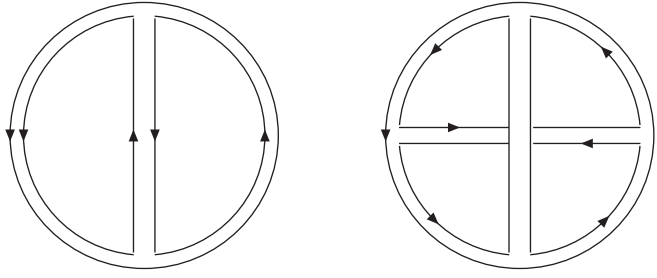
\begin{figure}
  \hfill
    \begin{center}  
     \begin{picture}(300,100)
             \ArrowArc(60,50)(50,0,360)
             \ArrowArc(60,50)(45,95,265)
             \ArrowArc(60,50)(45,275,445)
             \ArrowLine(56,6)(56,94)
             \ArrowLine(64,94)(64,6)

             \ArrowArc(200,50)(50,0,360)
             \ArrowArc(200,50)(45,95,175)
             \ArrowArc(200,50)(45,185,265)
             \ArrowArc(200,50)(45,275,355)
             \ArrowArc(200,50)(45,365,85)
             \Line(196,6)(196,94)
             \Line(204,94)(204,6)
             \Line(155,47)(196,47)
             \ArrowLine(155,53)(196,53)
             \ArrowLine(245,47)(205,47)
             \Line(205,53)(245,53)
      \end{picture}
\caption{Two examples of contributions to the vacuum amplitude in double line notation. The diagram on the left is planar and hence contributes $N^2$. The diagram on the right is of genus one and hence contributes $N^0$} \label{hooftexpan}
    \end{center}
 \hfill
\end{figure}

A $U(N_c)$ theory with matter in the adjoint representation can be represented as a direct product of a fundamental and an anti-fundamental field $\Phi^i_j$, as in figure \ref{hooftexpan}, where we have drawn two contributions to the vacuum amplitude.

What is the power of $N$ and $\lambda$ associated with each diagram? From (\ref{hooftlag}) we can see that each vertex will have a coefficient proportional to $N/\lambda$, and each propagator will have a coefficient proportional to $\lambda/N$. In addition, each closed loop will a factor of $N_c$ since we have to sum over all indices in the loop. Hence we have{\bea \mathrm{Vertex :}&& \frac{N_c}{\lambda} \\
\mathrm{Propagator :}&& \frac{\lambda}{N_c} \\
\mathrm{Loop :} && N_c \eea}So a diagram with V vertices, E propagators and F loops will include a factor{\be N^{\mathrm{N-E+F}}\lambda^{E-V}=N^{\chi}\lambda^{E-V} \ee}where $\chi$ is the \textbf{Euler characteristic} of the diagram. It is a topological invariant, depending only on the genus (number of holes), $g$, of the surface:{\be \chi = 2-2g \ee}Therefore the perturbative expansion of any diagram in field theory can be written as{\be \sum_{g=0}^{\infty} N^{2-2g} \sum_{i=0}^{\infty} c_{g,i} \lambda^i = \sum_{g=0}^{\infty} N^{2-2g} f_g(\lambda) \label{stringexpan}\ee}where $f_g$ is some polynomial in $\lambda$. In the large $N$ limit, we see that any computation will be dominated by those diagrams which are topologically equivalent to a sphere or plane (ie. no holes, genus = 0).

If we identify $1/N$ as some generic coupling constant, $g_p$, we can see that (\ref{stringexpan}) looks like a perturbative expansion about $g_p$. The resemblance between (\ref{stringexpan}) and perturbative string theory is one of the strongest motivations for believing that string theories and field theories are related \cite{bible}, in particular by a $1/N$ expansion.

In particular, the AdS/CFT correspondence, which is the inspiration for this thesis, and which we turn our attention to in section \ref{ch:AdS-CFT}, is formulated at large $N$. It is hoped that in the future $1/N$ corrections will make the correspondence even more applicable to $SU(3)$ quantum chromodynamics.

\subsection{Anomalous Dimensions}\label{sch:anom}

In section \ref{betasec} we looked at $\beta(g)$ when $g$ was small. Now we turn our attention to $g >1$, the strongly coupled regime. We can no longer calculate $\beta(g)$ explicitly, but we can consider the qualitative possibilities.\\

\begin{figure}[h]
  \hfill
  \begin{minipage}[t]{.45\textwidth}
    \begin{center}  
      \includegraphics[scale=0.45]{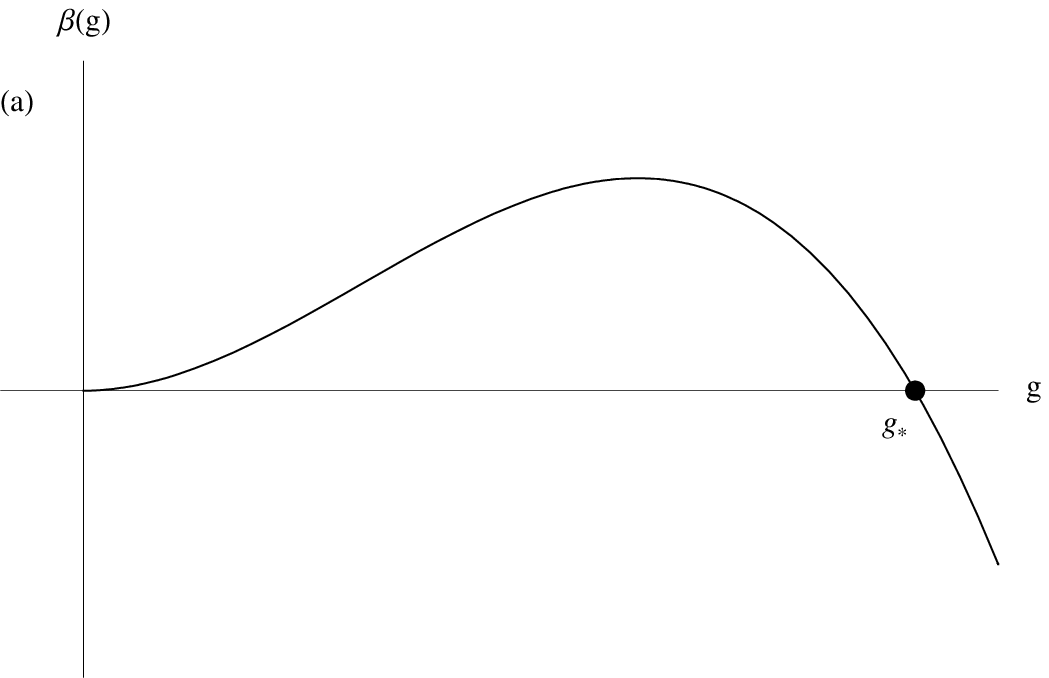}
    \end{center}
  \end{minipage}
  \hfill
  \begin{minipage}[t]{.45\textwidth}
    \begin{center}  
      \includegraphics[scale=0.45]{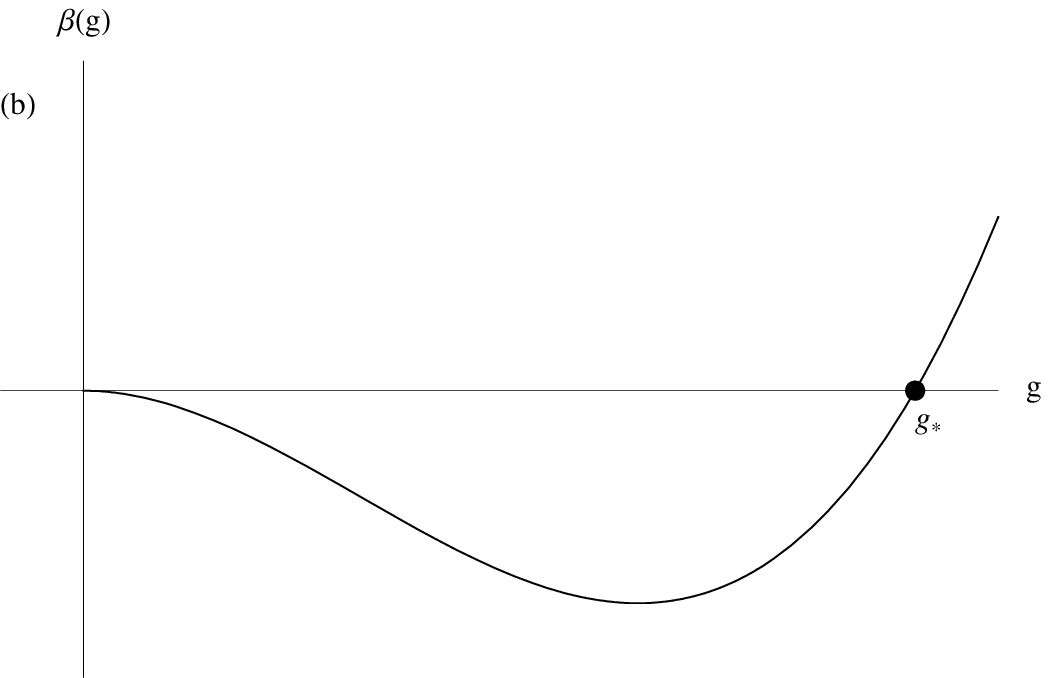}
    \end{center}
  \end{minipage}
 \hfill 
\caption{Possible forms of the $\beta$ function with non-trivial zeros: (a) ultraviolet-stable fixed point; (b) infrared-stable fixed point} \label{nontriv}
\end{figure}

Of most interest is when $\beta(g)$ is either positive or negative in the weakly coupled regime, but higher order corrections mean that $\beta(g)$ has non-trivial zeros: the two possibilities are shown in figure \ref{nontriv}.

A beta function of the form \ref{nontriv}(a) is weakly coupled at low energies. At higher energies the coupling constant grows, but only upto $g_*$. Once it reaches $g_*$, the coupling constant stays constant. This is a potential, but unproven resolution of the Landau pole\footnote{A Landau pole means that the coupling constant becomes infinite at finite energy. Technically QCD has a Landau pole at $\sim \Lambda_{\mathrm{QCD}}$, but the phrase is usually used to refer to non-asymptotically free theories only} in QED: the Standard Model has such a pole at $10^{34}\mathrm{MeV}$ \cite{Gockeler:1997dn}. However this analysis assumes a monotonic beta function. If the beta function of QED has a non-trivial zero, then the Landau pole would not exist.

Explicitly, near the fixed point of figure \ref{nontriv}(a), the $\beta$ function can be approximated by{\be \beta \sim -B (g-g_*) \ee}or{\be \frac{d}{d \log p}g \sim -B(g-g_*) \ee}which has the solution{\be \bar{g}(p) = g_* + C \left(\frac{M}{p} \right)^B \ee}This has important implications for the exact solution of the renormalization group equation (\ref{RGsoln}). For sufficiently large $p$, the integral in the exponential of (\ref{RGsoln}) will be dominated by those values of $p$ where $g(p)$ is close to $g_*$. Then{\bea G(p) &\approx& G(g_*)\exp\left[-(\log \frac{p}{M}2(1-\gamma(g_*)) \right] \\
&\approx& C \left(\frac{1}{p^2}\right)^{1-\gamma(g_*)} \eea}So the two-point function returns to a simple scaling law, as might be expected in a naive dimensional analysis argument. But there is an important difference: we would expect the power law to be $p^{-2}$, but instead it is $p^{-2(1-\gamma(g_*))}$. The complex interactions of the quantum field theory have affected the law of rescaling. Finally, since the fixed point is reached at large $p$, it is known as an ultraviolet fixed point.

An analysis of graph \ref{nontriv}(b) follows similar lines. Near the fixed point the $\beta$ function can be approximated by {\be \beta \sim +B (g-g_*) \ee}which has the solution{\be \bar{g}(p) = g_* + C \left(\frac{p}{M} \right)^B \ee}This means that for sufficiently small $p$, the integral in the exponential of (\ref{RGsoln}) will be dominated by those values of $p$ where $g(p)$ is close to $g_*$, giving{\bea G(p) &\approx& G(g_*)\exp\left[-(\log \frac{p}{M}2(1-\gamma(g_*)) \right] \\
&\approx& C \left(\frac{1}{p^2}\right)^{1-\gamma(g_*)} \eea}Because this occurs at small $p$, this fixed point is called an infrared fixed point.

Once again, the complex interactions of the field theory have affected the law of rescaling. For this reason, the $\gamma(g)$ function is commonly known as the {\it anomalous dimension}, even if there is no fixed point in the theory.

\subsection{Mass gap}

Another feature of QCD, and yet to be proved rigorously by anybody, is that of a mass gap. A quantum field theory is said to have a \textbf{mass gap} if the energy spectrum has a positive greatest lower bound, but does not include zero. This is thought to be very closely linked to the property of \textbf{confinement}, which is simply a statement that gluons cannot exist on their own, but only in colourless bound states. Lattice gauge theories have shown to the satisfaction of most that quarkless QCD exhibits such a phenomena, but solving this problem with due mathematical rigour still remains one of the Millennium Prize Problems. Experimentally it is certainly true that there is a mass gap in QCD.

\subsection{Chiral Symmetry Breaking}\label{ch:pert}

We have already seen in section \ref{betasec} that massless QCD anomalously breaks its classical scale invariance. There is a second classical symmetry to the massless QCD Lagrangian: that of chiral symmetry, meaning that the left and right handed quarks transform independently.

The Lagrangian of massless QCD is given by taking equation (\ref{QCDlag}) and simply setting $m=0$:{\be \mathcal{L} = \bar{\psi}\left(i \partial_{\nu}\gamma^{\nu}+ g A^a_{\mu}\gamma^{\mu}t^a \right)\psi  -\frac{1}{4}F^a_{\mu\nu}F^{\mu\nu a}\label{QCDnomass} \ee}(\ref{QCDnomass}) possesses a chiral symmetry which can be seen if we write $\psi$ as $ \left( \begin{array}{c}
\psi_L  \\
\psi_R  \end{array} \right)$ and choose the chiral representation of the gamma matrices:

 \begin{center}
    \begin{minipage}[t]{0.5\linewidth}
      
\[ \gamma^0 = \left( \begin{array}{cc}
0 & \mathbbm{1}  \\
\mathbbm{1} & 0  \end{array} \right)\] 
    \end{minipage}\hfill
    \begin{minipage}[t]{0.5\linewidth}
    \[ \gamma^i = \left( \begin{array}{cc}
0 & {\bf \sigma^i}  \\
{\bf-\sigma^i} & 0  \end{array} \right)\] 
    \end{minipage}
  \end{center}then (suppressing the gauge field which doesn't affect the analysis), equation (\ref{QCDnomass}) becomes{\be \mathcal{L} = i\psi^{\dagger}_L\left( \partial_0 - \sigma . \bigtriangledown \right)\psi_L +  i\psi^{\dagger}_R \left( \partial_0 + \sigma . \bigtriangledown\right)\psi_R \ee}In this form it is clear that we can perform separate global flavour transformations on $\psi_L$ and $\psi_R$. Since the quarks are in an SU(2) isospin multiplet, we write these transformations as $SU(2)_L \otimes U(1)_L$ and $SU(2)_R \otimes U(1)_R$ respectively. The Lagrangian is unchanged by these transformations: it is chirally invariant. (The analysis still holds if we turn on the gauge field $A_{\mu}$.)\\

However, this is a symmetry of the Lagrangian which is broken spontaneously by the choice of vacuum. The most obvious manifestation of this is the absence of a parity-doubled spectrum. If the chiral symmetry was respected, there would be a positive parity hadron for every negative parity hadron. This is not what is seen in nature. For example, the proton has isospin $\frac{1}{2}$, spin $\frac{1}{2}$ and a positive parity. Its mass is 938 MeV. Its parity partner, catchingly called the N(1535) $S_{11}$, has a mass of 1535 MeV \cite{PDG06}. We conclude that chiral symmetry is broken at a quantum level, and the easiest way to account for this is by assuming that QCD spontaneously forms a quark condensate. In other words, the QCD vacuum gives a non-zero vacuum expectation value to the scalar operator{\be \langle 0| \bar{Q}_L Q_R + \bar{Q}_R Q_L |0\rangle \neq 0 \label{VEV}\ee}What is the physical interpretation of this VEV? The up and down quarks are very light, and so it costs little energy to create a quark-antiquark pair. However, the binding energy released by a bound quark-antiquark pair is large. All (\ref{VEV}) is saying is that the process of forming a quark-antiquark pair and then binding them together is exothermic: we get out more energy then we put in.

The vacuum expectation value (\ref{VEV}) means that we can no longer perform independent gauge transformations on the left and right handed spinors: the only flavour transformation we still can make is one where we simultaneously make the same transformation on both $\Psi_L$ and $\Psi_R$. We say that the $SU(2)_L \times SU(2)_R$ flavour symmetry has been spontaneously broken to $SU(2)_V$.

Goldstone's theorem \cite{Goldstone:1962es} states that whenever a continuous symmetry of a quantum field theory is spontaneously broken, a massless particle will appear. In breaking $SU(2)_L \times SU(2)_R$ to $SU(2)_V$, we started with six group generators, and ended up with three: according to Goldstone's theorem we would expect three massless particles to appear in the spectrum of QCD.

However, the Lagrangian of (\ref{QCDnomass}) is not exactly that of real QCD. The proper Lagrangian of QCD has massive quarks and is given in (\ref{QCDlag}). Hence QCD does not have an exact $SU(2)_L \times SU(2)_R$ symmetry QCD. But because the up and down quarks are {\it almost} massless, there is an approximate symmetry, so a perturbative approach about $m_q \approx 0$ may still be useful.

Looking at the hadron spectrum there are three suspiciously light hadrons, namely the pions, whose masses are about one fifth of the next lightest hadron. Furthermore they have the correct parity to be created by the axial isospin current, $j^{\mu5a}$:{\be j^{\mu5a} = \bar{Q} \gamma^{\mu}\gamma^5\tau^a Q \ee}The coupling between the pions and the vector axial current is defined as{\be \langle\pi^b(p)|\bar{Q}\gamma_{\mu}\gamma_5T^aQ(0)|0\rangle=-if_{\pi}p_{\mu}\delta^{ab} \label{pion-axial-coup}\ee}where $T^a$ is a generator of the broken symmetry group and $f_{\pi}$ is a number with dimensions of mass. It is called the pion decay constant. For an SU(2) isospin symmetry, meaning two flavours of quark, we expect three types of pion.

$f_{\pi}$ in equation (\ref{pion-axial-coup}) is a measure of the extent of the chiral symmetry breaking. If there is no chiral symmetry breaking its value is zero. In QCD it can be determined from measuring the decay rate of $\pi^0$ particles \cite{PDG06}. It is found to be $92$ MeV.

\subsubsection{Meson decay constants}

\begin{figure}
\begin{center}
     \begin{picture}(210,100)
                \Photon(10,50)(70,50){4}{4}
                \Photon(130,50)(200,50){4}{4}
                  \Text(0,50)[l]{$\mu$}
                  \Text(205,50)[l]{$\nu$}
                  \CCirc(100,50){30}{Gray}{Gray}
\LongArrow(20,40)(50,40) \Text(35,30)[]{$q$}
\Text(100,50)[]{\large{1PI}}
      \end{picture}
\caption{The vector current two-point function, $\Pi_V(-q^2)$. The grey blob represents all possible one particle irreducible diagrams that can be inserted.} \label{vectFey}
\end{center}
\end{figure}
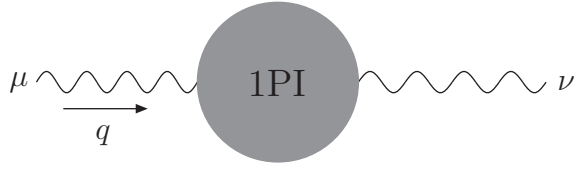

It is not just the pion that has a decay constant associated with it. All pseudoscalar mesons, P, have a decay constant $F_P$, defined by{\be \langle 0|A_{\mu}(0)|P(q)\rangle =  iF_Pq_{\mu} \label{axialcurrent1}\ee}And the vector meson decay constants\footnote{vector mesons have two decay constants: the transverse polarization decay constant, $F_V^T$, and the longitudinal polarization decay constant $F_V$. Only the longitudinal decay constant can be determined experimentally.} are defined as:{\be \langle 0|\bar{Q}(0)\gamma^{\mu}Q(0)|V(p;\lambda)\rangle = iF_Vm_V\epsilon_{\lambda}^{\mu} \ee}where $p$ and $\lambda$ are the momentum and polarization state of the vector meson $V(p;\lambda)$. $\epsilon_{\mu}$ is the corresponding polarization vector.

Experimentally, $F_P$ can be measured from leptonic decays of the appropriate meson \cite{PDG06}. $F_V$ can be measured from the decays of tau leptons \cite{Donoghue:1992dd,Isgur:1988vm,Son:2003et}.

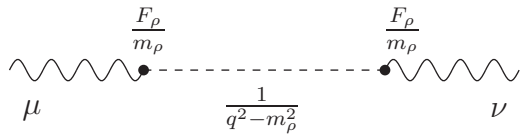
\begin{figure}
\begin{center}
     \begin{picture}(210,100)
                \Photon(10,50)(60,50){4}{4}
                \Photon(150,50)(200,50){4}{4}
                \DashLine(60,50)(150,50){3}
                \Vertex(60,50){2} \Vertex(150,50){2}
                  \Text(55,65)[l]{$\frac{F_{\rho}}{m_{\rho}}$}
                  \Text(90,35)[l]{$\frac{1}{q^2-m_{\rho}^2}$}
                  \Text(150,65)[l]{$\frac{F_{\rho}}{m_{\rho}}$}
                  \Text(15,35)[l]{$\mu$}
                  \Text(190,35)[l]{$\nu$}
      \end{picture}
\caption{One contribution to the vector current two-point function}\label{decayconstFey}
\end{center}
\end{figure}

If we represent the vector current two-point function as figure \ref{vectFey}, then one contribution to the vector two-point current will be that of figure \ref{decayconstFey}.

It can be shown \cite{Witten:1979kh,Manohar:1998xv} that at large N, the mesons form an infinite, stable spectrum and that the vector two-point current is given by the sum over all meson and glueball resonances (with the correct quantum numbers), such as that in figure \ref{decayconstFey}:{\bea \Pi_V(-q^2) &\equiv& -\displaystyle\sum_{\rho}\frac{\langle 0|J_{\mu}|\rho\rangle \langle \rho|J^{\mu}|0\rangle}{(q^2-m_{\rho}^2)} \nonumber\\
&=&-\displaystyle\sum_{\rho}\frac{F_V^2}{(q^2-m_{\rho}^2)m_{\rho}^2} \label{vectdecay}\eea}The axial vector current two-point function at large N, $\Pi_P(-q^2)$, is almost identical:{\be \Pi_P(-q^2) =-\displaystyle\sum_{a_1}\frac{F_P^2}{(q^2-m_{a_1}^2)m_{a_1}^2} \label{pseudodecay}\ee}

\subsubsection{Axial Current Anomaly}\label{ch:axialsym}

(\ref{QCDnomass}) also possesses a $U(1)_L \otimes U(1)_R$ symmetry which we have yet to mention. Whereas the $SU(2)$ symmetries mixed the up and down quarks, the $U(1)$ symmetries act to add a phase to the entire spinor. Classically, we have independent U(1) symmetries for both the left and right handed symmetries. These phases always cancel out since the fermions are always bilinear. The presence of the non-zero VEV (\ref{VEV}) also breaks this classical $U(1)_L \otimes U(1)_R$ symmetry to leave a single $U(1)_V$ symmetry. And once again, Goldstone's theorem applies: a spontaneously broken symmetry implies the presence of a massless boson. Here the broken symmetry has one only one generator, so we'd expect one massless boson. A look at the known QCD mass spectrum shows that the only suitable candidate, in terms of quantum numbers, is the $\eta'(958)$ meson. However it has a disappointingly high mass. Even allowing for the small masses of the quarks, we'd expect our boson to have a mass similar to that of the pions.

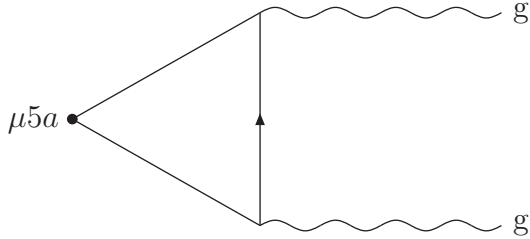
\begin{figure}
\begin{center}
     \begin{picture}(210,100)
              \Line(20,50)(90,90)
              \Line(20,50)(90,10)
              \ArrowLine(90,10)(90,90)
              \Photon(90,90)(180,90){2}{4}
              \Photon(90,10)(180,10){2}{4}
\Text(185,10)[l]{g}
\Text(185,90)[l]{g}
\Vertex(20,50){2}
\Text(15,50)[r]{$\mu5a$}
      \end{picture}
\caption{One of the diagrams that leads to the axial vector anomaly in QCD} \label{triangle}
\end{center}
\end{figure}

The explanation of this dichotomy puzzled physicists for a long while, and wasn't explained until 1986 \cite{'tHooft:1986nc}. It is quite easy to show \cite{peskin,Forkel:2000sq} that due to the necessity of regularization and renormalization (in particular triangle diagrams such as figure \ref{triangle}) the classical axial symmetry of QCD ($\partial_{\mu} j^{\mu 5}=0$) is broken by quantum effects to{\be \partial_{\mu} j^{\mu 5} =-\frac{e^2}{16 \pi^2} \epsilon^{\alpha\beta\mu\nu}F_{\alpha\beta}F_{\mu\nu} \ee}
The obvious question to ask is what is the source of such an anomaly, and the answer lies in the complicated field of instantons \cite{'tHooft:1986nc,Forkel:2000sq,Diakonov:2002fq,Percacci:1991hf}. There are an infinite number of distinct QCD vacua, whom differ from one another by their topological equivalence class: $\pi_3(S^3)=\mathbb{Z}$. Furthermore, it can be shown that whenever a correlation function tunnels in and back out of a vacuum in a different topological class, it picks up a contribution to the axial symmetry:{\be Q= -\frac{1}{2n_f}\int \partial_{\mu}j^{\mu5}\ee}where Q is defined as the topological charge, and can be shown to be equal to the number of left handed fermions minus the number of right handed fermions: $Q=n_L-n_R$. An instanton is defined as those solutions for which $Q=1$.

Probably the simplest way to think of instantons is as isolated sources of axial symmetry breaking, and they are responsible for making the $\eta'(958)$ meson heavy. More detailed analysis can be found in the sources quoted earlier.

\section{Solving QCD}

Now that we've investigated some of the complex features of QCD, a prudent question would be to ask whether we have the theoretical tools required to calculate experimental quantities such as hadron masses and decay constants. These are very difficult quantities to determine. Current approaches can be divided into four classifications.

\begin{itemize}

\item Perturbative QCD

Feynman diagrams are used highly successfully in QED to calculate cross-sections. The correlation function is expanded in terms of a small parameter, $\alpha \approx 1/137$, and then each contribution is calculated order-by-order. Beyond one loop the calculations can get very intractable, and it requires a great deal of effort and care. Fortunately, $\alpha$ is very small, and so higher order contributions are also small. The equivalent expansion in QCD is $\alpha_s \approx 0.118$ at $91.2$ GeV \cite{Schmelling:1996wm}, which means that the expansion does not converge as rapidly. In addition, as we discussed earlier, we can only use perturbative QCD at high energies, so for many interactions, it is not valid.

\item Lattice QCD \cite{Rothe:2005nw,DeGrand:2006zz}

The best established approach to non-perturbative QCD is the use of huge supercomputers. Continuous spacetime is represented by a discrete four dimensional lattice, with lattice spacings of $a$. The QCD Lagrangian (\ref{QCDlag}) is then reformulated for a discrete spacetime, and the calculations are made for as small a lattice spacing as possible. For the final answer, the limit $a \rightarrow 0$ is taken. Lattice QCD takes a lot of human time and effort, and it is not always clear what happens when the limit $a \rightarrow 0$ is taken.

\item Effective Theories

Mirroring the techniques of perturbation theory, theories are written down which mimic QCD for some limited parameter space. And within this parameter space, there will be a parameter which can be used as a perturbation parameter. Examples are chiral perturbation theory \cite{Ecker:1994gg,Naboulsi:2003ap,Ananthanarayan:2003eb}, which uses the light quark masses (at low momentum) as perturbation parameters, and heavy quark effective theory \cite{Naboulsi:2003ap,Mannel:1997ky}, which uses the inverse of a heavy quark mass as a perturbation parameter. These approaches can be very useful, but are limited by the extent of parameter space to which their model is valid.\pagebreak

\item 1/N expansion

Once again, mirroring the techniques of perturbation theory, experimental quantities are calculated as a perturbative expansion in terms of the $N$ in the $SU(N)$ of the gauge group of QCD, in line with the argument presented in section \ref{sch:hooft}. In theory this is applicable to all of QCD, but the perturbation parameter, $(1/3)^2$, is not very small, so the series does not converge as quickly as one might like. More importantly, it is not clear how to go about calculating the 1/N corrections. Hence there is little knowledge of the error in a tree-level calculation. The AdS-CFT correspondence falls under this classification. Before we investigate it we need to go over some basic aspects of string theory.

\end{itemize}

\section{String Theory} \label{sch:string}

The AdS/CFT correspondence is discussed in detail in chapter \ref{ch:AdS-CFT}. It describes a duality between a gauge theory and a string theory, so we take this opportunity to review some of those aspects of string theory relevant to this thesis.

\subsection{Premise}

String theory starts by assuming that fundamental particles are not point-like, but have a finite length. In other words, they are little strings.

We reach the Euler-Lagrange equations of motion for point-like particles by minimizing their world-lines. By analogy, we assume that the equations of motion for strings will be found by minimizing their world areas. The original formula for the area of the sheet, known as the Nambu-Goto action \cite{polchinski,Green:1987mn}, is{\be S=T \int d\sigma d\tau \sqrt{\dot{X}^2X'^2 -(\dot{X}.X')^2} \label{nambu-goto}\ee}where{\be \dot{X}^{\mu}=\frac{\partial X^{\mu}(\sigma,\tau)}{\partial \tau}, \hspace{1cm} X'^{\mu}=\frac{\partial X^{\mu}(\sigma,\tau)}{\partial \sigma}  \ee}The $X^{\mu}$ are the bosonic fields in the spacetime. (\ref{nambu-goto}) does not contain fermionic fields; this is addressed later.

The square root in (\ref{nambu-goto}) makes the mathematics of optimization awkward. Instead we can introduce a new variable, $h_{\alpha\beta}$ which will be the metric on the string world sheet. This leads to the conventional{\be S=-\frac{T}{2} \int d^2 \sigma \sqrt{h} h^{\alpha\beta}\eta_{\mu\nu}\partial_{\alpha}X^{\mu}\partial_{\beta}X^{\nu} \label{conven}\ee}Since the derivatives of $h$ do not appear in (\ref{conven}), it is not a real field, and so can be eliminated, leading back to (\ref{nambu-goto}) if required.\\

(\ref{conven}) is not satisfactory to describe the real world. Firstly, we need fermions in the theory, and secondly it turns out that (\ref{conven}) contains a tachyon. However it is possible to supersymmetrize (\ref{conven}) \cite{polchinski}: {\be S[X,\psi] = \frac{1}{4\pi\alpha'} \int d^2 \sigma (\eta^{ab}g_{\mu\nu}\partial_a X^{\mu}\partial_b X^{\nu})+ \alpha'(\eta^{ab}g_{\mu\nu}\bar{\psi}^{\mu}\gamma_a\partial_b\psi^{\nu}) \label{supersym} \ee}And with a lot of work \cite{polchinski} it is possible to show that so long as $X^{\mu}$ and $\Psi^{\mu}$ exist in ten dimensions, (\ref{supersym}) contains no tachyons or anomalies.

\subsection{World-sheet Boundary Conditions}

There are several types of string: their properties depend on the nature of the boundary conditions of the fields on the world-sheet. For the moment we only consider closed strings (open strings enter when we consider D-branes in section \ref{sch:d-branes}). We can choose to make our strings orientable or non-orientable. If a string is orientable, we can tell which way we are travelling along a string. If it is non-orientable, we can't. There are further choices we can make with respect to the fermionic boundary conditions, and in fact going through every possibility took researchers a lot of time. In the end, there are five consistent superstring theories, summarised in table \ref{strings}.

\begin{table}
\begin{center}
    \begin{tabular}{ | l |p{9cm}|}
    \hline
    Type & Details \\ \hline
    I     & All strings are charged under an SO(32)
gauge symmetry\\ \hline
 IIA    &   Chiral, meaning that parity is a good symmetry. 
No gauge symmetry either, so theory only contains gravity\\ \hline
 IIB    &  Non-chiral, both in the fermionic sector and the gauge sector \\ \hline
 HO    &   Heterotic, meaning that
 the left-movers are bosonic, and the right-movers are fermionic. Has 
 an SO(32) gauge symmetry\\ \hline
 HE   &   Heterotic, with an 
$E_8 \times E_8$ gauge symmetry\\\hline
    \end{tabular}
\end{center}
\caption{The five consistent types of superstring theory} \label{strings}
\end{table}

What is the spectrum of closed string theory? This is a detailed and complex subject that is visited by many sources (see \cite{polchinski, Green:1987sp, Howe:1983sra,Schwarz:1983wa,D'Hoker:2002aw,Shock:thesis} for example) so here we just briefly summarise the results.

The spectrum contains the metric, $G_{\mu\nu}$, the scalar dilaton $\Phi$, and a two index antisymmetric tensor $B_{\mu\nu}$:

\begin{itemize}

\item The metric needs little explanation, and plays the usual role as the device for measuring distance and angles.

\item The dilaton is a very important field in string theory: it generically appears in the action as $e^{\Phi}$, and measures the Euler characteristic (the number of holes and handles) of the string worldsheet. Hence the quantity $e^{\Phi}$ plays the role of the theory's coupling. When open and closed string sectors are
combined the Yang-Mills coupling from the open string sector has
$g^2_{YM} = e^\Phi$.

\item $B$, known as the Kalb-Ramond or NS-NS B-field, is the generalization of the one-form potential $A_{\mu}$ of electrodynamics. For point-like particles moving in an electromagnetic potential the action is given by $\int A_{\mu}dx^{\mu}$. For closed strings moving in a Kalb-Ramond potential, the action is given by $\int B_{\mu\nu}dx^{\mu}dx^{\nu}$.

\end{itemize}

All of the fields listed above are NS-NS fields (Neveu-Schwarz-Neveu-Schwarz). R-R (Ramond-Ramond), NS-R and R-NS fields also exist. The NS-R and R-NS fields are not a concern to us - they are simply the fermionic partners of the bosonic sector. However the R-R fields are sourced by D-branes \cite{polchinski} and so are also present in the spectrum. They enter as field strengths of antisymmetric fields as shown in (\ref{fullIIB}) and (\ref{fullIIBexp}).

For the AdS/CFT correspondence applied to $3+1$-dimensional field
theories, ten-dimensional type IIB string theory is of central
importance, and in particular its low-energy limit where strings
become point-like and string theory becomes supergravity. There
exists no completely satisfactory action for the type IIB
supergravity, since it involves an antisymmetric field $C_4$ with
self-dual field strength $F_5$. However, it is possible to write
an action involving both dualities of $C_4$, and then impose the
self-duality as a supplementary field equation.  In this way one
obtains (see for example
\cite{Howe:1983sra,Schwarz:1983wa,D'Hoker:2002aw})
\begin{eqnarray}
\label{IIBaction}
S_{IIB} &=& \phantom{+}  \frac{1}{4 \kappa_B ^2} \int \sqrt G e^{-2\Phi} (2R_G + 8
\partial_{\mu} \Phi \partial^{\mu} \Phi -  |H_3|^2 )\label{fullIIB}
\\
&& - \frac{1}{4 \kappa_B ^2} \int \biggl [ \sqrt G (|F_1|^2 +
|\tilde F_3|^2 + \frac{1}{2}  |\tilde F_5|^2) + C_4 \wedge H_3 \wedge F_3
\biggr ] + {\rm fermions} \, ,
\nonumber
\end{eqnarray}
where the field strengths are defined by
\begin{equation}\begin{gathered}
F_1  = dC \, , \quad  H_3 = dB \, , \quad  F_3  = dC_2 \, , \quad  F_5  =
dC_4\, , \\
\tilde F_3  = F_3 - C H_3 \, , \quad
\tilde F_5  = F_5 -\frac{1}{2} A_2 \wedge
H_3 + \frac{1}{2} B \wedge F_3 \, , 
\end{gathered}\label{fullIIBexp}\end{equation}
and we have the additional self-duality condition $* \tilde F_5 = \tilde
F_5$.

However, for the AdS/CFT correspondence, there are only three fields switched on: the graviton, the dilaton, and the five-form field strength $F_5$, so we can disregard most of the fields in (\ref{fullIIB}) and instead use:

{\be S_{\mathrm{IIB}}= \frac{1}{2\kappa^2} \int d^{10}x\:\sqrt{-G}\left[R-\frac{1}{2}(\partial\Phi)^2 -\frac{1}{2.5!}F_5^2 \right] \label{IIbact}\ee}
where we have redefined the metric ($\sqrt{G} e^{-2\Phi}\rightarrow \sqrt{G} $) to transform the action from the string frame to the Einstein frame, giving an action in the usual Einstein-Hilbert form \cite{Johnson:2000ch}. It is in this frame that the stress-energy tensor has its usual meaning.

There is more to the string theory story than that indicated by table \ref{strings}. As well as the 1+1 dimensional strings, there are higher dimensionful objects too, called D-branes. Their importance was first recognised in the mid-1990s, and they form an important part of the AdS-CFT correspondence. We look at them next.

\subsection{D-branes} \label{sch:d-branes}

Upto this point we have been conducting a purely mathematical exercise. It was hoped that string theory would provide a unique theory of the four fundamental forces of nature. However, we have already stated that there are five consistent versions of superstring theory.  This is disillusioning, and was the point at which the string community had reached in the early 1990s. Fortunately, it was shown that all five theories were linked by a series of simple dualities, meaning that there was only one theory after all, called M-theory. One of those dualities is T-duality.

T-duality is a symmetry between small and large distances. The `T' stands for {\it topology}, since T-duality transformations work on spaces in which at least one dimension has the topology of a circle.

Let us consider a string that is wrapped around the circle, ie. it is closed. It's energy spectrum is given by{\be m^2 \sim  \frac{n^2}{R^2} + \frac{\omega^2R^2}{\alpha'^2} \label{enspec}\ee}The first term is the Kaluza-Klein energy, and the second term is the winding number. (\ref{enspec}) is invariant under the following transformation:{\be R \rightarrow \alpha'/R, \hspace{1cm} n \rightarrow \omega \label{invar}\ee}Therefore a theory with only closed modes will be invariant under (\ref{invar}). But what if a theory contained open strings too? These aren't invariant under (\ref{invar}). The open strings will have a Kaluza-Klein contribution to their energy spectrum, but no winding term. Hence T-duality will change the spectrum of the open strings.

Under T-duality, the boundary conditions of the open string change:{\be \partial_{\sigma_2}X^{\mu}(\sigma_1,(0,l))=0 \longleftrightarrow X^{\mu}(\sigma_1,(0,l)) = c^{\mu} \ee}
We go from Neumann boundary conditions (LHS) to Dirichlet boundary conditions (RHS). For a long time this was result was ignored by the string community because the Dirichlet boundary condition is not Lorentz invariant - a special surface in the 10D spacetime has been picked out. It was Polchinski who first embraced this result, and extolled the modern explanation of this effect. The surface $c^{\mu}$ is indeed special; it represents the surface of another object, hitherto undiscovered in string theory. These objects are called {\bf D-branes}, the `D' standing for `Dirichlet', and the `brane' as in membrane, or surface. By the T-duality, all open strings must start and finish on a D-brane.

Just as closed strings can be considered fluctuations of the background geometry, open strings can be considered as fluctuations of the D-branes. The D-branes are solitonic objects, and thus exist in the low energy limit of string theory - supergravity.

\subsection{D-brane action}

It was shown \cite{Leigh:1989jq} in 1989 that the action{\be S \sim \int_M e^{\Phi/2}\sqrt{\mathrm{det}(G'+B'+F)} \label{braneact}\ee}for some arbitrary D-brane gives the equations of motion for all the background fields associated with any D-brane. This form of an action was first written down by Born and Infeld when considering a particular nonlinear generalization of electromagnetism \cite{Born:1934gh}, and was expanded upon by Dirac \cite{dirac}. All this occurred a long time before the advent of string theory and D-branes, but for historical reasons (\ref{braneact}) is still called the Dirac-Born-Infeld (DBI) action. Let us briefly explain each field in turn:

\begin{itemize}
\item $\Phi$ is the usual dilaton.
\item $G'$ is the induced metric on the brane's surface. The precise definition of this concept is given in section \ref{sch:quarkmass}, but it is the equivalent of the $\sqrt{-\mathrm{det}g}$ term in the standard Einstein-Hilbert action for general relativity.
\item  $B'$ is the induced Kalb-Ramond field on the brane's surface. The precise co-ordinate definition can be found in \cite{polchinski}, but all it represents is the two-form field which has been induced on the surface of the D-brane as a result of the Kalb-Ramond field in the bulk.
\item $F$ is the usual electromagnetic field potential for a single D-brane, and is attributable to the open string sector. In fact the $B'$ fields and $F$ fields interact, and it can be shown \cite{Zwiebach:2004tj} that only the combination $F+B'$ is gauge invariant.
\end{itemize}

In many publications, it has been conventional to re-define $F$ as $2\pi\alpha'F$, and for clarity we follow this convention in this thesis. Furthermore, the Kalb-Ramond field will be switched off for all the examples in this thesis. Hence the DBI action for the rest of this thesis is taken to mean{\be S \sim \int_M e^{\Phi/2}\sqrt{\mathrm{det}(G+2\pi\alpha'F)} \label{simpbraneact}\ee}There will be cases though when we turn off the electromagnetic field tensor $F$ in (\ref{simpbraneact}) too.

For completeness, we add that it is possible to add a Chern-Simons term to (\ref{braneact})\cite{Karch:1999pv,Gawedzki:1999bq}:{\be S \sim \int_M e^{\Phi/2}\sqrt{\mathrm{det}(G+B'+F)} + i \int_{\delta M} \mathrm{Tr}\:\left[\mathrm{exp}(2\pi\alpha'F_2 + B_2) \wedge \displaystyle\sum_q C_q \right]\label{braneact2}\ee}However this is a topological term that is only relevant if we are considering vacuums with non-trivial topologies. In all that follows, we implicitly assume that the vacuum is topologically trivial, and hence we will not include a Chern-Simons term in any subsequent action.

%% file: Files/AdS-CFT.tex
\chapter{AdS-CFT}
\label{ch:AdS-CFT}

The discovery of the AdS/CFT correspondence was built upon a whole raft of previous discoveries and conjectures. Here we ignore the historical steps that led to its discovery, and simply introduce it, as if by magic, as it stands today. On one side we have a type IIB string theory in the geometry induced by an infinite stack of D3 branes. The other side is a conformal gauge theory with an infinite number of colours. First we address the D3-brane construction.

\section{String description}\label{sch:stringdes}

Our starting point is to put $N$ D3-branes in the centre of a ten dimensional space, which is otherwise empty. There will be three types of interaction: those amongst the open strings on the branes; those amongst the closed strings in the bulk; and those between the open and closed strings. We can write the effective action as{\be S= S_{\mathrm{bulk}}+S_{\mathrm{brane}}+S_{\mathrm{int}} \ee}Now we consider the limit in which we send the string length to zero; $l_s \rightarrow 0$ ($\alpha' \rightarrow 0$). All other dimensionless parameters (the string coupling constant, $g_s$, and $N$) we keep fixed.  Doing this means the coupling of the strings to each other ($\sim g_s \alpha'^2$) goes to zero, and hence the strings do not interact: $S_{\mathrm{int}}$ is turned off, and we are left with two decoupled systems:  classical ten dimensional gravity in the bulk, $S_{\mathrm{bulk}}$, and a four dimensional gauge theory on the surface of the branes, $S_{\mathrm{brane}}$.

What is the exact nature of the four dimensional gauge theory? The ends of the open strings are labelled by the branes they are attached to. The strings are orientated, so there are two types of strings stretching between any two branes: one whose left end is on the first brane and whose right end is on the second brane, and vice versa. These labels are called Chan-Paton factors, and in the case of a coincidental stack of branes, as we have here, there is a clear re-labelling symmetry of the Chan-Paton factors.

There are $N^2$ possible string relabellings, which fills the adjoint representation of the U(N) Lie group. The strings are free to move around on the surfaces of the branes, so the relabelling can be performed at any point: hence the U(N) symmetry is a local one. In other words it is a gauge theory, and the Chan-Paton labels can be interpreted as labelling the gauge charge of each end of the string.

In the large $N$ limit, the propagators of adjoint fields in a $U(N)$ and $SU(N)$ gauge theory differ by a vanishing term, proportional to $N^{-1}$ \cite{bible}, so when $N \rightarrow \infty$, we can treat $U(N)$ and $SU(N)$ gauge theories as identical. The spectrum of strings is given by the dimensional reduction of an $\mathcal{N}=1$ gauge multiplet in ten dimensions to four dimensions:{\be  \underbrace{\left( \begin{array}{c}
A^{\mu}  \\
\Psi  \end{array} \right)}_{\mathrm{D=10},\: \mathcal{N}=1}  \longrightarrow  \underbrace{\left( \begin{array}{c}
A^{\mu}  \\
\Psi  \end{array}  \right) \left( \begin{array}{c}
\lambda \\
\Psi  \end{array}  \right)\left( \begin{array}{c}
\lambda  \\
\Psi  \end{array}  \right)\left( \begin{array}{c}
\lambda  \\
\Psi  \end{array}  \right)}_{\mathrm{D=4},\: \mathcal{N}=4}\label{dimred}\ee}where the left hand side of (\ref{dimred}) is defined in ten dimensions with one supersymmetry generator ($\mathcal{N}=1$), and the right hand side is defined in four dimensions, with four supersymmetry generators ($\mathcal{N}=4$). The $A^{\mu}$ are gauge fields, $\Psi$ are gauginos, and $\lambda$ represent complex scalars. Six of the ten-dimensional gauge field degrees of freedom become the three complex scalars in the four-dimensional reduction.

The one-loop $\beta$ function for this theory is {\be \beta = \left(\frac{11}{3}N-\frac{8}{3}N - N\right)=0 \ee}with the first term coming from the gauge fields, the second from the gauginos, and the third from the scalar fields. Supersymmetric non-renormalization theorems \cite{Buchbinder:1998qv} can be used to show that there are no higher order contributions to the $\beta$ function, and thus it is zero for the full quantum theory. A theory which has no massive fields, and whose coupling constant doesn't undergo dimensional transmutation, is a conformal field theory (CFT). This means the theory is invariant to rescalings of the metric.

To conclude: on the string side, we have two decoupled systems: ten-dimensional supergravity; and an $SU(N)$ conformal gauge theory.

\section{Supergravity description}

Let us consider the system from a different point of view. D-branes are massive objects, and hence warp the spacetime that they are put in. It is possible to show that there is a D3 brane solution to supergravity \cite{Horowitz:1991cd} which is{\be ds^2 = Z^{-1/2} \eta_{\mu\nu} dx^{\mu}dx^{\nu} + Z^{1/2}\displaystyle\sum_{a=1}^6 dx^a dx^a \nonumber\ee}{\be Z = 1+ \frac{R^4}{r^4} \hspace{1cm} R^4 = 4\pi g_s l_s^4 N \hspace{1cm} \int_{S^5}{}^*F_5=N\label{branesoln} \ee} $R$ characterises the typical curvature scale of the gravitational solution, and $N$ is the number of D3-branes in the spacetime.

Now let's take the same low energy limit as we did before ($l_s \rightarrow$ 0). From the point of view of an observer at infinity, there will be two types of low energy excitations. The first type would be any massless excitation in the bulk with a suitably large wavelength. The second is a bit more subtle, and is a result of $g_{tt}$ not being constant. All observations made at infinity will be redshifted, by a factor $Z^{-1/4}$, when compared to the observations made at some fixed position $r$. So we can have very highly excited states in this low energy system, just so long as they are close enough to $r=0$, because by the time they have reached some observer at infinity the redshift factor will be so large that all excitations will appear low energy to that observer. And as we tend closer and closer to the limit $l_s \rightarrow 0$, the two systems will decouple: the excitations near $r=0$ because the gravitational well is insurmountable; and the supergravity excitations because the wavelengths will be very large compared to the gravitational extent of the branes. 

In conclusion, we have supergravity in the bulk, and in the near horizon we have{\be ds^2 = \frac{r^2}{R^2} \left(-dt^2+dx_1^2+dx_2^2+dx_3^2 \right) + R^2 \frac{dr^2}{r^2} + R^2 d\Omega_5^2 \label{ads5}\ee}where we have been able to go from (\ref{branesoln}) to (\ref{ads5}) because in the limit we are considering, $r \ll R$. Equation (\ref{ads5}) is simply the geometry of $AdS_5 \times S^5$, with the radius of the 5-sphere given by $R$.

We see that from both the string description, and the supergravity description, the system decouples into supergravity in flat space, and something else. This suggests the conjecture that the \textit{`something else'} are alternative descriptions of exactly the same thing: this is the essence of the AdS/CFT correspondence:

\begin{center}
    \begin{tabular}{ | c |}
    \hline
$\mathcal{N}=4$ U(N) SYM in 3+1 dimensions \\
 is dual to type IIB superstring theory on $AdS_5 \times S^5$  \\ \hline
    \end{tabular}
\end{center}

It is sometimes easy to get lost in the mathematical details of what we are doing, but it is worth pausing to consider exactly what we are claiming. The SYM side of the duality is defined in 3+1 dimensions: the supergravity side exists in 9+1 dimensions! In some sense, six of the ten dimensions on the supergravity side are superfluous! This type of duality is classed as a holographic duality, taking its name from holograms which use a two dimensional object to depict a three dimensional image. The first holographic dual was proposed for black-holes, with the postulate that the entropy of a black-hole is given by its area. This meant that all the information contained within a black hole is encoded in a lower dimensional surface. Similarly we are claiming here that all the information in the ten dimensional supergravity can be equally well explained by a four dimensional field theory.

\newpage
\section{Matching the Symmetries}

What symmetries do each side of the description have? On the $AdS_5 \times S^5$ side we clearly have two symmetry groups: the $SO(6)$ of the 5-sphere, and the symmetry of the $AdS_5$ spacetime is SO(4,2) \cite{bible}. What does this relate to on the super Yang-Mills side? Since it is $\mathcal{N}=4$ there are four gauginos, hence there is an $SU(4)_R \cong SO(6)$ symmetry. It is also conformal in all four dimensions: the conformal group in four dimensions has a symmetry group SO(4,2). So we can identify corresponding symmetries on either side of the correspondence:
\begin{table}[h]
\begin{center}
    \begin{tabular}{c|c|c}
& $\mathcal{N}=4$ SYM & $AdS_5 \times S^5$\\ \hline
SO(4,2) & conformal group & spacetime symmetry  \\ \hline
SU(4)$\cong$SO(6) & R-symmetry & $S^5$ isometry
    \end{tabular}
\end{center}
\end{table}
\section{Regime of usefulness}

In what regime are our approximations valid? On the supergravity side, the two regimes of interest decoupled when the wavelength of the supergravity excitations were much larger than the string length:{\be 1 \ll\frac{R^4}{l_s^4}  \sim g_{\mathrm{YM}}^2 N \sim g_sN \ee}But on the field theory side we need the 't Hooft coupling to be much less than one if the theory is to be in the perturbative regime:{\be 1 \gg  g_{\mathrm{YM}}^2N  \sim g_sN \sim \frac{R^4}{l_s^4}  \ee}Clearly the two inequalities are incompatible, which leads us to the conclusion that when the gravity dual is weakly coupled, the field theory must be strongly coupled. We can be bolder still, and make the conjecture that when the gravity side is strongly coupled, the field side will be weakly coupled. We propose that the duality is a \textbf{strong-weak duality}.

This is why this duality is so exciting to those studying strongly coupled gauge theories. As explained earlier, solutions to the non-perturbative regime of such theories have evaded capture for decades. This duality gives us hope that instead of having to try and solve the nightmare of a strongly coupled theory, we can instead re-parameterise the theory in terms of a weakly coupled supergravity theory, which we can solve using perturbation theory!

\section{Energy-radius Duality}

Ignoring for the moment the $SU(4)_R \cong SO(6)$ part of the duality, we have a 4+1 spacetime dual to a 3+1 field theory. Conventionally the `extra' fifth dimension of the spacetime is denoted with $r$, and here we consider its physical interpretation.

The $AdS$ side has the metric{\be ds^2 = r^{-2} dr^2 + r^2\eta_{\mu\nu}dx^{\mu}dx^{\nu} \ee}But the CFT is invariant under $x^{\mu} \rightarrow e^{\alpha} x^{\mu}$. If the CFT side is invariant under this transformation, then so too must be the $AdS$ side. Hence $r$ must scale as $r \rightarrow re^{-\alpha}$. It transforms as an energy.

This can be understood from the point of view of an observer at infinity (the boundary of $AdS_5 \times S^5$). To her, a photon emitted from the centre of the space is red-shifted. If the observer moves in from infinity to some finite $r$, the photon is less red-shifted. So the energy-radius duality indicates that instead of just considering the duality as being constructed on the boundary of the $AdS_5 \times S^5$ space, we are free to look at the duality at any value of $r$: each different choice of $r$ simply represents the same field theory at a different energy scale.\\

From here onwards, the $r$ coordinate is referred to as the energy scale. The small $r$ limit is called the infrared (IR), and the high $r$ limit is called the UV (ultraviolet).

\section{Explicit Correspondence}\label{sch:explic}

The correspondence will only be useful if there is a prescribed way of calculating physical quantities on each side. The way to do this was first suggested in \cite{Witten:1998qj}, and the postulate is that the generating functional for correlation functions with some source $\phi_0(x)$ is equal to the supergravity partition function whose fields $\phi(x)$ tend to $\phi_0(x)$ at the boundary of the $AdS_5 \times S^5$ space:{\be \langle\mathrm{Exp}\left(\int d^4x \phi_0(x)\mathcal{O}(x) \right)\rangle_{AdS_5 \times S^5}\Big|_{\phi(x,\infty)=\phi_0(x)} = \mathcal{Z}_{\mathrm{string}}[\phi_0(x)] \label{genfn}\ee}So, in simple terms, the boundary values for supergravity fields correspond to sources for the field theory operators.

The LHS side of (\ref{genfn}) is independent of the global symmetries of the conformal field theory. So, the entity $\int d^4x\:\phi_0(x)\mathcal{O}(x)$ must be a singlet under all the symmetries of the dual field theory. This is how we go about identifying which fields in the supergravity match with which operators in the CFT, which we address next.
\newpage
\section{Field-operator Matching} \label{sch:field-op}

We showed in the previous section that an operator $\mathcal{O}$ on the conformal side corresponds to putting in a field $\phi$ on the gravity side.

The example canonically presented is to consider the action of a scalar field, $\phi$, of mass $m$ on the $AdS_5 \times S^5$ side:{\be S=\int d^{4}x\:dr\:d\Omega_5\: \sqrt{-g}\left(\eta^{\mu\nu}\partial_{\mu}\phi\partial_{\nu}\phi -m^2 \phi^2 \right) \label{matchact}\ee}The simplest solution is obtained by assuming that $\phi$ is a function of $r$, the radial direction of the $AdS_5 \times S^5$ space only. Then the solution is given by{\be \phi(r) = A r^{-(4-\Delta)} + B r^{-\Delta} \label{matchsoln}\ee}with $m^2=\Delta(\Delta-4)$\footnote{This formula can be generalized to $m^2=(\Delta-p)(\Delta+p-4)$ in the more general case of the action of a p-form \cite{bible}}.

In fact in general, a bulk scalar $\phi$ with mass $m^2$ corresponds to a scalar operator on the conformal side with dimension $\Delta=-2 \pm \sqrt{4+m^2}$\footnote{$\Delta$ refers to the \textbf{full} dimension of the operator (i.e. classical + anomalous)}. Solving $\Box \phi=0$ gives two independent solutions, $r^{\Delta-4}$ and $r^{-\Delta}$: $\phi(r) \sim Ar^{\Delta-4} + Br^{-\Delta}$. We can then identify \cite{Gubser:1998bc,Klebanov:1999tb, Witten:1998qj, Rattazzi:2000hs} $B$ as the source of the operator $\mathcal{O}$ and $A$ as the VEV for the same operator. As discussed in the previous section, the most natural way \cite{Witten:1998qj} to combine these quantities to form a dimensionless quantity is $\int d^4x\:\left(A\times B\right) \equiv\int d^4x\:\phi_0\mathcal{O}$.

We also stated in the last section that  $\int d^4x\:\phi_0(x)\mathcal{O}(x)$ must be a singlet under all symmetries of the conformal field theory. Furthermore, it must have mass dimension 0 since it is part of an exponential, so $\phi_0(x)\mathcal{O}(x)$ must have mass dimension 4. Clearly such a combination of $A$ and $B$ satisfies these requirements.

Let us consolidate these ideas by looking at the example of the gaugino bilinear $\bar{\lambda}\lambda$. It is a Lorentz scalar of dimension 3: therefore we need a supergravity scalar field with mass dimension 1. It is also a ten dimensional representation of the SU(4) R-symmetry. We find such a field in the reduction of type IIB strings on $AdS_5$ \cite{Witten:1998qj} if we set $m^2=-3$\footnote{AdS has negative curvature, and so we can have $m^2<0$ without faster-than-light travel} in (\ref{matchact}). Then the solution becomes{\bea \phi(r) &=& Ar^{-1} + Br^{-3} \\
&\equiv& \frac{m}{r} + \frac{c}{r^3} \eea}where we have just redefined the constants $A$ and $B$. Supergravity fields do not scale under four-dimensional conformal transformations, so $m$ must have dimension 1 and $c$ must have dimension 3, which corresponds with $m$ being a source (i.e. a mass) for  $\bar{\lambda}\lambda$ and $c$ being its VEV (or condensate).

The careful reader may ask at this point about existence and uniqueness when it comes to operator-field matching. Fortunately that question has been answered in \cite{Kim:1985ez}: for every operator there exists a corresponding field which is unique. However, as is the way with existence-uniqueness theories, there is no easy way to match operators with fields: each one has to be done by hand. And in any gauge theory there are an infinite number of operators. When we come to consider duals to QCD, a lot of the work is in deciding which operators of QCD, from an infinite choice, are to be included in the holographic field theory.

\section{Deformations and Adding Flavour}

So far we have explored the AdS/CFT duality as it was first presented, basically as a mathematical construct. However the promise of describing a strongly coupled gauge theory in terms of a weakly coupled one was very exciting to students of QCD. Perhaps, it was thought, this would be a useful way to make calculations in the strongly coupled regime of QCD. However, there were some immediate obstacles to overcome: the CFT field theory present in the duality had many characteristics which makes it distinctly different to QCD. It is conformal, supersymmetric, strongly coupled in the UV, and has an infinite number of colours. In addition there aren't any quarks. The addition of quarks is addressed in the next chapter.

%% file: Files/Flavour.tex
\chapter{Flavouring the $\mathrm{AdS_5 \times S^5}$}\label{ch:flavour}

All the matter discussed in chapter \ref{ch:AdS-CFT} was in the adjoint representation of the $\mathrm{SU(N)}$ colour group. This is because both ends of the open strings start and end on the same type of brane - a D3-brane - and so are indistinguishable. All the Chan-Paton factors are in effect colour indices. Quarks have both colour and flavour indices, and are in the fundamental representation of the gauge group. So we need strings in our geometry which can take on these properties. A lot of work, by a lot of contributers, showed that this is achieved by introducing D7 branes \cite{Karch:2002sh,Karch:2002xe,Kruczenski:2003be,Sakai:2003wu,Babington:2003vm,Nunez:2003cf,Ouyang:2003df,Wang:2003yc,Hong:2003jm,Evans:2004ia,Burrington:2004id,Erdmenger:2004dk,Kruczenski:2004me,Kuperstein:2004hy,Paredes:2004is,Sakai:2004cn,Evans:2005ti,Shock:thesis}. In this section we summarise the salient points of their work.

\section{The D7-Brane Probe}

A brane is massive, and will distort the spacetime in which it is placed. At the moment this is an unwanted complication for us, so we shall use the \textbf{probe} limit, which means that we ignore the gravitational effects of the new brane. This approximation, known as the quenched approximation, holds if $N_f \ll N_c$, where $N_f$ is the number of probe branes, and $N_c$ is the number of D3 branes. This approximation is the same as introducing a test charge in electrodynamics. Work has been done on unquenched spacetimes \cite{Aharony:1998xz,Grana:2001xn,Bertolini:2001qa,Burrington:2004id,Kirsch:2005uy}, but this is not too much of a concern for us here.

The $\mathrm{AdS_5} \times \mathrm{S^5}$ metric can be written as {\be ds^2 = \frac{r^2}{R^2} \eta_{\mu\nu}dx^{\mu}dx^{\nu} + \frac{R^2}{r^2} \displaystyle\sum_{i=1}^6 dr_i^2 \ee}A D7-brane probe can be chosen to fill the $x$ directions and four of the $r$ directions, as illustrated in table \ref{probetab}.

\begin{table}
    \begin{center}  
       \begin{tabular}{c|c|c|c|c|c|c|c|c|c|c}
          &$x_0$     & $x_1$    & $x_2$    & $x_3$     & $r_1$    & $r_2$    & $r_3$    & $r_4$    & $r_5$    & $r_6$ \\
          \hline
       D3 & $\times$ & $\times$ & $\times$ & $\times$ & .        &.         &.         &.         &.        &.\\
       D7 & $\times$ & $\times$ & $\times$ & $\times$ & $\times$ & $\times$ & $\times$ & $\times$ &.        &.\\
         \end{tabular} 
    \end{center}
\caption{Choice of D3 and D7 embedding. Filled directions are marked with a cross. Unfilled directions by a dot.} \label{probetab}
  \end{table}

Introducing the D7 branes has allowed two new types of string to exist in this system:

\begin{itemize}
\item Strings stretched between the D3 and D7 branes - these have one colour index and one flavour index, and so can be identified as quarks
\item Strings with both ends on a D7 brane - these have two flavour indices, and so are identified as mesons
\end{itemize}

This is, of course, in addition to the strings with both ends on a D3 brane (`gluons') and closed strings (`gravitons') which were present before the D7 branes, and are still there.

\begin{figure}[t]
\begin{center}
\begin{picture}(150,150)
\Line(75,75)(120,120)
\LongArrow(75,0)(75,150)
\LongArrow(0,75)(150,75)
\SetColor{Red}
\Vertex(75,75){3} \Text(79,71)[lt]{D3}
\Vertex(120,120){3} \Text(125,122)[lt]{D7}
\SetColor{Black}
\Text(75,155)[cb]{$r_5$}
\Text(155,75)[lc]{$r_6$}
\end{picture}
\caption{If we place the D7 brane(s) at any non-zero distance from the D3 branes, we break the conformal symmetry}\label{U1symm}
\end{center}
\end{figure}
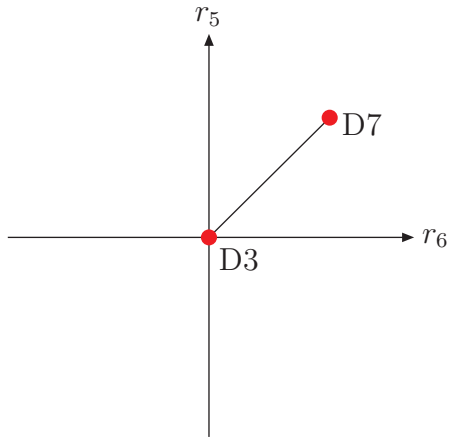

If the D3 and D7 stack are coincidental in the $r_5-r_6$ plane, there is a conformal symmetry in the system. If they are not coincidental, then we have introduced a scale into the system, and the conformal symmetry is broken; this is shown in figure \ref{U1symm}. In breaking the conformal symmetry, all the fermions and scalars of the $\mathcal{N}=2$ multiplet acquire a mass.

The exact embedding of $N_f$ D7-branes has been chosen so that the directions of $r_5$ and $r_6$ are orthogonal to the D7-branes' world-volume (table \ref{probetab}). There is now an $\mathcal{N}=2$ chiral supermultiplet connecting the D3 and D7 brane stack, and this interacts with the $\mathcal{N}=4$ supermultiplet on the surface of the D3 branes. How to know which symmetries are preserved and which are broken in some setup of Dp and Dp' branes is not an obvious one, and involves some detailed calculations \cite{Nastase:2003dd}. In this particular case, the $\mathcal{N}=4$ supermultiplet is broken completely by the presence of the D7 brane, leaving just the $\mathcal{N}=2$ supermultiplet. When the D7 brane is separated from the D3 stack, the scalar associated with the $r_5$ and $r_6$ directions gains a VEV and gives a mass to the quarks:{\be \bar{Q}\lambda_{56}Q \longrightarrow m\bar{Q}Q\hspace{1cm} \mathrm{as\;} \lambda_{56} \mathrm{\;acquires\; a \;VEV}\ee}
However when when we have separated the D3-D7 system, there is still a U(1) symmetry because of our freedom to re-parameterise the $r_5-r_6$ coordinates. This U(1) symmetry corresponds to the axial symmetry of a single-flavour QCD model. As explained in section \ref{ch:axialsym}, in QCD this symmetry is spontaneously broken by the chiral condensate $\langle\bar{q}q\rangle$. We don't see a Goldstone boson though, because of the presence of instantons. However, in the 't Hooft limit ($N \longrightarrow \infty$ with $g_{\mathrm{YM}}^2N$ fixed), the instanton effect is negated, and so the Goldstone boson becomes light once again.

\section{Particle spectrum in the probe limit}

The excitations of the D7-brane in the directions perpendicular to its world volume correspond to scalar fields, which are singlets under the colour group and are in the adjoint of the flavour group: these states include mesons.

Since the theory is supersymmetric, there will also be other states, including fermions, living on the brane, but these are not explored in this thesis.

The presence of supersymmetry also means that a fermion condensate cannot be expected to form (we check this explicitly in the next section). Hence chiral symmetry will not be spontaneously broken, as it is in QCD. Non-supersymmetric deformations of the correspondence can overcome this drawback, and in chapter \ref{ch:CM} we address a particularly important example of such a deformation.

\subsection{Quark mass and condensate}\label{sch:quarkmass}

In this section we analyse how the probe D7 brane lies when introduced into the D3-brane setup. We will show that its asymptotic solution is holographic to the quark mass and condensate.

The full action for a probe D7-brane in the D3-brane setup is{\be S_{D7}=-T_7 \int d^8 \xi \sqrt{-\mathrm{det}\left(P[G]+2\pi\alpha'F \right)}+ \frac{(2\pi\alpha')^2}{2} T_7\int P[C^{(4)}]\wedge F \wedge F \label{D7act}\ee}where $G$ is the ten dimensional metric, and $F$ is the field strength of the gauge field U(1) living on the D7 brane. $P[G]$ is the pullback of the ten dimensional metric onto the surface of the D7 brane. It is defined as{\be P[G] \equiv \frac{\partial x^{\alpha}}{\partial \xi^a}  \frac{\partial x^{\beta}}{\partial \xi^b} G_{ab} \ee}with $\alpha,\beta$ running over the coordinates on the D7 brane, and $a,b$ running over the whole spacetime's coordinates.

For now, we neglect the U(1) field and the Wess-Zumino term, leaving us with{\be S_{D7}=-T_7 \int d^8 \xi \sqrt{-\mathrm{det}\left(P[G] \right)} \label{simpD7act}\ee}We can neglect the Wess-Zumino term because it is only relevant for gauge fields with a vector index on the 5-sphere. In other words they carry a supersymmetry charge. Such states are important in the full dual, but for this thesis they are unimportant - we are trying to provide holographic descriptions of QCD, which does not (to the best of our knowledge) contain any supersymmetric states. We will re-introduce the U(1) field in section \ref{sch:vector}. It will turn out to be holographic to the vector meson sector.

The DBI action for (\ref{simpD7act}) is{\be -T_7 \int d^8 \xi \frac{\rho^3}{R^3} \sqrt{1+\left(\frac{\partial u_5}{\partial \rho}^2+\frac{\partial u_6}{\partial \rho}^2\right)+\frac{R^2}{u^2}\left(\frac{\partial u_5}{\partial x}^2+\frac{\partial u_6}{\partial x}^2\right)} \label{DBIact}\ee}where the ten-dimensional metric has been written as{\be ds^2 = \frac{u^2}{R^2}\eta_{\mu\nu}dx^{\mu}dx^{\nu} + \frac{R^2}{u^2}\left(d\rho^2+\rho^2d\Omega_3^2+du_5^2+du_6^2\right) \label{10Dmet}\ee}with $u^2=\rho^2+u_5^2+u_6^2$. The $x$ direction is perpendicular to the D7's world volume, so fluctuations in $x$ correspond to the meson spectrum. We come to this next, but first we address fluctuations in the $\rho$ direction - this governs how the probe brane lies within the spacetime. Using the U(1) freedom we have to re-parameterise $u_5$ and $u_6$, let us set $u_6=0$, and look at the equation of motion for $u_5$. Expanding the square root to second order, the Euler-Lagrange equation of motion is{\be \partial_{\rho} \left(\frac{\rho^3 \partial_{\rho}u_5}{\sqrt{1+\partial_{\rho}u_5^2}}\right) =0 \label{ELD7}\ee}whose solution tends to $m+c\rho^{-2}$ as $\rho \rightarrow \infty$.

From (\ref{DBIact}) we can see that $u_5$ has dimensions of mass. Hence the parameter $m$ corresponds to a mass for the $\bar{q}q$ operator, and the parameter $c$, with mass dimension three, corresponds to a VEV for the $\bar{q}q$ bilinear.

Now we can investigate whether this system does or does not allow a quark condensate. For the field theory to be consistent, we require two things of the D7 flow:

\begin{itemize}
\item The field theory must have a unique description at any given energy: hence the sum $u_5^2 + \rho^2$ must be monotonic as $\rho$ varies from 0 to $\infty$
\item The metric (\ref{10Dmet}) is symmetric under $\rho \rightarrow -\rho$, which implies that well-behaved solutions must obey $u_5'(0)=0$ and $u_5(0)\neq \pm\infty$. Were it otherwise, the brane would have a kink, which would give an infinite contribution to the action
\end{itemize}

\begin{figure}
\begin{center}
\includegraphics[height=6.4cm]{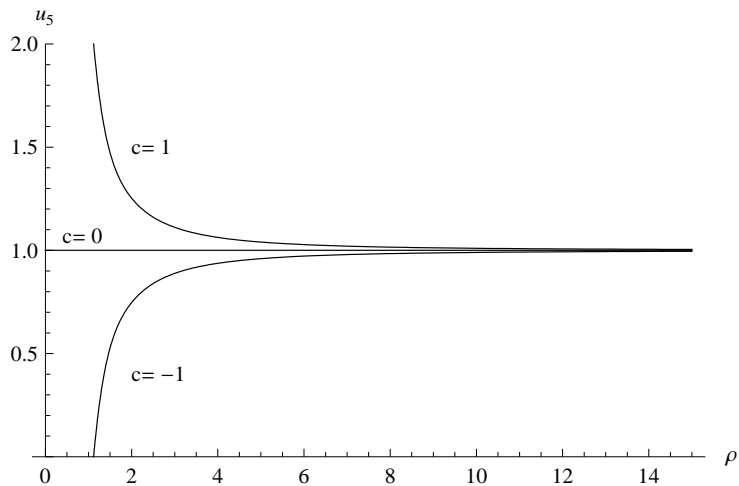}
\caption{D7-brane flows for $m=1$ with several different condensate values in $AdS_5 \times S^5$ background. Only the $c=0$ solution allows a consistent theory}\label{D7flow}
\end{center}
\end{figure}

Figure \ref{D7flow} shows three flows for $m=1$ with $c=-1,0$ and 1. As expected for a supersymmetric theory, a condensate is not allowed to form. 

If the reader is not satisfied with this numerical explanation, we can provide an analytical proof. (\ref{ELD7}) can be solved analytically, for all $\rho$. Its solution is{\bea u_5(\rho) =&& A + \frac{i^{2/3} B}{2.3^{1/4}\sqrt{B^6-r^6}} \:F(\phi,\frac{1}{4}(2+\sqrt{3}))\times \nonumber\\
&&\sqrt{(B^2+i^{2/3}r^2)(B^4-i^{2/3}B^2r^2+i^{4/3}r^4)}  \label{ellip}\eea}with $\phi = \arccos (\frac{B^2 - i^{2/3}(-1+\sqrt{3})r^2}{B^2+i^{2/3}(1+\sqrt{3})r^2})$. $F(\phi,m)$  is the incomplete elliptic integral of the first kind, using the convention defined in Mathematica version 5.2.

The function is complex unless $B=0$, meaning that the only well-behaved solution is when $u_5$ is equal to a constant. This is the solution with a zero valued condensate. By allowing $m \neq 0$ we have clearly broken the U(1) chiral symmetry that existed before. We showed in section \ref{ch:pert} that QCD breaks chiral symmetry both by the spontaneous formation of a chiral condensate, and to a lesser extent, by the mass of the quarks. In this pure AdS system, the only way we can break chiral symmetry is by making the quarks massive.

\subsubsection{Meson mass spectrum}\label{sch:mesonspec}

How to calculate the meson spectrum in this background has been detailed in \cite{Kruczenski:2003be}. We summarise the results here.

The meson spectrum corresponds to x-dependent fluctuations of the fields $u_5$ and $u_6$ on the D7-brane world volume. To find the spectrum, the mesons are treated as small fluctuations on the brane surface, meaning that the DBI action can be expanded to second order, with no interaction between the meson fields. Hence we can treat the fields as freely moving plane waves, which will allow us to explicitly calculate the spectrum.

There are two directions in which the fields $u_5$ and $u_6$ can fluctuate: perpendicular to the $\rho$-dependent flow of the brane, and parallel to the $\rho$-dependent flow. We use the U(1) symmetry to write this as{\be u_5=m+\chi(x,r) \hspace{1cm} u_6=\phi(x,r) \ee}The induced metric on the D7 brane can be written as{\be ds^2 = \frac{\rho^2+L^2}{R^2}\eta_{\mu\nu}dx^{\mu}dx^{\nu} + \frac{R^2}{\rho^2+L^2}d\rho^2+\frac{R^2\rho^2}{\rho^2+L^2}d\Omega_3^2 \ee}with $\rho^2=u^2-L^2$. $L$ is the distance between the centres of the D7 brane and the D3 brane stack. It is therefore proportional to the quark mass: the constant of proportionality is unimportant in this analysis. It just means that we will predict the masses of the mesons relative to one another, rather than relative to some absolute scale.

The DBI action is then given by{\be S_{\mathrm{DBI}} =-T_7 \int d^8\xi\: \sqrt{-\mathrm{det}G}\left(1+2(\pi\alpha'R)^2\frac{g^{cd}}{\rho^2+L^2}(\partial_c\chi \partial_d\chi+\partial_c\phi \partial_d\phi) \right)  \ee}The equations of motion for $\phi$ and $\chi$ are identical, so we write the generic field as $\Phi$, and its equation of motion is{\be \frac{R^4}{(\rho^2+L^2)^2}\partial^{\mu}\partial_{\mu}\Phi + \frac{1}{\rho^3}(\rho^3\partial_{\rho}\Phi)=0 \label{EoMmeson}\ee}We assume that the mesons will be non-interacting and plane-wave, so we make the ans\"atz $\Phi=f(\rho)e^{ik.x}$, with $k^2=-M^2$ ($M$ being the mass of the meson). We then look for normalisable solutions to (\ref{EoMmeson}): these correspond to consistent field theory solutions. 

(\ref{EoMmeson}) can be solved exactly by:{\be f(\rho)=(\rho^2+L^2)^{-\alpha} \hspace{0.1cm}{}_2F_1(-\alpha,-\alpha+1;2;-\rho^2/L^2) \ee}$_2F_1(a,b;c;z)$ is the regular hypergeometric function, defined as{\be _2F_1(a,b;c;z) \equiv \displaystyle\sum_{n=0}^{\infty}\frac{(a)_n(b)_n}{(c)_n}\frac{z^n}{n!} \ee}where $(a)_n$ is the rising factorial. $\alpha$ is as yet undetermined. We can determine it by considering our requirements on $f(\rho)$: the solution must be normalizable and real. To stop $(\rho^2+L^2)^{-\alpha} \hspace{0.1cm}{}_2F_1(-\alpha,-\alpha+1;2;-\rho^2/L^2)$ running off to $\pm \infty$ as $\rho \rightarrow \infty$ we need to terminate the series at order $\rho^{-2\alpha}$. A little maths shows that this can be done by setting{\be n=\alpha-1 \label{quant}\ee}in which case the hypergeometric function terminates at order $(\rho^2/L^2)^n$. Hence $\phi \sim \rho^{-2}$ as $\rho \rightarrow \infty$. So the final solution to (\ref{EoMmeson}) is{\be  f(\rho)=\frac{1}{(\rho^2+L^2)^{n+1}} \hspace{0.1cm}{}_2F_1(-n-1,-n;2;-\rho^2/L^2) \ee}The condition (\ref{quant}) means that the mass spectrum is discrete:{\be M^2_n=\frac{2L}{R^2}\sqrt{(n+1)(n+2)} \label{massspec}\ee}Equation (\ref{massspec}) shows that in pure $AdS_5 \times S^5$, the meson mass spectrum is proportional to the quark mass. This is at odds to what we'd expect in QCD: even if the quarks were massless we'd still expect the mesons of QCD to be massive. This is another sign that this geometry does not support a quark condensate. A meson in QCD gets the majority of its mass from interactions with the non-zero quark condensate: when this condensate is zero, its mass becomes proportional to the quark mass.

\subsubsection{Numerical Solutions}

We have been lucky in being able to solve (\ref{EoMmeson}) exactly: in the more complicated geometries to come, we will not be able to, and so it will be helpful to go over the alternative numerical technique, first detailed in \cite{Shock:thesis}, using an example where we can check that the answers are what we'd expect.

\begin{figure}
\begin{center}
\includegraphics[height=6.4cm]{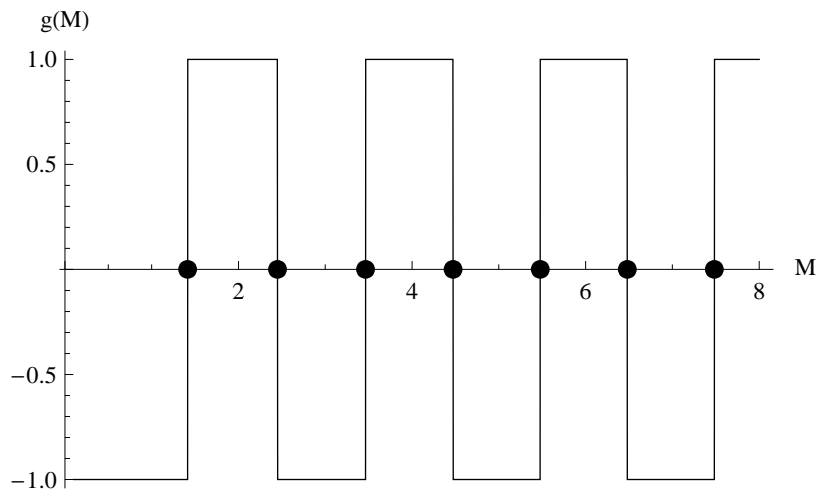}
\caption{Numerical calculation of the meson spectrum in $AdS_5 \times S^5$. The dots are the analytical values.}\label{numsol}
\end{center}
\end{figure}

The task is to find those values of $M_n$ in (\ref{EoMmeson}) for which $f_n(\rho)$ is a normalizable, real and smooth solution.
\begin{enumerate}
\item Define the numerical solution $f_n(\rho)$ as a function of $M_n$. To define the numerical solution requires us to set boundary conditions, so we make the appropriate choice in the UV: for the meson spectrum we require $f_n(\infty)=\rho^{-2}$. So now $f_n$ is a function of $M_n$ and $\rho$: $f_n=f_n(\rho,M)$
\item We now want those solutions which are finite in the IR and smooth across the origin: otherwise the solution will have a kink across $\rho=0$. We could do this by altering $M$ slightly each time, and then re-plotting the solution. A better method is to define a new function $g(M)= \mathrm{sgn}\:[\partial_{\rho}f(0,M)]$ and plot $g(M)$ over some suitable range.
\item Where the value of $g(M)$ changes sign indicates a stable flow: an example plot for (\ref{EoMmeson}) with $L=1, M=\sqrt{2}$ is shown in figure \ref{numsol}
\end{enumerate}

\begin{figure}
\begin{center}
\includegraphics[height=6.4cm]{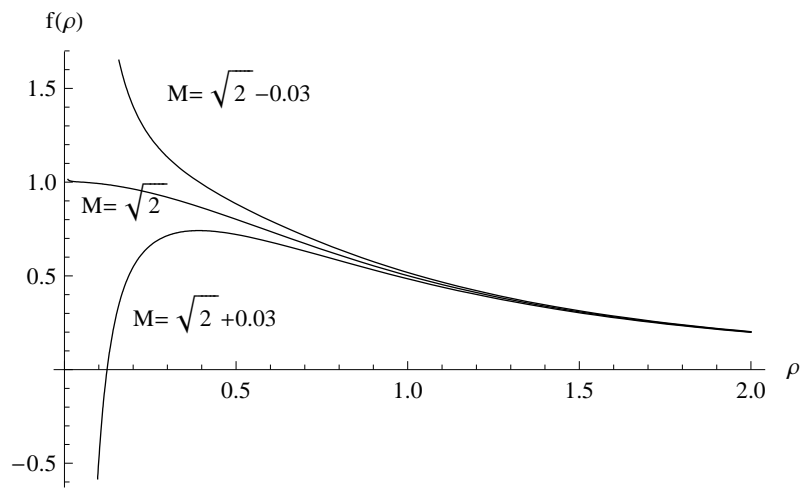}
\caption{Plot of three solutions around the $M=\sqrt{2}$ solution.}\label{truesol}
\end{center}
\end{figure}

Figure \ref{numsol} shows that the numerics agree with the analytical solution. This will be important - in future we will not be able to find an analytical solution, so we have to be confident that the numerical solutions can be trusted. To give an even clearer idea of what is happening around one of the normalizable solutions to (\ref{EoMmeson}) we plot $f(r)$ for three different values of $M$ around $\sqrt{2}$ in figure \ref{truesol}. It can be seen that as we pass through $M=\sqrt{2}$ the solution flips from $-\infty$ to $+\infty$, so the only normalizable solution in this neighbourhood is at $M=\sqrt{2}$.

\subsection{Vector Meson Spectrum} \label{sch:vector}

We analysed the scalar meson sector in section \ref{sch:mesonspec}. Now we re-introduce the U(1) gauge field that we neglected in equation (\ref{simpD7act}):{\be S_{D7}=-T_7 \int d^8 \xi \sqrt{-\mathrm{det}\left(P[g] +2\pi\alpha'F\right)} \ee}The process is identical to that in section \ref{sch:mesonspec} once we have made the following ans\"atz for the holographic gauge field:{\be A^{\mu}=g(\rho)e^{ik \cdot x}\epsilon^{\mu} \ee}$\epsilon^{\mu}$ is the polarization vector, defined only in the Minkowski directions. $g(\rho)$ is determined from the equations of motion, just as $f(\rho)$ was for the scalars.
In this geometry, the mass spectrum of the vector mesons is exactly the same as that for the scalars:{\be M^2_n=\frac{2L}{R^2}\sqrt{(n+1)(n+2)} \ee}This is to be expected since they both form part of an $\mathcal{N}=2$ hypermultiplet. In the more complicated geometries that we will study later the $\mathcal{N}=2$ hypermultiplet will be broken, and then the masses will differ.

\section{Breaking the Supersymmetry}
As it stands at the moment, supersymmetry is still exact, and hence a fermion bilinear condensate cannot be formed. In other words, chiral symmetry is not broken, which is a necessity for any holographic description of QCD. To induce a breaking of the supersymmetry, we must deform the gravity side in a non-supersymmetric and non-conformal way. There are many examples of such deformations \cite{Csaki:1998qr,Ooguri:1998hq,Zyskin:1998tg,Minahan:1998tm,Russo:1998by,Csaki:1999vb,Russo:1998mm,Csaki:1998cb,Hashimoto:1998if,Li:1998nw,Li:1998ce,Gross:1998gk,Klebanov:1998yya,Klebanov:1999ch,Ferretti:1998xu,Armoni:1999fb,Alishahiha:1999iz,Ferretti:1999gj,Minahan:1999yr,Kehagias:1999tr,Gubser:1999pk,Kehagias:1999iy,Nojiri:1999gf,Nojiri:1999sb,Nojiri:1999uh,deMelloKoch:1999hn,Nojiri:1998yx}. The one of most relevance to us is that considered in \cite{CM} and we address it next.

%% file: Files/CM.tex
\chapter{The Constable-Myers Geometry} \label{ch:CM}

We showed in chapter \ref{ch:flavour} that D7 probe branes can be used to include quarks in holographic duals. However, the resulting theory still had a massless quark condensate. Here we take a step back, and before introducing quarks, we consider a gravity dual where supersymmetry has already been broken.

\section{The Background}

The Constable-Myers geometry, first studied in \cite{CM}, is a consistent solution to the type IIB supergravity equations of motion. We state the solution first, and then describe its properties. In the Einstein frame, the geometry is given by:{\be ds^2 = H^{-1/2} f^{\delta/4} dx_{4}^2 + R^2 H^{1/2} f^{(2-\delta)/4} \frac{w^4 - b^4}{ w^4 } \sum_{i=1}^6 dw_i^2  \label{CMact}\ee}{\be \mathrm{where\;\;}  H =  f^{\delta} - 1,\hspace{0.5cm} f=\frac{w^4+b^4}{w^4-b^4} \ee}The dilaton and four-form are{\be e^{2 \phi} = e^{2 \phi_0} f^{\Delta}, \hspace{0.5cm} C^{(4)}=-\frac{1}{4} H^{-1} dt\wedge dx \wedge dy\wedge dz\ee}There are formally two free parameters, $R$ and $b$, since {\be \delta = R^4/ 2 b^4, \hspace{0.5cm}  \Delta^2 = 10 - \delta^2  \ee}The metric (\ref{CMact}) has a singularity at $\omega^4=b^4$; we loosely expect this to correspond to the presence of the central stack of D3 branes \cite{Evans:2004ia}. 

The dilaton and five-form field strength are a function of $r$, and in the UV limit ($r\rightarrow \infty)$ the geometry returns to $\mathrm{AdS}_5 \times \mathrm{S}^5$. However, due to the dependence on $r$, the interior of the field theory is strongly deformed.

The set of solutions are described by two dimensionful parameters, $b$ and $R$. However, by looking at (\ref{CMact}) we can see that if $b=0$, then the solution returns to $\mathrm{AdS}_5 \times \mathrm{S}^5$, with radius $R$. Hence $b$ determines the scale of the conformal symmetry breaking. We can identify $b$'s holographic partner explicitly by looking at the form of the dilaton in the UV limit ($r \rightarrow \infty$):{\bea e^{\phi} &\sim& e^{\phi_0}\left(1+\frac{\sqrt{40b^8-R^8}}{2r^4} \right) \nonumber\\
&\equiv&  e^{\phi_0}\left(\frac{A}{r^4} \right) \label{meanb}\eea}$A$ has scaling dimension 4 and is a singlet under the 5-sphere isometry. Therefore it must correspond to an operator of dimension 4 which is an R-singlet. There is only one candidate (as pointed out in section \ref{sch:field-op}, we are guaranteed to have such a unique candidate). It must be $\langle \mathrm{Tr}\:F^2\rangle$. This deformation completely breaks the supersymmetry of the gauge theory (ie. from $\mathcal{N}=4$ to $\mathcal{N}=0$). In addition, a complex VEV does not make much sense, so $b$ must be greater than or equal to $40^{-1/8}R$. Below this, the geometry is not well-defined.

Meanwhile $R$ determines $g_{\mathrm{YM}}^2N$ in the field theory, as is usual in the correspondence. So $R$ specifies if the field theory is in the strongly coupled or weakly coupled regime.

\section{Confinement and Glueballs in the Constable-Myers Geometry}

Before we introduce quarks into this geometry, it is worth pausing to consider some of the properties already present in this theory, namely confinement and the glueball spectrum.

\subsection{Confinement}
 
Confinement means that quarks are consigned to exist in colour-neutral combinations. A single quark cannot exist on its own. One way of showing this is to calculate the energy of a quark-antiquark as a function of their separation. In QED, the energy between a positron-electron pair would monotonically decrease to zero as we separated the two. In QCD, we'd expect the energy to increase as we tried to separate a quark-antiquark pair, showing that it would take an infinite amount of energy to produce a solitary quark.

In \cite{Maldacena:1998im} it is argued that this effect can be studied holographically by looking at the Nambu-Goto action for a fundamental string in the geometry dual to the field theory. Specifically, the expectation value of a Wilson loop in the field theory is dual to the fundamental string in the holographic background:

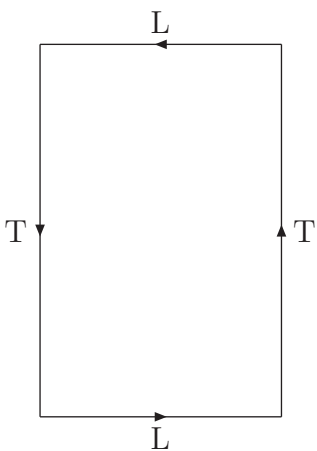
\begin{figure}
\begin{center}
     \begin{picture}(110,160) \Text(1,80)[]{T}
                \ArrowLine(10,10)(100,10)  \Text(109,80)[]{T}
                \ArrowLine(100,10)(100,150)  \Text(55,2)[]{L}
                \ArrowLine(100,150)(10,150)  \Text(55,159)[]{L}
                \ArrowLine(10,150)(10,10)  
      \end{picture}
\caption{Wilson loop used to calculate the quark-antiquark force. The vertical direction represents Euclidean time. The horizontal direction indicates one of the spatial coordinates} \label{wilson}
\end{center}
\end{figure}

{\be \langle W(\mathcal{C})\rangle \sim e^{-S} \label{q-aqsep}\ee}where $W(\mathcal{C})$ is the Wilson loop around some closed contour $\mathcal{C}$: $W(\mathcal{C})=\frac{1}{N}\mathrm{Tr}\mathcal{P}e^{i\int_{\mathcal{C}}A}$. $\mathcal{P}$ means that the calculation is path-ordered. $S$ is the action of the worldsheet of the string. 

The LHS of (\ref{q-aqsep}) needs further explanation. In Euclidean field theory with a rectangular contour with sides of length $T$ and $L$ (figure \ref{wilson}), the energy of a quark-antiquark pair can be found by calculating the expectation of the Wilson loop in the $T\rightarrow \infty$ limit \cite{Maldacena:1998im}:{\be \langle W(\mathcal{C})\rangle \sim e^{-TE(L)} \label{CMquark}\ee}where $E(L)$ is the energy of the quark-antiquark pair. A good analysis of equation (\ref{CMquark}) can be found in \cite{Kaku:1993ym}. Here we summarise the result. We start by defining the two quark state at time $t$ by:{\be |\Gamma(t,R)\rangle \equiv |\bar{q}(t,0)q(t,R)\rangle \ee}Then we need to consider{\bea \displaystyle\lim_{T\rightarrow \infty} \Omega(T,R) &\equiv& \displaystyle\lim_{T\rightarrow \infty} \langle q(T,0)\bar{q}(T,R)|\bar{q}(0,0)q(0,R)\rangle \nonumber \\
&=&   \displaystyle\lim_{T\rightarrow \infty} \langle \Gamma^{\dagger}(T,R)\Gamma(0,R)\rangle \nonumber \\
&=&  \displaystyle\lim_{T\rightarrow \infty} \displaystyle\sum_n |\langle \Gamma^{\dagger}(0,R)\rangle|^2 e^{-E_nT} \nonumber \\
&\sim& e^{-E_0(R)T} \label{wilsonloop}\eea}We have been able to pick out the smallest energy eigenvalue $E_0$ in (\ref{wilsonloop}) because in the $T\rightarrow \infty$ limit, it will dominate all the other terms.

All we have left to do to demonstrate (\ref{CMquark}) is to show that $\displaystyle\lim_{T\rightarrow \infty} \Omega(T,R) \sim \displaystyle\lim_{T\rightarrow \infty} W(\mathcal{C})$. However, the Wilson loop corresponds to an extended `source' in the fundamental representation (ie. quarks), separated by a distance $R$, being created at $T=0$ and annihilated at some general $T$, so we are done.

Let us apply equation (\ref{q-aqsep}) to the Constable-Myers geometry considered in this chapter. The Nambu-Goto action of the fundamental string is \cite{CM}:{\be S = \frac{1}{2\pi\alpha'} \int d^2 \sigma \sqrt{-\mathrm{det}G_{MN}\partial_ax^M \partial_b x^N} \ee}with $G_{MN}$ denoting the string frame metric, which is related to the Einstein metric by a scaling factor:{\bea ds^2_{\mathrm{string}}&=&e^{\phi/2}ds^2_{\mathrm{Einstein}} \label{CMnosphere}\\
&=& H^{-1/2} f^{(\Delta+\delta)/4} dx_{4}^2 + R^2 H^{1/2} f^{(\Delta+2-\delta)/4} \frac{w^4 - b^4}{ w^4 } \sum_{i=1}^6 dw_i^2 \nonumber\eea}It will be useful to recast (\ref{CMnosphere}) with an explicit 5-sphere symmetry:{\be ds^2=  H^{-1/2} f^{(\Delta+\delta)/4} dx_{4}^2 + R^2 H^{1/2} f^{(\Delta+2-\delta)/4} \frac{w^4 - b^4}{ w^4 } \left(dr^2 + r^2 d\Omega_5^2 \right)\ee}We want a static solution to the equations of motion, so it is convenient to set $\sigma^0=t$ and $\sigma^1=x$, and assume $y=z=0$. We also assume that the string is at a constant angular position in $S^5$. Then the action becomes{\be S=\frac{1}{2\pi\alpha'} \int dt\;dx\;n(r) \left[1+m(r)\left(\frac{dr}{dx}\right)^2\right]^{1/2} \label{CMoptim}\ee}with{\bea n(r)&=& G_{xx}= H^{-1/2} f^{(\Delta+\delta)/4} \nonumber\\
m(r) &=& G_{xx}^{-1}G_{rr} = H \frac{w^4 - b^4}{ w^4 } f^{(1-\delta+\Delta)/2} \nonumber\eea}Equation (\ref{CMoptim}) is very similar to the integral found when performing Fermat's Principle for the path of light rays \cite{Gubser:1999pk}. For large separation of endpoints, the trajectory which minimizes $S$ locates itself very nearly at the minimum of $n(r)$ for most of its length. Hence the complicated process of optimizing $S$ is reduced to simply optimizing $n(r)$. Furthermore, for small separations $\Delta x$ about its minimum, $S$ becomes{\be S \simeq \frac{n(r_{\mathrm{min}})}{2\pi\alpha'}\Delta x \ee}It is easy to show that $n(r)$ is optimized by {\be r^4_{\mathrm{min}}=\frac{2\omega^4}{\left(\frac{\Delta +\delta}{\Delta -\delta}\right)^{1/\delta}-1} \ee}which gives{\be V(\Delta x) = S(\Delta x)= \frac{\Sigma^{\frac{\delta-\Delta}{4\delta}}}{2\pi\alpha'\sqrt{1-\Sigma}} \Delta x \label{linearpot}\ee}with $\Sigma=\frac{\Delta-\delta}{\Delta+\delta}$. (\ref{linearpot}) shows us that this geometry gives us a linear quark-antiquark potential. In other words, it exhibits confinement.

\subsection{Glueballs}

The glueball spectrum for this geometry was investigated in \cite{CM}. The procedure for calculating holographic glueball spectra is quite straightforward \cite{Csaki:1998qr,Ooguri:1998hq,deMelloKoch:1998qs,Zyskin:1998tg}: it is found by solving the equation of motion for the scalar field $\Phi$:{\be \partial_{\mu}\left( \sqrt{g}g^{\mu\nu}e^{-2\phi}\partial_{\nu}\Phi\right) =0 \label{glueballs}\ee}with the metric given in equation (\ref{CMact}). Making the ans\"atz $\Phi=g(\rho)e^{ik.x}$ (since the scalar field is assumed to be a small perturbation) we then look for solutions to (\ref{glueballs}). This can be done numerically. As $r \longrightarrow \infty$, the metric asymptotes to $AdS$ space, and the normalizable solution to the wave equation goes like $r^{-4}$. Hence we use the boundary conditions $g(\Lambda)=\Lambda^{-4}, g'(\Lambda)=-4 \Lambda^{-5}$ we then look for those solutions which obey $f'(0)=0$ (for continuity). We find a discrete spectrum (table \ref{CMglueball}). In other words, this geometry has a mass gap in the glueball sector.

\begin{table}
    \begin{center}  
       \begin{tabular}{c|c}
            Glueball State   & Mass   \\
          \hline
       $0^{++}$ & 1.00 \\
 $0^{++*}$ & 1.56 \\
 $0^{++**}$ & 2.08 \\
 $0^{++***}$ & 2.60 \\
 $0^{++****}$ & 3.11 \\
         \end{tabular} 
    \end{center}
\caption{$0^{++}$ glueball masses from Constable Myers geometry. Normalization is such that the ground state mass is set to one (arbitrary units).} \label{CMglueball}
  \end{table}

\section{Chiral Symmetry Breaking}

The most exciting aspect of this geometry is that once we introduce quarks, a quark condensate forms spontaneously. This is an essential characteristic of QCD.

We start by introducing quarks in the same way we did in chapter \ref{ch:flavour}, by adding $N_f$ probe D7 branes to the geometry. The exact dimensions occupied by the D7 and D3 branes are shown in table \ref{CMtab}.

We make a change of variables in the two directions perpendicular to both branes:{\be \Phi = r_5 + i r_6 = \sigma e^{i\theta} \ee}
and introduce one flavour of quark via a D7 brane probe in the geometry. The Dirac Born Infeld (DBI) action for the probe is{\be S_{D7} = -\frac{1}{(2\pi)^7\alpha'^4 g_s} \int d^8 \xi e^{\phi}\sqrt{-\mathrm{det(P[G_{ab}])}} \label{DBIact-theory}\ee}where P indicates the pullback of the spacetime metric onto the D7 world volume. The pullback of a metric is defined as{\be \mathrm{P[G_{ab}]}= \frac{\partial x^{\alpha}}{\partial \xi^a}\frac{\partial x^{\beta}}{\partial \xi^b} G_{\alpha\beta} \ee}
with $x_0 \rightarrow x_9$ being the coordinates in the bulk, and $\xi_0 \rightarrow \xi_7$ being the coordinates on the D7 brane. We choose the static gauge, which simply means $x_i=\xi_i$ for $i=1\rightarrow 7$, and $x_8=x_8(\xi_0,\ldots,\xi_7),x_9=x_9(\xi_0,\ldots,\xi_7)$. This action will determine how the D7 lies in the remaining $r_5-r_6$ directions.\\

Performing the necessary calculation we get{\bea S_{\mathrm{D7}}&=&-\frac{1}{(2\pi)^7\alpha'^4g_s} \int d^8 \xi\left(\frac{\omega^4+b^4}{\omega^4-b^4} \right)^{\frac{\Delta}{2}} \rho^3 \left(\frac{\omega^8-b^8}{\omega^8}\right) \nonumber \\\nonumber\\
&\times& \sqrt{1+|\partial_{\rho}\Phi|^2 + H \left(\frac{\omega^4+b^4}{\omega^4-b^4} \right)^{\frac{4-\delta}{8}} \frac{\omega^4-b^4}{\omega^4} |\partial_x\Phi|^2} \eea
with $\omega^2 = \rho^2 + |\Phi|^2= \rho^2 +\omega_5^2 +\omega_6^2$.

\begin{table}
    \begin{center}  
       \begin{tabular}{c|c|c|c|c|c|c|c|c|c|c}
           & $x_0$   & $x_1$    & $x_2$    & $x_3$    & $r$     &  $\omega_1$ & $\omega_2$& $\omega_3$& $r_5$& $r_6$\\
          \hline
       D3 & $\times$ & $\times$ & $\times$ & $\times$ & .        &.         &.         &.         &.      &.\\
       D7 & $\times$ & $\times$ & $\times$ & $\times$ & $\times$ & $\times$ & $\times$ & $\times$ &.      &.\\
         \end{tabular} 
    \end{center}
\caption{Choice of D3 and D7 embedding. Filled directions are marked with a cross. Unfilled directions by a dot.} \label{CMtab}
  \end{table}

Applying the Euler-Lagrange equation of motion for $\Phi$ we find{\be \frac{d}{d \rho} \left(\frac{e^{\phi}\mathcal{G}(\rho,\Phi)}{\sqrt{1+|\partial_{\rho}\Phi|^2}}\partial_{\rho}\Phi \right) - 2\sqrt{1+|\partial_{\rho}\Phi|^2} \frac{d}{d \bar{\Phi}} \left(e^{\phi} \mathcal{G}(\rho,\Phi)\right)=0 \label{CMprobe}\ee}{\be \mathrm{with} \hspace{0.2cm} \mathcal{G}(\rho,\Phi)= \rho^3 \frac{((\rho^2+|\Phi|^2)^2+1)((\rho^2+|\Phi|^2)^2-1)}{(\rho^2+|\Phi|^2)^4} \label{CMfuncG}\ee}There are two important points to be noted in equation (\ref{CMprobe}).

Firstly there is a an explicit $\Phi \rightarrow e^{i\alpha} \Phi$ symmetry. In the massless case this is the $U(1)_A$ symmetry on the quarks.\\

Secondly, in the UV limit ($ r \rightarrow \infty$), the equation of motion simplifies to{\be \frac{d}{d\rho} \left( \rho^3 \frac{d\Phi}{d \rho} \right)=0 \ee}which has the simple solution $\Phi = m + c \rho^{-2}$. The two integration constants, $m$ and $c$, correspond to a quark mass and a VEV for the quark bilinear $\bar{q}q$ respectively.

Returning to the full equation of motion, we look for regular solutions that have the UV boundary condition $\sigma = m + c \rho^{-2}$. (For the moment $\theta$ is turned off: it corresponds to the pion fields, which we will demonstrate later). A few of the solutions are plotted in figure \ref{chiralCM}.

\begin{figure}
\begin{center}
\includegraphics{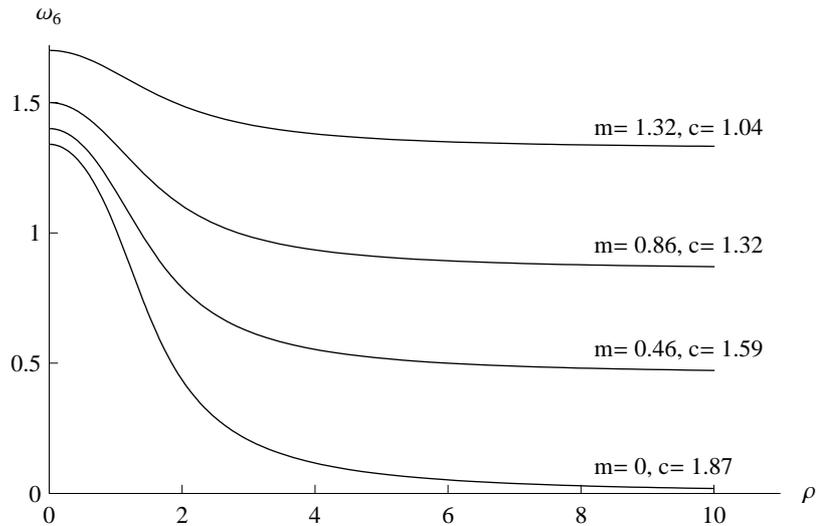}
\caption{Solutions for the $\omega_6$ flow when $\omega_5=0$ showing the dependence of the condensate on the quark mass}\label{chiralCM}
\end{center}
\end{figure}

The key feature of figure \ref{chiralCM} is the $m=0, \:c=1.86$ curve. This shows that this geometry demonstrates chiral symmetry breaking: there is a quark condensate even if the quarks are massless! This makes the Constable-Myers geometry a very interesting model to study other properties of QCD.

\section{Pions and their Interactions}

We now turn our attention to the presence of Goldstone bosons. Numerical solutions discussed in the last section have demonstrated that this system dynamically breaks the classical $U(1)_A$ symmetry by forming a quark condensate. We first met Goldstone's theorem \cite{Goldstone:1962es} in chapter \ref{ch:pert}, and in that section we analysed the spontaneous breaking of the $U(1)_L \times U(1)_R$ symmetry to $U(1)_V$. We identified the $\eta'(958)$ meson as the relevant Goldstone boson. However, matters were complicated by the presence of instantons, which made the $\eta'(958)$ heavier than would otherwise be expected. Here, at large N, there are no instantons and so the equivalent $\eta'$ meson should still be light.

Henceforth we will describe the pseudo-Goldstone boson resulting from the spontaneous breaking of the $U(1)_A$ symmetry of this geometry as a pion.

\subsection{The Pion Equation Of Motion}

In the case of one flavour, the pion is simply a bound $\bar{q}q$ state, and will be described by the field $\Phi$. The Goldstone fluctuation corresponds to oscillations of the field along the vacuum manifold, which in the chiral limit means fluctuations in the angular $\theta$ direction of $\Phi$. To leading order, if we have a background configuration for the D7 brane described by $\sigma_0$, we can look at small fluctuations in the $\theta$ direction.\\

The action for these fluctuations (to quadratic order) is given by expanding the DBI action (\ref{DBIact-theory}):{\bea S &=& \frac{2\pi^2}{(2\pi)^7\alpha'^4g_s} \int d\rho \: d^4x \: e^{\phi} R^4 \mathcal{G}  \sqrt{1+(\partial_{\rho}\sigma_0)^2} \nonumber\\
&\times& \left(1+ \frac{1}{2}\frac{g^{\rho\rho}g_{\theta\theta}(\partial_{\rho}\theta)^2}{1+(\partial_{\rho}\sigma_0)^2}+\frac{1}{2}\frac{g^{\mu\nu}g_{\theta\theta}\partial_{\nu}\theta\partial_{\mu}\theta}{1+(\partial_{\rho}\sigma_0)^2} \right) \eea}And the resulting Euler-Lagrange equation of motion is:{\bea \frac{M^2R^2 e^{\phi}\mathcal{G}}{\sqrt{1+(\partial_{\rho}\sigma_0)^2}}H\left(\frac{(\rho^2+\sigma_0^2)^2+1}{(\rho^2+\sigma_0^2)^2-1} \right)^{(1-\delta)/2} \frac{(\rho^2+\sigma_0^2)^2-1}{(\rho^2+\sigma_0^2)^2}\sigma_0^2 f &&\nonumber \\
 +\frac{d}{d\rho} \left(\frac{e^{\phi}\mathcal{G}}{\sqrt{1+(\partial_{\rho}\sigma_0^2)^2}}\sigma_0^2 \partial_{\rho}f \right)=0&&\label{CMpion}\eea}We look for values of $MR$ as a function of $m$ that give regular solutions to (\ref{CMpion}). The UV boundary condition we impose is $f=1/\rho^2$, which reflects the fact that pion has the same UV scaling dimension as $\bar{q}q$. The results are plotted in figure \ref{goldCM}. There is a massless pion when $m_q=0$, in accordance with Goldstone's theorem.

\begin{figure}
\begin{center}
\includegraphics{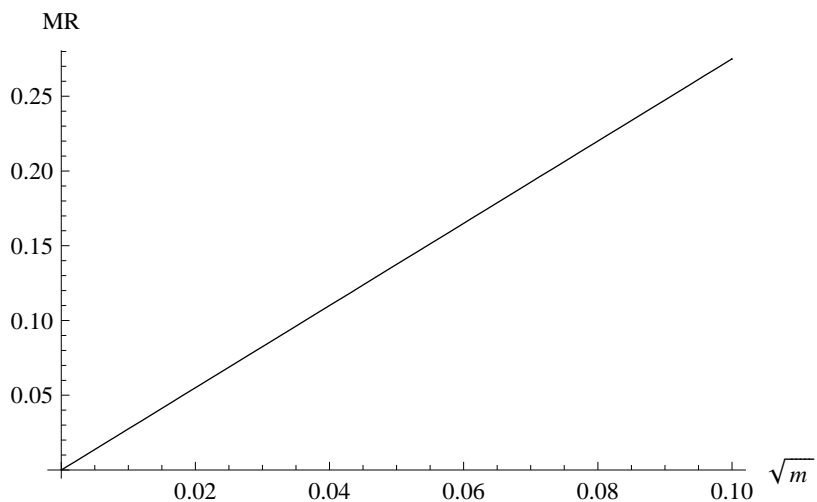}
\caption{Quark mass versus meson mass, showing the massless pion when $m_q=0$. We plot the square root of the quark mass to demonstrate the linear relationship.}\label{goldCM}
\end{center}
\end{figure}

\subsection{The Vector Meson Equation Of Motion}

There is an additional field present on the D7 brane which we have left undiscussed: it is the gauge field partner of the scalar $\Phi$. The field strength of this gauge field enters in the standard way as $2\pi\alpha'F^{ab}$ in the square root. The leading Lagrangian for this field, in a background configuration $\sigma_0$ is{\bea \mathcal{L} &=& \frac{2\pi^2R^4}{(2\pi)^7\alpha'^4g_s}\int d\rho \: e^{\phi}\:\mathcal{G} \Big( \sqrt{1+(\partial_{\rho}\sigma_0)^2}H \left(\frac{(\rho^2+\sigma_0^2)^2-1}{(\rho^2+\sigma_0^2)^2+1} \right)^{\frac{1}{4}}(2\pi\alpha')^2\frac{1}{4}F^{\mu\nu}F_{\mu\nu} \nonumber \\
&+& \frac{1}{2R^2} \frac{1}{\sqrt{1+(\partial_{\rho}\sigma_0)^2}}\left( \frac{(\rho^2+\sigma_0^2)^2-1}{(\rho^2+\sigma_0^2)+1} \right)^{\frac{1}{2}}\frac{(\rho^2+\sigma_0^2)^2}{(\rho^2+\sigma_0^2)^2-1}F^{\mu\rho}F_{\mu\rho} \label{CMvectEoM} \eea}There is an additional term that could be added to the DBI action: the Wess-Zumino term, which gives the coupling of the four-form $C^{(4)}$ to the gauge fields. It is not included here because when we calculate the equations of motion for the gauge fields, this term is only relevant for the gauge fields with a vector index on the $S^3$. We are only interested in states that carry no $SO(4)$ R-charge. We write the gauge field as{\be A^{\mu} = g(\rho) \sin(kx)\epsilon^{\mu} \ee}The resulting Euler-Lagrange equation of motion is then{\bea && e^{\phi}\mathcal{G} \sqrt{1+(\partial_{\rho}\sigma_0)^2} M^2R^2g(\rho)H \left(\frac{\omega^4-1}{\omega^4+1} \right)^{\frac{1}{4}} \nonumber\\
&+& \partial_{\rho} \left(\frac{e^{\phi}\mathcal{G}}{\sqrt{1+(\partial_{\rho}\sigma_0)^2}}\partial_{\rho}g(\rho)\frac{\omega^4}{\sqrt{(\omega^4-1)(\omega^4+1)}} \right) =0 \label{CMvect}\eea}There are two asymptotic solutions to equation (\ref{CMvect}). The simplest is $M=0, \:g(\rho)=\mathrm{\:constant}$. This corresponds to introducing a background gauge field associated with $U(1)$ baryon number in the field theory. It is of little interest to us here.

The second solution is $g(\rho) \sim 1/\rho^2$, and this has the right dimension and symmetries to be identified as dual to the operator $\bar{q}\gamma^{\mu}q$. By seeking smooth solutions to the equation of motion we can determine the vector mass spectrum. The results are shown in table \ref{CMvecttab}, and compared to the equivalent result in the pure $AdS_5 \times S^5$ spacetime, which we calculated in section \ref{sch:vector}.

\begin{table}[t]
    \begin{center}  
       \begin{tabular}{|c|c|c|}\hline
$n$ & AdS case &CM case \\ \hline
0&2.83&2.16\\
1&4.90&4.85\\
2&6.93&7.05\\
3&8.94&9.20\\
4&11.0&11.3\\  \hline 
         \end{tabular} 
    \end{center}
\caption{Vector meson spectrum comparing CM and pure AdS backgrounds} \label{CMvecttab}
  \end{table}

\section{$m_{\rho}$ vs. $m_{\pi}^2$}

We can compare the dependence of the rho meson mass on the pion mass squared. This is an interesting thing to do because the same result has been computed for large $N$ in lattice simulations \cite{Erdmenger:2007cm}, so a direct comparison of gauge/gravity and lattice results is possible. We show the result in figure \ref{mpivsmrho}.

\begin{figure}
\begin{center}
\includegraphics[scale=0.55]{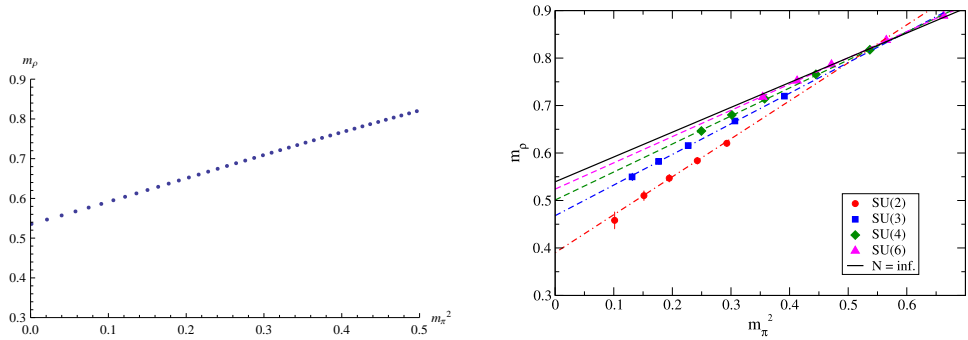}
\hspace{0.5cm}
\includegraphics[scale=0.25]{mrhompilat.epsi}
\caption{A plot of $m_\rho$ vs $m_\pi^2$ in the Constable-Myers background on the left. Lattice data \cite{Erdmenger:2007cm} (preliminary, quenched and at finite spacing) for the same quantity is also shown on the right.}
\label{mpivsmrho}
\end{center}
\end{figure}

We have chosen units in the Constable-Myers geometry such that we can compare directly with the lattice results of \cite{Erdmenger:2007cm}. It is striking that not only does the lattice data display the same linearity as the gauge/gravity result, but that the slopes are so similar (0.52 and 0.57 respectively). This is one of many examples where AdS predictions match QCD much better than one would naively expect. In the next section we explore some more predictions.

%% file: Files/AdS-QCD.tex
\chapter{AdS-QCD}
\label{ch:AdS-QCD}

\section{Premise}

In 2005 two very interesting papers \cite{son1,DaRold:2005zs} were published that for the first time dispensed with trying to deform supergravity solutions to mimic QCD. Instead the authors aim was to ``start from QCD and attempt to construct its five-dimensional holographic dual''. Since QCD has an infinite number of operators, they had to be careful about which operators they were going to try and include. \cite{son1} focused on holographically modelling the dynamics of chiral symmetry breaking in QCD. To this end, we would expect only a small number of operators to be influential. These would be

\begin{itemize}
\item the left-handed and right-handed currents corresponding to the $SU(N_f)_L \times SU(N_f)_R$ chiral flavour symmetry
\item the chiral order parameter
\end{itemize}

So in keeping with the unique operator-field matching, this would require three fields on the five-dimensional holographic side. Their details are given in table \ref{oper-field}.

\begin{table}
\begin{center}
\begin{tabular}{ccccc}
4D:$\mathcal{O}(x)$ &  5D: $\phi(x,r)$    &   p  & $\Delta$   & $m_5^2$  \\
	\hline
$\bar{q}_L\gamma^{\mu}t^a q_L$  & $A^a_{L\mu}$ & 1 & 3  & 0 \\
$\bar{q}_R\gamma^{\mu}t^a q_R$  & $A^a_{R\mu}$ & 1 & 3  & 0\\
$\bar{q}^{\alpha}_Rq^{\beta}_L$  & $(2/z) X^{\alpha\beta}$ & 0 & 3 & -3 \\
\hline
\end{tabular}
\caption{Operators/fields of the model of \cite{son1}}\label{oper-field}
\end{center}
\end{table}

The choice of $m_5$ for each field is determined by the relation $(\Delta-p)(\Delta+p-4)=m_5^2$ which we discussed in section \ref{sch:field-op}. The metric, also an ans\"atz, is simply $AdS_5$, which is the simplest choice we could make. In keeping with \cite{son1}, in this section we will write the fifth dimension in terms of $z$, instead of the usual $r$. The two are related by $z=r^{-1}$. Hence $z\sim0$ is the ultraviolet part of the spacetime, and $z\sim z_m$ is the infrared of the spacetime.{\be ds^2 = - z^{-2}dz^2 + z^{-2} \eta_{\mu\nu}dx^{\mu}dx^{\nu} \label{holomet}\ee}The use of an AdS geometry implies that the dual theory will be conformal - not something we want in a dual to QCD. So to give the theory a mass gap, the spacetime is defined to end at some IR scale $z=z_m$. In the window $0 < z \leqslant z_m$, the background gauge is conformal, and the coupling doesn't run.

The 5D action is{\be S=\int d^5x \: \sqrt{g}\: \mathrm{Tr} \left\{|DX|^2+3|X|^2-\frac{1}{4g_5^2}(F_L^2+F_R^2) \right\} \label{holoact}\ee}where $D_{\mu}X=\partial_{\mu}X-iA_{L\mu}X+iXA_{R\mu}$, $A_{L,R}=A^a_{L,R}t^a$, and $F_{\mu\nu}=\partial_{\mu}A_{\nu}-\partial_{\nu}A_{\mu}-i[A_{\mu},A_{\nu}]$.$X$ is a Higgs-like field, holographically dual to the quark mass and condensate. Its value in the UV will be:{\be X_0(\epsilon) \longrightarrow 2 m_q\epsilon + 2\epsilon^3\sigma \;\;\;\mathrm{as}\;\;\; \epsilon \longrightarrow 0 \ee}where $m_q$ is the quark mass matrix, and $\sigma$ is the quark condensate matrix. Both are diagonal matrices.

There are now four free parameters: $z_m,m_q,\sigma,g_5$. $g_5$ can be fixed by operator product expansion for the product of currents, and we come to this a little later. We also take this opportunity to focus on just two flavours of quark, so in table \ref{oper-field}, $\alpha,\beta=1,2$; $a,b=1,2,3$; and $t^a=\sigma^a/2$, where $\sigma^a$ are the Pauli matrices. This model with three flavours of quark have been investigated \cite{Shock:2006qy}, and the results are good. Here though, the third flavour would provide an unnecessary complication. Models beyond three flavours would be futile because the lightest three quark flavours are of roughly the same mass and thus have an approximate SU(3) symmetry. The charm quark however is much heavier than the previous three quarks, and so the SU(4) symmetry is badly broken.

As with the AdS/CFT correspondence, we claim that precise correspondence is obtained by equating the generating functional of the connected correlators in the four-dimensional theory with the effective action of the five-dimensional theory, with the UV boundary values of the 5D bulk fields set to the value of the sources in the 4D theory:{\be \langle\mathrm{Exp}\left(\int d^4x \phi_0(x) \mathcal{O}(x) \right)\rangle\Big|_{\phi(x,0)=\phi_0(x)} \cong \mathcal{Z}_{4D} \label{partfn}\ee}

\newpage
\section{Fixing the $g_5$ coupling} \label{g5match}

The idea in this section is to calculate the vector current two-point function in the holographic scheme, and then by equating the result to that obtained from perturbative QCD we can hence fix $g_5$.

We introduce the vector field as $V=(A_L+A_R)/2$, choose the gauge $V_r(x,r)=0$, and by using the Euler-Lagrange equation on the holographic action (\ref{holoact}), we obtain an equation of motion:{\be \big[\partial_z\left(\frac{1}{z}\partial_zV^a_{\mu}(q,z) \right)+ \frac{q^2}{z}V^a_{\mu}(q,z) \big]_{\perp} =0 \label{sonvect}\ee}We then substitute (\ref{sonvect}) back into the action (\ref{holoact}), and we are left with the boundary term{\be S=-\frac{1}{2g_5^2} \int d^4 x \left(\frac{1}{z} V_{\mu}^a \partial_z V^{\mu a} \right)\Big|_{z=\epsilon} \label{bdyact}\ee}So, by the claim made above (\ref{partfn}), equation (\ref{bdyact}) is the generating functional for QCD. To obtain the vector current two-point function, we simply functionally differentiate (\ref{bdyact}) twice with respect to the source $V_0$, with $V_0$ defined in $V^{\mu}(q,z)=V(q,z)V_0^{\mu}(q)$. ($V^{\mu a}_0(q)$ is the Fourier transform of the source of the vector current $J^a_{\mu}=\bar{q}\gamma_{\mu}t^aq$).Doing the necessary maths, we get{\bea \int d^4x \langle J_{\mu}^{a}(x) J_{\nu}^b(0)\rangle &=& \delta^{ab}\left(q_{\mu}q_{\nu}-q^2g_{\mu\nu} \right) \Pi_V(Q^2) \\
\Pi_V(-q^2)&=& -\frac{1}{g_5^2 Q^2}\frac{\partial_z V(q,\epsilon)}{\epsilon} \label{propo}\eea}For large Euclidean $Q^2 (\equiv-q^2)$ we only need to know $V(q,r)$ near the boundary,{\be V(Q,z)=1+\frac{Q^2z^2}{4} \ln(Q^2z^2) + \ldots \ee}which upto contact terms gives{\be \Pi_V(Q^2)=-\frac{1}{2g_5^2}\ln(Q^2) \label{vectholo}\ee}We can calculate exactly the same quantity using Feynman diagrams \cite{Shifman:1978bx,Shifman:1978by}. The leading-order diagram is the quark bubble{\be \Pi_V(Q^2)=-\frac{N_c}{24\pi^2}\ln(Q^2) \label{pertholo}\ee}And so comparing (\ref{vectholo}) and (\ref{pertholo}) we can fix $g_5$ as{\be g_5^2 = \frac{12\pi^2}{N_c} \label{fixg5}\ee}
\vspace{-1cm}
\subsubsection{A cautionary note}

The matching we have just performed is a little naive. One should match the gravitational theory to QCD only at the point where the QCD coupling becomes non-perturbative, for this is the point where the gravitational dual will be weakly coupled. And of course, when the QCD coupling is non-perturbative, gluonic contributions become important. In all of the analysis to follow, we keep $g_5$ fixed as in equation (\ref{fixg5}). However we have performed an analysis that treats $g_5$ as a free parameter: this indicated that the optimal value of $g_5$ is 5.19, rather than the 3.54 that (\ref{fixg5}) suggests: this is a discrepancy of 32\%! So for the rest of this section we add the disclaimer that non-perturbative effects could have a significant effect on our results. We also remind the reader that AdS/CFT correspondence is a large $N_c$, large `t Hooft limit ($g_{YM}^2N \equiv \lambda \gg 1$). In calculating $g_5$ we have just assumed $N_c=3$; this is a further source of error. We return to the matter of $g_5$ in chapter \ref{ch:IRimp}.

\section{Calculations}\label{sch:QCDcalc}

\subsection{Vector Mesons} \label{sch:AdSQCDvect}

There are now only three parameters in our model: $z_m, m_q$ and $\sigma$. The hadrons of QCD correspond to the normalizable modes of the 5D fields. Requiring the modes to be normalizable means that when we substitute the chosen modes back into the 5D action, the action remains finite. As usual, this means that the solution must die away sufficiently rapidly in the UV, and be smooth over the IR boundary (here given by $z=z_m$). This gives us two explicit conditions for a general solution $\psi(z)$.{\be \psi(0)=0 \hspace{2cm} \partial_z\psi(z_m)=0 \label{BCson1}\ee}There is also a third constraint: that of orthogonality. $\psi_n$ and $\psi_m$ must be orthogonal to one another if they are to represent distinct mesons. So we also have $\int (dz/z)\psi_{\rho}^2=1$. Two constraints is sufficient to solve a second order differential equation: having three constraints guarantees us a discrete spectrum. This is clearly a pre-requisite for a sensible model of QCD.\\

A simple check shows that{\be G(q;z,z')= \displaystyle\sum_{\rho} \frac{\psi_{\rho}(z)\psi_{\rho}(z')}{q^2-m_{\rho}^2+i\epsilon} \ee}is the Green's function to (\ref{sonvect}). It can also be shown that $V(q,z')$ of equation (\ref{propo}) is given by $-(1/z)\partial_zG(q;z,z')$ at $z=\epsilon$. Putting this together with equation (\ref{propo}) we find{\be \Pi_V(-q^2) = -\frac{1}{g_5^2} \displaystyle\sum_{\rho} \frac{(\psi'(\epsilon)/\epsilon)^2}{(q^2 -m_{\rho}^2+i\epsilon)m_{\rho}^2} \ee}which on comparison with equation \ref{vectdecay} of section \ref{ch:pert}, in the limit $\epsilon \rightarrow 0$, allows us to extract the decay constant $F_{\rho}$:{\be F_{\rho}^2 = \frac{1}{g_5^2} \left(\psi_{\rho}'(\epsilon)/\epsilon \right)^2 \label{decayvect}\ee}\vspace{-1cm}

\subsection{Axial Vector Mesons}\label{sch:AdSQCDaxial}

Our analysis of the axial vector mesons closely follows that of the vector mesons. However things are complicated by the mixing of the pion field $\pi$ and the longitudinal axial gauge field $\phi$. We start with the action, expanded to quadratic order:{\be S= \int d^5x \left(-\frac{1}{4g_5^2z}F^a_AF^a_A+ \frac{v(z)^2}{2z^3}(\partial\pi^a-A^a)^2 \right) \ee}where we have defined $v(z)=m_qz+\sigma z^2, A=(A_L-A_R)/2$ and $X=X_0\exp(i2\pi^at^a)$. We make the gauge choice $A_z=0$ and $A_{\mu}=A_{\mu\perp}+\partial_{\mu}\phi$. Then the equations of motion become{\bea \left[\partial_z \left(\frac{1}{z}\partial_zA^a_{\mu}\right)+\frac{q^2}{z}A^a_{\mu}-\frac{g_5^2v^2}{z^3}A^a_{\mu} \right]_{\perp} &=&0 \label{a1eom}\\
\partial_z\left(\frac{1}{z}\partial_z\phi^a\right)+ \frac{g_5^2v^2}{z^3}(\pi^a-\phi^a) &=&0 \label{pion1}\\
-q^2\partial_z\phi^a+\frac{g_5^2v^2}{z^2}\partial_z\pi^a &=&0 \label{pion2}\eea}The $a_1$ meson is a spin-1 particle, and so is the solution to equation (\ref{a1eom}) with the same boundary conditions we imposed for the vector mesons, namely{\be \psi_{a_1}(0)=0 \hspace{2cm} \partial_z\psi_{a_1}(z_m)=0 \ee}The normalization condition is also the same: $\int (dz/z)\psi_{a_1}^2=1$. Similarly, the decay constant $F_{a_1}$ is given by (\ref{decayvect}), but with $\rho$ replaced by $a_1$:{\be F_{a_1}^2 = \frac{1}{g_5^2} \left(\psi_{a_1}'(\epsilon)/\epsilon \right)^2 \ee}The pion can be found by solving (\ref{pion1}) and (\ref{pion2}) simultaneously, subject to the boundary conditions $\phi'(z_m)=0, \phi(0)=0$ and $\pi(0)=0$. The pion decay constant can be found by considering the axial current correlator of (\ref{axialcurrent1}):{\be \langle 0|A_{\mu}(0)|\pi(q)\rangle =  if_{\pi}q_{\mu} \label{axialcurrent2}\ee}According to (\ref{axialcurrent2}), $\Pi_A(-q^2)$ will have a pole at $q^2=m_{\pi}^2$. In the limit $m_{\pi}=0$, $\Pi_A(-q^2) \sim -f_{\pi}^2/q^2$. Hence, on comparison with (\ref{decayvect}), we conclude{\be f_{\pi}^2=-\frac{1}{g_5^2}\frac{\partial_z A(0,\epsilon)}{\epsilon}\ee}\vspace{-1cm}

\section{Results} \label{sonresult}

From this simple model we can predict six quantities\footnote{in \cite{son1} they claim to predict seven quantities; the seventh being the $\pi-\rho$ coupling, $g_{\rho\pi\pi}$. However this term would presumably receive an important contribution from $F^3$ terms in the holographic action, which haven't been included. It could be argued, though, that a good agreement between the prediction for $g_{\rho\pi\pi}$ and its experimental value in the absence of $F^3$ terms indicates that these terms have little influence}: the mass and decay constants of the $\rho$ mesons, the pions, and the $a_1$ mesons. We still have three inputs: the quark condensate, the quark mass, and the position of the infrared cutoff. Performing a best-fit analysis gives a rms error of 9 \%.(Note $\epsilon_{rms} = \sum_O ((\delta O/ O)^2/n)^{1/2}$ with $O$ the
observable and $n$ equal to the number of predictions minus the number of inputs, so here $n=3$).The results are displayed in table \ref{son1res}.

\begin{table}
\begin{center}
\begin{tabular}{c|c|c}
Observable & Measured (MeV) & Best Fit (MeV) \\
\hline
$m_{\pi}$  & 140 & 141 \\
$m_{\rho}$ & 776 & 832 \\
$m_{a_1}$ & 1230 & 1220 \\
$f_{\pi}$ & 92 & 84 \\
$F_{\rho}^{1/2}$ & 345 & 353 \\
$F_{a_1}^{1/2}$ & 433 & 440 \\
\hline
\end{tabular}
\caption{Results of model from \cite{son1} for QCD observables. $m_q=2.3 \mathrm{MeV}$, $\sigma=(327 \mathrm{MeV})^3$, $z_m = 323 (\mathrm{MeV})^{-1}$} \label{son1res}
\end{center}
\end{table}

\section{Improvements} \label{sch:improve}

\begin{figure}
  \hfill
    \begin{center}  
      \includegraphics[scale=0.5]{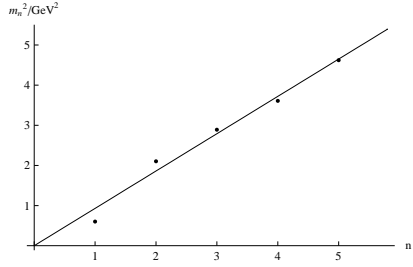}
    \end{center}
 \hfill 
\caption{Experimentally, the $\rho$ mesons follow very closely the relationship $m_n^2 \sim n$. The dots are the physical values} \label{linear}
\end{figure}

Since the publication of \cite{son1} in December 2005, there have been many further publications, suggesting improvements to the simple model presented in the preceding parts of this chapter. We consider here one of the more well-known extensions, achieved by including a dilaton in the action (\ref{holoact}) and generalizing the metric (\ref{holomet}) to:{\bea S&=&\int d^5x \: e^{-\Phi(z)}\:\sqrt{g}\: \mathrm{Tr} \left\{|DX|^2+3|X|^2-\frac{1}{4g_5^2}(F_L^2+F_R^2) \right\} \label{holoact2}\\
 ds^2 &=& e^{2A(z)}\left( dz^2 +  \eta_{\mu\nu}dx^{\mu}dx^{\nu} \right)\label{holomet2}\eea}As yet, $\Phi(z)$ and $A(z)$ are undetermined. We can constrain them with the following pieces of physics:

\begin{enumerate}
\item The geometry must return to AdS in the UV ($z \rightarrow \epsilon$), so the linear combination $\Phi-A$ must asymptote to $\log z$ for small $z$
\item A simple flux model \cite{son2} shows that meson masses obey the relation $m_n^2 \sim n$ for each radial excitation $n$. This agrees with experiment (figure \ref{linear}). To reproduce this phenomenon, $\Phi -A$ must asymptote to $z^2$ for large $z$. It is worth noting that the model \cite{son1} we analysed in the previous section predicts a meson mass spectrum with $m_n^2 \sim n^2$
\end{enumerate}

There are still an infinite number of choices we could make for $A(z)$ and $\Phi(z)$ that would obey these constraints. We make the simplest choice, and set $A(z)=-\log(z)$ and $\Phi=z^2$. Hence the action and metric become{\bea S&=&\int d^5x \: e^{-z^2}\:\sqrt{g}\: \mathrm{Tr} \left\{|DX|^2+3|X|^2-\frac{1}{4g_5^2}(F_L^2+F_R^2) \right\} \label{holoact3}\\
 ds^2 &=& \frac{1}{z^2}\left( dz^2 +  \eta_{\mu\nu}dx^{\mu}dx^{\nu} \right)\label{holomet3}\eea}Now we repeat the procedure of section \ref{sch:QCDcalc} to find the meson masses and decay constants.

\subsection{Vector Mesons} \label{sch:son2vector}

Repeating the process of section \ref{sch:AdSQCDvect}, the equation of motion becomes{\be \partial_z \left(\frac{e^{-z^2}}{z}\partial_z v_n\right) + m_n^2\frac{e^{-z^2}}{z}v_n = 0 \label{dilQCD}\ee}As usual, we look for normalizable solutions to (\ref{dilQCD}). However, now our choice of boundary conditions is much simpler than when there was no dilaton term. There is no hard IR cutoff at $z=z_m$ (as there was previously), and the action is defined for all $z$. So the ambiguity of (\ref{BCson1}) is removed when we have a theory with a smooth wall cutoff (although we have re-introduced some ambiguity through our choice of $A(z)$ and $\Phi(z)$). Requiring that the solution is simply normalizable gives us a discrete spectrum:{\be v_n(z) = z^2 \sqrt{\frac{2}{n+1}}L_n^1(z^2) \ee}where $L_n^m$ are the associated Laguerre polynomials. The masses and decay constants are found using the same formulae in section \ref{sch:AdSQCDvect}, the only difference being the presence of the non-zero dilaton, which changes the results to:{\bea m_n^2 &=&4(n+1) \\
F^2_{\rho_n}&=&\frac{8(n+1)}{g_5^2} \eea}Note that the mass spectrum goes like $m_n^2 \sim n$, as we set out to achieve.

\subsection{Axial Vector Mesons}

The treatment of the axial vector mesons follows that of section \ref{sch:AdSQCDaxial}. We define the axial gauge field as $A\equiv \frac{1}{2}\left(A_L-A_R\right)$, and the equation of motion is found to be{\be \partial_z \left(\frac{e^{-z^2}}{z}\partial_z a_n \right)+ \left(m_n^2-\frac{g_5^2}{z^2}X(z)^2\right)\frac{e^{-z^2}}{z}a_n = 0 \ee}
with the linearized equation of motion for the Higgs-like field $X(z)$ given by{\be \partial_z \left(\frac{e^{-z^2}}{z^3}\partial_z X(z) \right) + 3 \frac{e^{-z^2}}{z^5}X(z) =0 \label{Higgslin}\ee}and, of course, in the UV the Higgs-like field must asymptote to the usual holographic quark mass ($m_q$) and condensate ($\sigma$) terms:{\be X(z)  \stackrel{z\to0}{\longrightarrow} \frac{1}{2}m_q z + \frac{1}{2}\sigma z^3 \ee}Unfortunately, although it is not yet clear, including the dilaton term in the action (\ref{holoact3}) ruins the predictability of the axial vector meson sector. We demonstrate why in the next section.

\subsubsection{Loss of predictability}

For large $z$, the equation of motion for $X(z)$ (\ref{Higgslin}) becomes{\be X'' -2 zX' + \frac{3}{z^2}X = 0 \hspace{2cm} (z\gg 1)\label{asymHiggs}\ee}(\ref{asymHiggs}) has two linearly independent solutions. One asymptotes to $e^{z^2}$ in the IR, whilst the other tends to $e^{-0.75z^{-2}}$. The solution must be finite for all $z$, so we must discard the first solution, and hence conclude that $m_q$ and $\sigma$ are proportional to one another. This is not what one wants in a theory with anomalous symmetry breaking. We conclude that including a dilaton term $\Phi(z)=-z^2$ greatly improves the vector meson sector, but removes anomalous symmetry breaking from the holographic field theory.

\subsubsection{Further extensions}

We only included terms quadratic in $X$ in the holographic action (\ref{holoact3}). The previous section showed that this is not sufficient to model the axial vector meson well. Including such terms would introduce nonlinearity into equation (\ref{Higgslin}). This in turn would make the relationship between $m_q$ and $\sigma$ nonlinear: an essential result if we are to holographically model spontaneous symmetry breaking. This avenue is left to other authors.

Many other aspects of QCD phenomenology have been successfully broached by AdS-QCD models. Baryons, strange quarks and higher spin mesons have all been addressed \cite{Shock:2006qy,Katz:2005ir,deTeramond:2005su}. Four-point current-current correlators relevant to the $\Delta I= 1/2$ rule
and the $B_K$ parameter for K-meson mixing are analyzed
in \cite{Hambye:2005up, Hambye:2006av}. Heavy quark potentials are computed
in \cite{White:2007tu}. The AdS/QCD model is related to light
cone QCD in \cite{Brodsky:2006uqa, Brodsky:2007hb} allowing form
factor computations. Form factors for mesons can also be found in
\cite{Grigoryan:2007vg,Grigoryan:2007wn}.

Properties of QCD at high temperature and density and the
deconfinement transition have been analyzed in this context in
\cite{Ghoroku:2005kg,Ghoroku:2006cc,Nakano:2006js,Kajantie:2006hv,Kim:2006ut,Cai:2007zw,Kim:2007gq}.

Such models have also been adapted to describe walking
\cite{Holdom:1981rm} technicolour \cite{Weinberg:1979bn,Susskind:1978ms} dynamics for electroweak symmetry breaking in
\cite{Hong:2006si,Hirn:2006nt,Piai:2006hy,Carone:2006wj,Carone:2007md}.

%% file: Files/IRimproving.tex
\chapter{Improving the IR of Holographic Descriptions of QCD}
\label{ch:IRimp}

As we saw chapter \ref{ch:AdS-QCD}, a surprisingly good description of QCD can be obtained from a simple five dimensional gauge theory on a truncated AdS space \cite{son1}. However, there are many short-comings of the model:

\begin{itemize}
\item The use of an AdS geometry implicitly means that the background gauge configuration is conformal
\item Hence the UV does not become asymptotically free
\item The existence of a mass gap is imposed by hand by the introduction of an infrared cutoff. It is not the result of a running coupling
\item The fields that holographically describe the quark bilinears are included phenomenologically
\item The solution for the field holographic to the quark mass and condensate is put in by hand - it is not dynamically determined by either the gauge configuration or the quark mass.
\item The excited meson mass spectrum scales like the excitation number $n$. Experimentally, the masses scale like $\sqrt{n}$, a result which a simple flux model can reproduce (see \cite{son2})
\end{itemize}

It would be useful to address the above points, and some progress has been made \cite{me1, son2, Ghoroku:2005vt}. Here we take the opportunity to examine the approach taken in \cite{me1}. The idea was to take the well-documented Constable-Myers supergravity solution of chapter \ref{ch:CM}, whose dual describes many of the characteristics of QCD, most crucially, chiral symmetry breaking. The salient features of the model are

\begin{itemize}
\item The background gauge configuration in which the quarks live is non-supersymmetric, and has a running coupling
\item The mass gap arises naturally as a result of the non-supersymmetric gauge configuration
\item The holographic dual of the quark bilinear is explicit in the string configuration
\item The quark condensate is a prediction of the gauge configuration, and is determined by the quark mass
\end{itemize}

These points go a considerable way towards addressing the short-comings mentioned earlier. We will, however, continue to adopt the phenomenological approach of treating the background as describing an $\mathrm{N}=3$ theory rather than $\mathrm{N}=\infty$. Furthermore, the string theory construction can only realise a U(1) axial symmetry and does not  provide a holographic description of the axial vector mesons. We artificially include the appropriate fields to provide a non-Abelian chiral symmetry and axial vector states. This is in keeping with the phenomenological spirit of \cite{son1,DaRold:2005zs}.

In this chapter we compute with our phenomenological model the
masses and decay constants for the pion and the rho and $a_1$
vector mesons. We find
that the model gives comparable predictions to the pure AdS models
within 12\% of the QCD values. We believe these results provide
support for the robustness of the predictions of these holographic
models.

The geometry we propose returns to pure AdS space in the
ultra-violet, so we do not address here the absence of asymptotic
freedom in the gravity description.  As pointed out in
\cite{Evans:2005ip}, the gravity theory should only be used up
to a UV cut off, corresponding to the scale at which QCD switches
from perturbative to non-perturbative behaviour. Above that cut
off the gravitational dynamics must become non-perturbative with
its loop corrections completely dominating the classical results.
The correct UV dynamics should be encoded at that cut off by
correcting the values of higher dimension operator couplings. In
principle, these can be tuned in the AdS/CFT approach to produce
the holographic equivalent of a perfect lattice action.

As a small example of these ideas we consider the matching of the
five dimensional gauge coupling in the UV. In
\cite{son1, DaRold:2005zs} this coupling is matched to
the perturbative result for the vector vector correlator in QCD.
The AdS gravitational dual presumably describes a strongly coupled
conformal theory in the UV and so the correlator behaviour matches
the logarithmic result of the conformal but weakly coupled UV
behaviour of QCD. It is surprising that the numerical coefficient
of the log term can be matched though. In chapter \ref{ch:AdS-QCD} we indicated that this matching might underpredict $g_5$ by as much as 30\%. We perform a similar analysis here, and find that once again the ideal value of $g_5$ is higher than that indicated in \cite{son1, DaRold:2005zs}, this time by 18\%. This provides a measure of non-perturbative corrections at the scale of matching to the strongly coupled
regime of QCD. We leave attempts to further improve the UV of the
theory to chapter \ref{ch:UVimp}.

Finally, as we showed in section \ref{sch:improve},
an appropriate change to the IR behaviour of the dilaton can
correct the $n$ scaling of the tower of excited $\rho$ meson
states. We have tested our model in this respect but find only a
marginal improvement over the pure AdS case. This is a sign that,
although our geometry describes a non-supersymmetric gauge
configuration, it is still not a perfect description of QCD and
work remains to be done on improving the geometric background.

\section{The generic 5D holographic model}

For a generic five dimensional holographic model of QCD, we can write the action as{\be S \sim \int d^4x \: dr \: e^{\phi} \sqrt{-g} \left(\mathcal{L}_{\sigma} +  \mathrm{Tr}|DX|^2 - \frac{1}{4g_5^2}\mathrm{Tr}(F_L^2+F_R^2) \right) \label{gen5D}\ee}with $D_{\mu}U = \partial_{\mu}U - i A_{L\mu}U + iUA_{R\mu}$. The field $U(x,r)=\exp(i\pi^a(x,r)T^a)$ describes the pions produced by the breaking of a $U(N_f)$ chiral symmetry with generators $T^a$. The non-Abelian gauge fields $A_L$ and $A_R$ couple by left and right action on $U$ and holographically describe the axial and vector mesons (we describe this in more detail later). The $\sigma$ is a Higgs-like field and we only consider fluctuations in its $r$ direction. It holographically describes the quark mass and condensate. A non-zero value for this field will break the $U(N_f)_L \times U(N_f)_R$ symmetry of the action down to $U(N_f)_V$.

If we set $\phi=0$ and $\mathcal{L}_{\sigma}=3|X|^2$, we reproduce the model discussed in chapter \ref{ch:AdS-QCD}.

However, in this chapter, we are going to use a compactified model of the Constable-Myers solution, presented in chapter \ref{ch:CM}.

\subsection{Truncating the Constable-Myers Geometry}

The truncation we propose is incredibly simple. For convenience we reproduce the equations of motion (\ref{CMpion}), (\ref{CMvect}) for the eight dimensional fields defined on the surface of the D7 brane. The functions $f_n$ represents the pion fields, and $g_n$ the vector fields.
\begin{multline}
 \frac{R^2 M_n^2 e^{\phi}\mathcal{G}}{\sqrt{1+(\partial_{\rho}\sigma_0)^2}}H\left(\frac{(\rho^2+\sigma_0^2)^2+1}{(\rho^2+\sigma_0^2)^2-1} \right)^{(1-\delta)/2} \frac{(\rho^2+\sigma_0^2)^2-1}{(\rho^2+\sigma_0^2)^2}\sigma_0^2 f_n  \\
 +\partial_{\rho} \left(\frac{e^{\phi}\mathcal{G}}{\sqrt{1+(\partial_{\rho}\sigma_0^2)^2}}\sigma_0^2 \partial_{\rho}f_n \right)=0\label{CMpion2}\end{multline}
\begin{multline}  e^{\phi}\mathcal{G} \sqrt{1+(\partial_{\rho}\sigma_0)^2} M_n^2R^2g_n(\rho)H \left(\frac{\omega^4-1}{\omega^4+1} \right)^{\frac{1}{4}} \\
+ \partial_{\rho} \left(\frac{e^{\phi}\mathcal{G}}{\sqrt{1+(\partial_{\rho}\sigma_0)^2}}\partial_{\rho}g_n(\rho)\frac{\omega^4}{\sqrt{(\omega^4-1)(\omega^4+1)}} \right) =0 \label{CMvect2}\end{multline}

$\mathcal{G}$ is defined as in (\ref{CMfuncG}). The reader is reminded that the metric is also defined in eight dimensions:
\begin{multline} ds^2_{8D}= \\
H^{-1/2} \left(\frac{\omega^4+b^4}{\omega^4-b^4}\right)^{\delta/4}\eta_{\mu\nu}dx^{\mu}dx^{\nu} + R^2 H^{1/2}\left(\frac{\omega^4+b^4}{\omega^4-b^4}\right)^{(2-\delta)/4}\frac{\omega^4-b^4}{\omega^4}\displaystyle\sum^{4}_{i=1} d\omega_i^2 \label{8Dmet}\end{multline}

Since we want to propose a theory that is dual to QCD, the fields must not have any components on the three sphere. The simplest way to do this is to claim that all of the fields $f_n, \: g_n, \:\phi$ and $\sigma$ are constants on the 3-sphere. They are only functions of $r$ and $x$ only.

On making this choice, we can then either work directly with (\ref{CMpion2}) and (\ref{CMvect2}), whilst remembering that $\phi,\pi$ and $\sigma$ are constants under the SO(3) isometry, or we can explicitly recast the entire ensemble in five dimensions. This is how it was presented in \cite{me1}, and we reproduce the five dimensional theory here.

The five dimensional metric is{\be ds^2 = H^{-1/2}f^{1/8}\displaystyle\sum_{i=0}^3 \eta_{\mu\nu}dx^{\mu}dx^{\nu} + H^{1/2}f^{3/8}h dr^2 \label{IRmet}\ee}
with{\be f=\frac{(\sigma(r)^2+r^2)^2+1}{(\sigma(r)^2+r^2)^2-1},\hspace{1cm} h=\frac{(\sigma(r)^2+r^2)^2-1}{(\sigma(r)^2+r^2)^2}, \hspace{1cm}H=f^{1/2}-1 \ee}
The explicit values of $\phi,g_5$ and $\mathcal{L}_{\sigma}$ are{\bea e^{\phi} &=& H^{5/4}f^{15/16-\sqrt{39}/4}h^{5/2}r^3(1+\dot{\sigma}^2)^{-1/2} \\
g_5^2 &=& 4\pi^2 H^{1/2}f^{3/8+\sqrt{39}/4}h \\
\mathcal{L}_{\sigma} &=& \sqrt{-g} f^{\sqrt{39}/4}g_{rr}^{3/2}\sqrt{1+\dot{\sigma}^2} \label{5Dlag}\eea}So on substituting (\ref{5Dlag}) into (\ref{gen5D}) we get the five dimensional action. Note that as $r \rightarrow \infty$, the entire model reverts back to the simple AdS model considered in section \ref{ch:AdS-QCD}.

We now go on to calculate the masses and decay constant of the vector and axial meson sectors.

\section{Vector and Axial Mesons}

\subsection{Vector sector}

In exactly the same way that we calculated the vector and axial meson masses and decay constants in chapter \ref{ch:AdS-QCD}, we repeat here. 

Just as in section \ref{sch:AdSQCDvect}, we look for solutions to the vector equation of motion that are of the form $V^a_{\mu}(x,r)=V^a_{\mu}(r)\exp(iqx)$ with the gauge $V_r^a(x,r)=0$. We find{\be \partial_r\big(K_1(r)\partial_r V^a_{\mu}(r) \big) + q^2 K_2(r)V^a_{\mu}(r)=0 \label{vecEoM}\ee}
with $K_1=f^{1/2}hr^3(1+\dot{\sigma}^2)^{-1/2}$ and $K_2=Hf^{1-\delta/2}h^2r^3(1+\dot{\sigma}^2)^{-1/2}$

The $\rho$ mesons are interpreted as the normalizable modes of (\ref{vecEoM}), meaning that we impose the boundary conditions $V'(0)=0$ to ensure the smoothness of the solution, and $V(\infty)=0$ to ensure the normalization is finite.

Our analysis for the decay constants follows the same path as that laid out in section \ref{sch:AdSQCDvect}. For large N, the vector current correlator can be written as the sum over $\rho$ resonances:{\be  \Pi_V(-q^2)=-\sum_\rho\frac{F_\rho^2}{(q^2-m_\rho^2)m_\rho^2} \label{vectcurr}\ee}In order to find $F_\rho$, we proceed by finding the Green's function solution to (\ref{vecEoM}). Imposing the completeness relation $\sum_\rho K_2(r)\psi_\rho(r)\psi_\rho(r')=\delta(r-r')$ on the set of eigenfunctions one finds{\be  G(q;r,r')=\sum_\rho\frac{\psi_\rho(r)\psi_\rho(r')}{q^2-m_\rho^2} \ee}The generalisation of equation (\ref{propo}) is{\be  \Pi_V(-q^2)=\left[\frac{1}{ g_5^2q^2} K_1(r)\partial_rv(q,r)\right]_{r=\infty} \ee}Using the fact  $v(q,r')=\left[K_1(r)\partial_rG(q;r,r')\right]_{r=\infty}$, we find{\be  \Pi_V(-q^2)=-\lim_{r\rightarrow\infty}\frac{1}{ g_5^2}\sum_\rho\frac{(K_1(r)\psi_\rho'(r))^2}{(q^2-m_\rho^2)m_\rho^2} \ee}Comparing this to (\ref{vectcurr}) we identify the rho decay constant as:{\be  F_\rho^2=\lim_{r\rightarrow\infty}\frac{1}{ g_5^2}\left(K_1(r)\psi_\rho'(r)\right)^2 \ee}\vspace{-1cm}

\subsection{Axial sector}

The axial vector field is defined as $A^a_\mu=(A^a_{L\mu}-A^a_{R\mu})$. We choose the gauge such that $A_{a r}(x,r)=0$ and $A^\mu_a= A^\mu_{a \bot} + \partial^\mu \phi$. There are three equations of motion: one for each of the $a_1$ meson field, the $\pi$ field and the longitudinal axial gauge field $\phi$:{\bea \big[\partial_r(K_1(r)\partial_r A^a_\mu(r))+q^2K_2(r)A^a_\mu(r)- g_5^2
  \sigma(r)^2K_3(r)A^a_\mu(r)\big]_{\perp}&=&0 \label{a1mes}\hspace{0.5cm}\\
 \partial_r ( K_1(r) \partial_r \phi) + { g_5}^2 \sigma(r)^2 K_3(r) (\pi^a - \phi^a)&=&0 \hspace{0.5cm}\label{pionlong1}\\
 - q^2 K_1(r) \partial_r \phi + { g_5}^2 K_4(r) \sigma(r)^2 \partial_r \pi &=&0\hspace{0.5cm}\label{pionlong2}\eea}where $K_3(r)=Hf^{3/2-\delta/2+\Delta/2}h^3r^3(1+\dot\sigma^2)^{-1/2}$ and $K_4(r)=f^{1+\Delta/2}h^2r^3(1+\dot\sigma^2)^{-1/2}$, and we have separated $A_{\perp}$ as $A_{i\perp}(x,r)=A(q,r)\exp(iqx)$.

Looking for the normalizable solutions to (\ref{a1mes}) requires us to set the boundary conditions $\psi_{a1}(\infty)=0$ and $\partial_r\psi_{a1}(0)=0$. We can then predict the $a_1$ meson mass spectrum. And in an almost identical analysis to that we performed for the vector sector, the $a_1$ decay constants are given by{\be  F_{a_1}^2=\lim_{r\rightarrow\infty}\frac{1}{ g_5^2}\left(K_1(r)\psi_{a_1}'(r)\right)^2 \ee}Looking at equations (\ref{pionlong1}) and (\ref{pionlong2}) we see that the pion and longitudinal gauge fields mix. This makes finding numerical solutions to (\ref{pionlong1}) and (\ref{pionlong2}) very hard. In principle we'd have to search for solutions with two unknowns: the pion mass, $-q^2$, and the ratio of the $\phi$ and $\pi$ fields at $r=0$. So instead, we use the Gell-Mann-Oakes-Renner relation ($m_{\pi}^2f_{\pi}^2=2m_qc$) \cite{DeGrand}, to fix $m_{\pi}$. Here we introduce this relation by hand, but it is possible to show that in the geometry studied in chapter \ref{ch:AdS-QCD}, the equation can be derived explicitly \cite{son1}. This relation eliminates one of our free parameters, and makes the numerical search much easier.

This model has two free parameters: $m$ and $b$. $m$ represents the quark mass (this is a model with only two flavours of quark, which we have assumed to be of equal mass), and $b$, which in the original supergravity solution represented the size of the conformal symmetry breaking. Here then it roughly corresponds to $\Lambda_{QCD}$. Note that this model has the same number of free parameters as real QCD.

\begin{center}
\begin{table}
\begin{tabular}{c|c|c|c}
Observable & Measured (MeV) & New Model (MeV) & Previous Model (MeV) \\
\hline
$m_{\pi}$  & 140 & 139 & 141 \\
$m_{\rho}$ & 776 & 743& 832 \\
$m_{a_1}$ & 1230 & 1337& 1220 \\
$f_{\pi}$ & 92 & 84&84 \\
$F_{\rho}^{1/2}$ & 345&297 & 353 \\
$F_{a_1}^{1/2}$ & 433 &491& 440 \\
\hline
\end{tabular}
\caption{Comparing the best fit results in \cite{son1} and the model presented here }\label{IRtab}
\end{table}
\end{center}

\subsection{Results}

The results of the model are displayed in Table \ref{IRtab}.  We compute six
QCD meson parameters for our fits (we do not include $g_{\rho \pi
\pi}$). Our model has two free parameters (after fixing $g_5$
phenomenologically as discussed above), $b$ corresponding roughly
to the strong coupling scale $\Lambda$, and $m$ corresponding to
the light quark mass. The model therefore has the same number of
free parameters as real QCD.

In the first model, $A$, we match $b$ and $m$ by demanding that we
correctly reproduce $m_\pi$ and $m_\rho$. In order to do this, we
must set $\Lambda_b=264.5$ MeV and $m=2.16$ MeV. This gives a
prediction of $325.8$ MeV for the scale of the quark condensate.
The overall rms error for this model is 12.8\%. For comparison we also reproduce the
pure AdS fit to the same parameters found in
\cite{son1}. That model has three free parameters, the
value of the IR cut off, the quark mass and the quark condensate
and is therefore less predictive.

In model B, we perform a global fit to all observables. This gives
$\Lambda_b=253.2$ MeV and $m=2.24$ MeV, with the characteristic
scale for the quark condensate $311.9$ MeV. The overall rms error
for this model is 11.6\%. Again we reproduce the equivalent pure
AdS model fit for comparison.

It is also possible to calculate the rho-pion-pion coupling constant in this geometry \cite{me1}. This was also done in \cite{son1}. However, as we noted in section \ref{sonresult}, this term would receive important contributions from $F^3$ terms in the holographic action, which we have thus far neglected. Nevertheless, for comparison with \cite{son1} we do perform this result, and find that in this geometry $g_{\rho\pi\pi}=4.81$ MeV. This compares well with the experimental value of $6.03 \pm 0.07$ MeV \cite{PDG06}.

\subsubsection{$g_5$ coupling}

In section \ref{g5match} we pointed out that the perturbative matching of $g_5$ may not be ideal. So once again, we have performed a fit with $g_5$ being considered as a extra free parameter. This indicated that the optimal value of $g_5$ is 4.36, which is 18\% higher than 3.54, the value suggested by the perturbative-matching result in equation (\ref{fixg5}).

\subsection{Conclusions}

We have adapted a string theoretic model of chiral symmetry
breaking to a phenomenological description of QCD.  The model we
have proposed goes some way towards addressing the inconsistencies
of simple AdS slice holographic QCD models \cite{DaRold:2005zs, son1}.
The background geometry of our model is non-supersymmetric, and it is the smooth
variation of this geometry with the radial direction $r$ that
provides a mass gap, without the need for an artificial hard IR
cut-off. In addition, the dual field to the quark mass/condensate
operator is a natural part of the geometrical set-up with the
value of the condensate being determined by the quark mass.

However, this is still a phenomenological approach in that we
introduce extra fields and symmetries by hand into the model in
order to describe the full pion and axial vector sectors. Formally
there is no geometric string interpretation for this system. We
also treat the background as though it describes an $N=3$ rather
than an $N=\infty$ field theory by matching the 5D gauge coupling
to QCD.

We find that the predictions of this model match experimental
results to within 12\%. This model is a little more predictive
than the pure AdS slice models since the condensate is dynamically
determined by the geometry. The best fit is in fact a few percent
worse than the AdS slice models but hopefully the theoretical
improvements represent at least a moral victory. In any case one
would naively have expected errors of order a few 100\% in all of
these models so the closeness to QCD across a range of holographic
models supports the robustness of the approach.

A drawback of these models to date has been that the geometry
returns to AdS for large $r$, meaning that the field theory is not
asymptotically free in the UV. Incorrect physics in the UV may
affect the strong coupling regime in the IR \cite{Evans:2005ip}.
Here we investigated corrections to the matching of the 5D gauge
coupling to naive perturbative QCD results. We found that this
coupling's value should be changed at the 18\% level indicating
the size of non-perturbative effects. In the future one might hope
to study the importance of higher dimension operators in the IR
physics as well.

%% file: Files/UVimproving.tex
\chapter{Improving the UV of Holographic Descriptions of QCD}
\label{ch:UVimp}

This thesis has investigated the possibility of finding a holographic description of QCD. However, QCD is asymptotically free and thus its dual is strongly coupled in the UV. We don't know how to solve string theory in the strongly coupled regime, and so a naive step is to simply eliminate this part of the duality by cutting off the gravity dual in the UV at some as yet undetermined energy $\Lambda$. The dual field theory is then only defined upto $\Lambda$.

In this chapter we investigate the effect of including such a sharp UV cutoff in the gravity dual. The first task is to identify operator-field matches at $r=\Lambda$ rather than at $r=\infty$, which we discuss next.

\newpage
\section{Operator-field matching} \label{sch:op-fi}

The matching of operator dimensions to field solutions is complicated by the presence of a UV cutoff, $\Lambda$. Without it the matching is performed at $r=\infty$: now the matching must be performed at $r=\Lambda$. This makes things more difficult for two reasons:

\begin{enumerate}
\item As we discussed in section \ref{sch:anom}, fields can have anomalous dimensions in the bulk which become small as $r \rightarrow \infty$, and so don't require consideration when matching at $r=\infty$. By introducing a finite cutoff $r=\Lambda$, we need to match not to the classical dimension, but to the anomalously-adjusted dimension of the fields.
\item The presence of large couplings for higher dimension operators before QCD can be matched to the perturbative gravity theory would be another signal of non-perturbative phenomena. Formally there are an infinite set of such couplings. They are called irrelevant perturbations (because they become small in the perturbative regime), and will appear in the gravitational dual as deformations of the metric which grow at large $r$.
\end{enumerate}

These two points seem almost unsurmountable, but all is not lost. There are an infinite number of operators in QCD anyway, yet we have shown in chapters \ref{ch:AdS-QCD}-\ref{ch:IRimp} that including just a handful gives us very close matches to experimental values. In addition, we do have some idea of what to expect from lattice QCD results from a subject called `perfect actions' \cite{Bietenholz:1995cy}.\\

For the rest of this section, we first see the effect of imposing a hard UV cutoff, without worrying about the effects of anomalous dimensions or higher dimension operators. The results are very good. We then consider the possible effect of anomalous dimensions and higher dimension operators, and find that for the cases we consider, the results are changed minimally.

\section{A UV Cutoff}

\begin{figure}
\begin{center}
\includegraphics{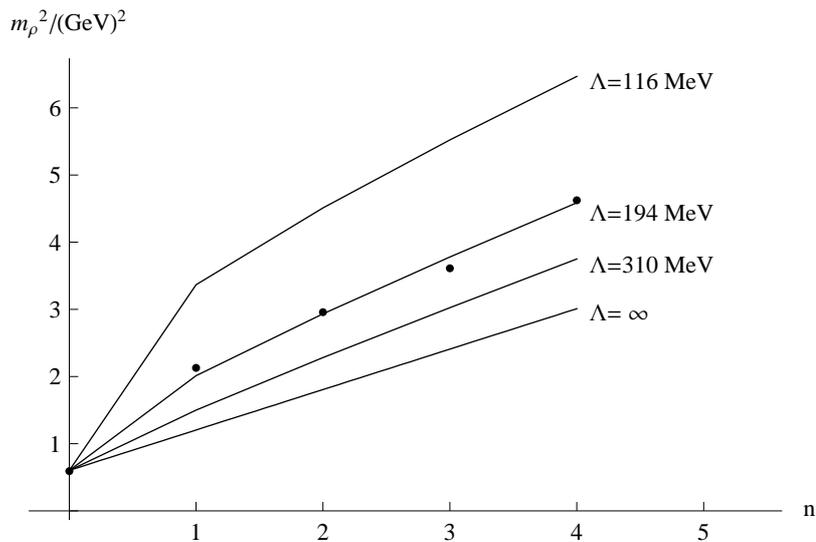}
\caption{The mass squared of the $n$ lightest $\rho$ meson excitations for different values of the UV cutoff. The dots represent the experimental values.}\label{UVfig}
\end{center}
\end{figure}

Here our starting point is the model described in section \ref{sch:improve}, chosen both because of its simplicity, and because the model is smooth for the entire IR spectrum. Whereas before we were looking for solutions to the equations of motion that went like $f_n(r) \sim r^{-2}$ as $r \rightarrow \infty$, we now look for solutions that go like $f_n(r) \sim r^{-2}$ at $r = \Lambda$. $\Lambda$ is the cutoff, above which QCD becomes perturbative or, in the holographic dual, the scale below which the supergravity approximation is trustable. $\Lambda$ will be determined phenomenologically. For convenience, we reproduce the equation of motion for the vector meson spectrum (\ref{dilQCD}) here:{\be \partial_z \left(\frac{e^{-z^2}}{z}\partial_z v_n\right) + m_n^2\frac{e^{-z^2}}{z}v_n = 0 \ee}The results for the first five radial excited  states with several different values of $\Lambda$ are shown in figure \ref{UVfig}. Note that the $\Lambda = \infty$ reproduces that of section \ref{sch:improve}, as expected.

The best phenomenological fit is when the cutoff is set at 194 MeV, and gives an rms error of just 1.8\%. This is a startling result, not just because the fit is so good, but also because it suggests that the regime of validity of a gravity dual to QCD is rather small. Similar conclusions were made in \cite{Evans:2005ip}.

It is worth pointing out that with this value of $\Lambda$, we
predict the next 3 excited $\rho$ mesons to be at 2320 MeV, 2475
MeV and 2626 MeV. Experimental searches have reached up to 2510
MeV, and so far the highest excited $\rho$ meson found is the
final one listed above at 2150 MeV.

\section{Decay constants}

The decay constants $F_{\rho}$ can be found by substituting the regular solutions $f_n(r)\rho^{\mu}(x)$ back into the five dimensional action, and integrating over $r$, in exactly the same manner as section \ref{sch:AdSQCDvect}. Repeating the result here for convenience, the decay constants are then given by{\be F_{\rho}^2=\frac{1}{g_5^2}\left(\Lambda^3 f'_n(\Lambda)\right)^2 \ee}Since the large $r$ behaviour of $f_n(r)$ is the same for all $n$ ($f_n \sim r^{-2}$ as $r\rightarrow \infty$), the different excited states only differ in their decay constants as a result of the different normalizations of the $f_n$.

We require that the kinetic terms for the different rho excitations
are all canonical which implies imposing{\be \int_0^{\Lambda} dr  \frac{e^{-r^{-2}}}{rg_5^2}f_n^2 = 1 \ee}In the original AdS/QCD model (section \ref{sch:son2vector}) with the UV cutoff at infinity, one
finds the decay constants grow as the square root of the
excitation number $n$

{\be F_{\rho_n}^2 = \frac{8(n + 1)}{g_5^2} \ee}If one matches $g_5$ to the perturbative high energy vector
correlator \cite{son1,DaRold:2005zs}, giving $g_5^2 = 12
\pi^2/ N_c$ then the $\rho_0$ meson has a decay constant
$F_{\rho_0}^{1/2} = 260$ MeV compared to the physical value of
$345$ MeV.

\begin{figure}[ht]
  \hfill
  \begin{minipage}[t]{.45\textwidth}
    \begin{center}  
      \includegraphics[scale=0.5]{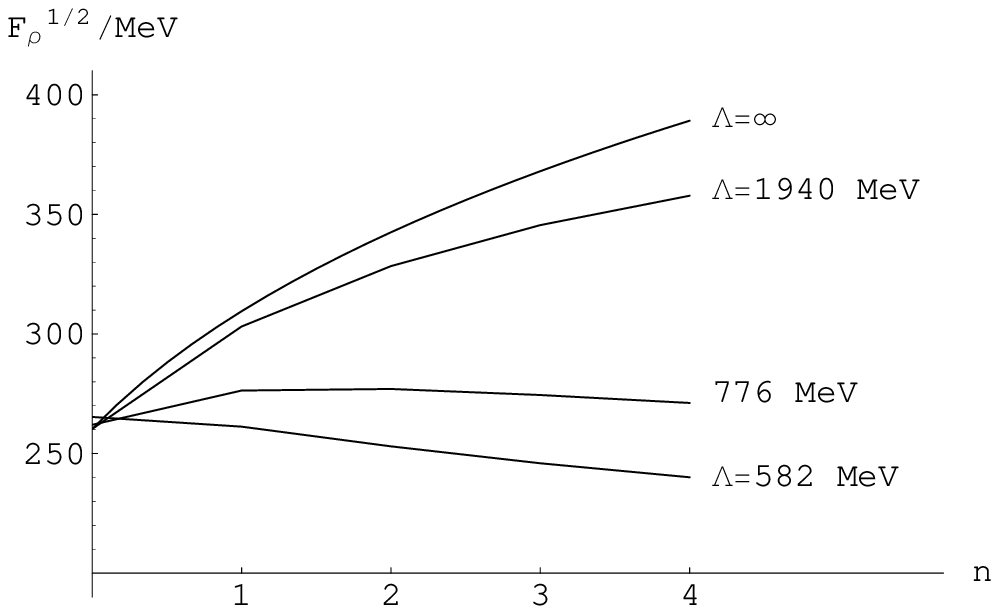}
    \end{center}
  \end{minipage}
\hfill
  \begin{minipage}[t]{.45\textwidth}
    \begin{center}  
      \includegraphics[scale=0.5]{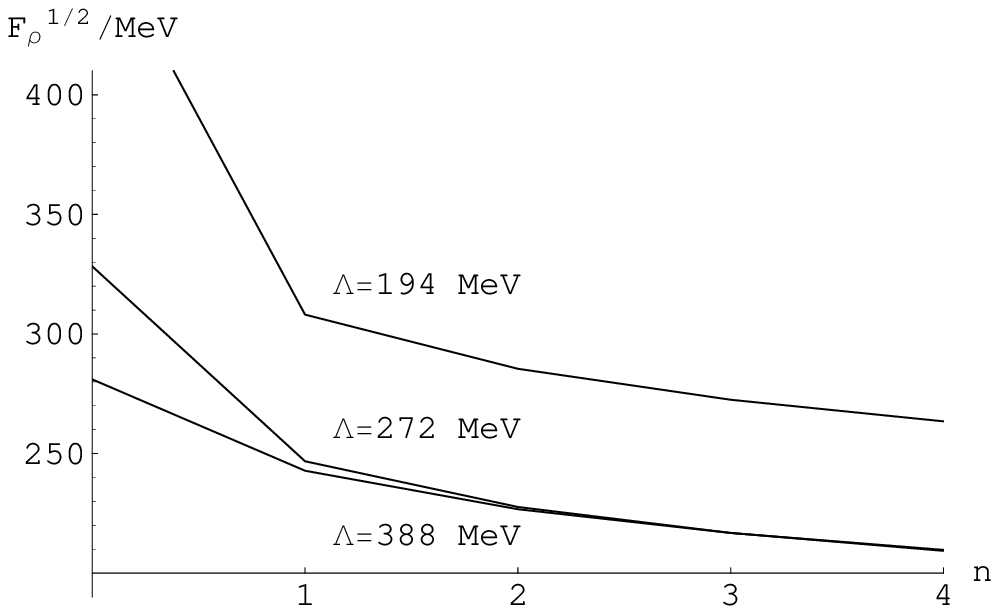}
    \end{center}
  \end{minipage}
 \hfill
\caption{The decay constant $F_{\rho}^{1/2}$ for different $\rho$ excitations, plotted for varying UV cutoffs}\label{UVfig2}
\end{figure}

In figure \ref{UVfig2} we display the results of the same computation with a
UV cutoff present. For low cutoffs $F_{\rho_0}$ rises: with
$\Lambda = 194$ MeV $F_{\rho_0}^{1/2}=478$ MeV. Comparison with
the physical value again hints that a low cutoff is appropriate.

On the other hand, as the cutoff is brought down the $\sqrt{n}$
behaviour (argued for in \cite{Shifman:2005zn}) is apparently lost and the
higher resonance decay constants fall relative to the $n=0$ case.
The reason for this is that the cutoff impedes on the values of
$r$ where the wave functions of the eigenstates are substantial.
By the time that the cutoff is of order a few hundred MeV the
integral for the normalization is dominated around the cutoff.
This makes the computation of the decay constant suspect -
formally one needs a description of the physics to higher energies
which may lie beyond the region of perturbative validity for the
supergravity. 

This contrasts with the computation of the
masses in the previous section - those values are determined by requiring regular
solutions in the infrared away from the cutoff.

\section{Anomalous Dimensions at the Cutoff Scale}

There remains the possibility that the classical dimension of any operator is not respected at the quantum level. Here we investigate possible effects to the operator $\bar{q} \gamma^{\mu}q$\footnote{There is a theorem \cite{coleman} that says that conserved currents are forbidden from acquiring anomalous dimensions. However this theorem is only true for theories that do not contain vector mesons with the same quantum numbers as the conserved current, hence the theorem is not applicable here}. Classically is has dimension three, and this is encoded in the gravity dual by fixing the UV boundary condition as{\be f_n(r) \sim r^{-2},\hspace{0.5cm} f_n'(r)\sim -2 r^{-3} \ee}In general though we should write{\be f_n(r) \sim r^{-\omega}, \hspace{0.5cm} f_n'(r)\sim -\omega r^{-\omega-1} \ee}and numerically investigate the preferred value of $\omega$. The result is that for all values of the cutoff, from $\infty$ down to our tuned value of 194 MeV, the preferred value of $\omega$ is two, so it appears that the classical value is conserved at the quantum level.

However, this is a rather naive analysis: if one evaluates the derivative of the flows $f_n(r)$ at finite values of $r$ in the case of the model with an infinite cutoff, the derivatives differ for each value of $n$. Hence we should be considering the possibility of a different value of $\omega$ for each $f_n$. Pragmatically, tuning each radial excitation is both time consuming, and means that we are introducing so many free parameters that we have lost all powers of prediction. However it is interesting to see how much the anomalous dimensions may contribute. The results for our best fit of $\Lambda = 194$ MeV is shown in table \ref{highdim}(a). All corrections are less than 10\%, which in AdS-QCD terms, is negligible.

\section{Coupling of Higher Dimension Operators}

The second cause for concern when introducing a finite UV cutoff in the strong coupling regime is the contribution of irrelevant operators. As we pointed out in section \ref{sch:op-fi}, there are an infinite number of such operators, and it is not known how to encode the majority into the gravity dual. However, one might hope the couplings of lower dimension operators would grow fastest as one moved into the
non-perturbative regime. We know how to
encode one simply \cite{Evans:2005ip,Evans:2001zn,Intriligator:1999ai,Constable:1999ch}, so we investigate its effect.

Let us rewrite the metric (\ref{holomet}) of model \cite{son1} as{\be ds^2 = H^{1/2}dr^2 + H^{-1/2}\eta_{\mu\nu}dx^{\mu}dx^{\nu} \ee}with $H$ equal to $r^{-4}$ in \ref{holomet}. We can deform the AdS space by allowing it to return to flat space asymptotically:{\be H(r) \rightarrow r^{-4}+\alpha = r^{-4}\left( 1+ \alpha r^{-4}\right) \ee}The parameter $\alpha$ is a symmetry singlet and has energy dimension minus four. Therefore it is identified with the coupling $G$ of the operator $\mathrm{Tr\:F}^4$.\\

We repeated the fit to the lightest five $\rho$ mesons masses in this deformed geometry. The results are displayed in table \ref{highdim}(b).

\begin{table}[t]
  \hfill
  \begin{minipage}[t]{.45\textwidth}
    \begin{center}  
\begin{tabular}{c|c|c}
$\Lambda$ /MeV & $\alpha$ & $\epsilon_{\mathrm{rms}}$ /\%\\
\hline
11000 & 0.0011 & 8.2 \\
5600 & 0.0045 & 7.5 \\
2700 & 0.019 & 6.8 \\
470 & 0.75 & 4.7 \\
194 & 0.08 & 1.8 \\
\end{tabular}
    \end{center}
  \end{minipage}
  \hfill
  \begin{minipage}[t]{.45\textwidth}
\begin{center}
\begin{tabular}{c|c}
meson & $\omega$ \\
\hline
$\rho^*$ & 1.85 \\
$\rho^{**}$ & 1.98 \\
$\rho^{***}$ & 2.12 \\
$\rho^{****}$ & 1.98 \\
\end{tabular}
\end{center}
  \end{minipage}
 \hfill
\caption{Preferred values for: a) the quantum dimension $\omega$ of the operator $\bar{q}\gamma^{\mu}q$ (left), and b) the field dual to $\mathrm{G\:Tr \: F}^4$ (right)}\label{highdim}
\end{table}

The results show two effects. Firstly that $\alpha$ grows as $\Lambda$ decreases. This is to be expected since $\alpha$ changes the large $r$ part of the metric. To change the results when only a small $r$ slice of the metric is present needs a large $\alpha$. Secondly, the result at $\Lambda=$ 194 MeV is an exception to this rule. In fact changing $\alpha$ in this case only has an effect on the third significant figure in the error. This reflects just how good the fit is from a simple cutoff. We conclude that at an appropriately low cutoff, the influence of the $\mathrm{G\:Tr}^4$ is a small effect.

\section{Discussion}

A perturbative gravitational dual of QCD should only be expected
to work at energies below a few GeV at best, where QCD is
non-perturbative. We have investigated imposing a UV cutoff on an
AdS/QCD model of the $\rho$ mesons and found that the data has a fit
at the $2\%$ level with a UV cutoff of a few hundred MeV
(compared to a fit of $21\%$ with an infinite cutoff). This leads us to the conclusion that
the holographic description of QCD should only be
used at low energies on a quite small radial interval. However this conclusion is countered by table \ref{highdim}(a): all the errors are small, and so it is also plausible that the holographic description of QCD could stretch upto several GeV. This would dovetail nicely with chiral perturbation theory and perturbative QCD.

We have also looked at fitting corrections to the anomalous
dimension of the operator $\bar{q} \gamma^\mu q$ and introducing a
coupling of the operator $\mathrm{Tr}\: F^4$. Although these corrections
could be used to fine tune the fit by a percent or so they do not
appear to be significant corrections to the model. Of course these
are only easily implementable examples from an infinite set of
possible corrections but finding the corrections to be small
provides further understanding of the success of the basic AdS/QCD
models. One could also try to include the vacuum expectation
values of more operators in the metric (see for example
\cite{Csaki:2006ji}) and a dynamical, predictive mechanism of
chiral symmetry breaking \cite{Evans:2006dj}. Such effects would
be important to study the pion and axial vector meson sectors of
the model. As explained in \cite{son2} the model used here
does not give a good prediction of these sectors because the
dilaton form, put in to give the $\sqrt{n}$ rise in masses, does
not lead to a sensible condensate prediction. If one attempted to
tackle all of these problems then most likely the number of free
parameters would rise faster than the number of available data
points. Of course this reflects the fact that a perfect action is
in the end just a reparameterization of the full QCD spectrum. We
hope though that we have identified the imposition of a UV barrier
as an important correction and that these other effects are
sub-leading in the $\rho$ sector. Putting together a complete model
of all sectors including the baryons remains as an important
challenge.

%% file: Files/Hadronisation.tex
\chapter{Hadronization}
\label{ch:Hadronisation}

We now turn to a notoriously difficult problem of strongly coupled QCD: hadronization. Hadronization is the process of the formation of hadrons from gluons and quarks. It is widely believed that this is how hadrons were formed in the very early stages of the universe, as the quark-gluon plasma cooled sufficiently. Nowadays, those kind of energies and densities are only reached in extreme natural conditions, such as in the cores of neutron stars. They can also be reproduced in the laboratory, typically at high energy particle colliders of the type found at CERN, FermiLab or DESY, and most recently with the Relativistic Heavy Ion Collider (RHIC) in Brookhaven, USA.

All these colliders are based on the simple premise of colliding two particles together at very high energies, and seeing what the result is. Richard Feynmann likened the process to smashing two Swiss watches together to figure out how the watches work. It is a crude method, but has been very successful. Of primary interest are the hard processes which occur immediately after the collision: these processes can be calculated using Feynman diagrams and have been used to verify the predictions of QCD to within an error of a few percent. However, the vast majority of the collisions after the initial hard collision are soft, and cannot be treated using perturbative QCD. The question we address in the this section is one of the most simple: ``If we annihilate an $e^{+}e^{-}$ pair at high energy, what particles, and in what numbers do we detect in our apparatus?'' This simple question hides a multitude of very difficult issues. If taken at its most fundamental interpretation, it would require us to predict the mass spectrum of all hadrons, their branching ratios, as well as the physics through a probable phase change.

Instead we limit our answer to just predicting the initial yield of hadrons following the (assumed) annihilation of the $e^{+}e^{-}$ pair. For the hadronic mass spectrum we use what has been phenomenologically observed \cite{PDG06}, and we use the same source to obtain the branching ratios of every hadron. Our treatment is not without precedent. \cite{thermo} assumed that after the quarks freeze into hadrons, they may be described as a hadron gas in thermodynamical equilibrium. \cite{thermo} made no attempt to predict branching ratios or mass spectra.

It will be prudent to first review the elementary facts of hadronization.

\section{Basics}

\begin{figure}[t]
\begin{center}
     \begin{picture}(300,100)
                \Photon(30,50)(70,50){4}{4}
                \ArrowLine(10,10)(30,50) \Text(5,30)[l]{$e^-$}
                \ArrowLine(30,50)(10,90)  \Text(5,70)[l]{$e^+$}
                \ArrowLine(70,50)(90,90)  \Text(85,30)[l]{$\bar{q}$}
                \ArrowLine(90,10)(70,50)  \Text(85,70)[l]{$q$}
                \LongArrow(120,50)(150,50) \Vertex(30,50){2} \Vertex(70,50){2}
                \Text(160,50)[l]{hadrons ($\pi^{\pm},K^{\pm},p,\Lambda^0$ etc.)}
      \end{picture}
\caption{Hadronization} \label{hadronpic}
\end{center}
\end{figure}
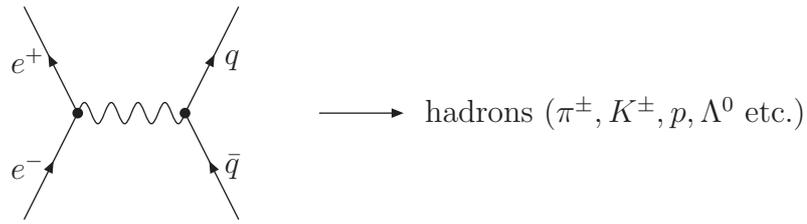

\begin{figure}[!b]
    \begin{center}  
      \includegraphics[scale=0.7]{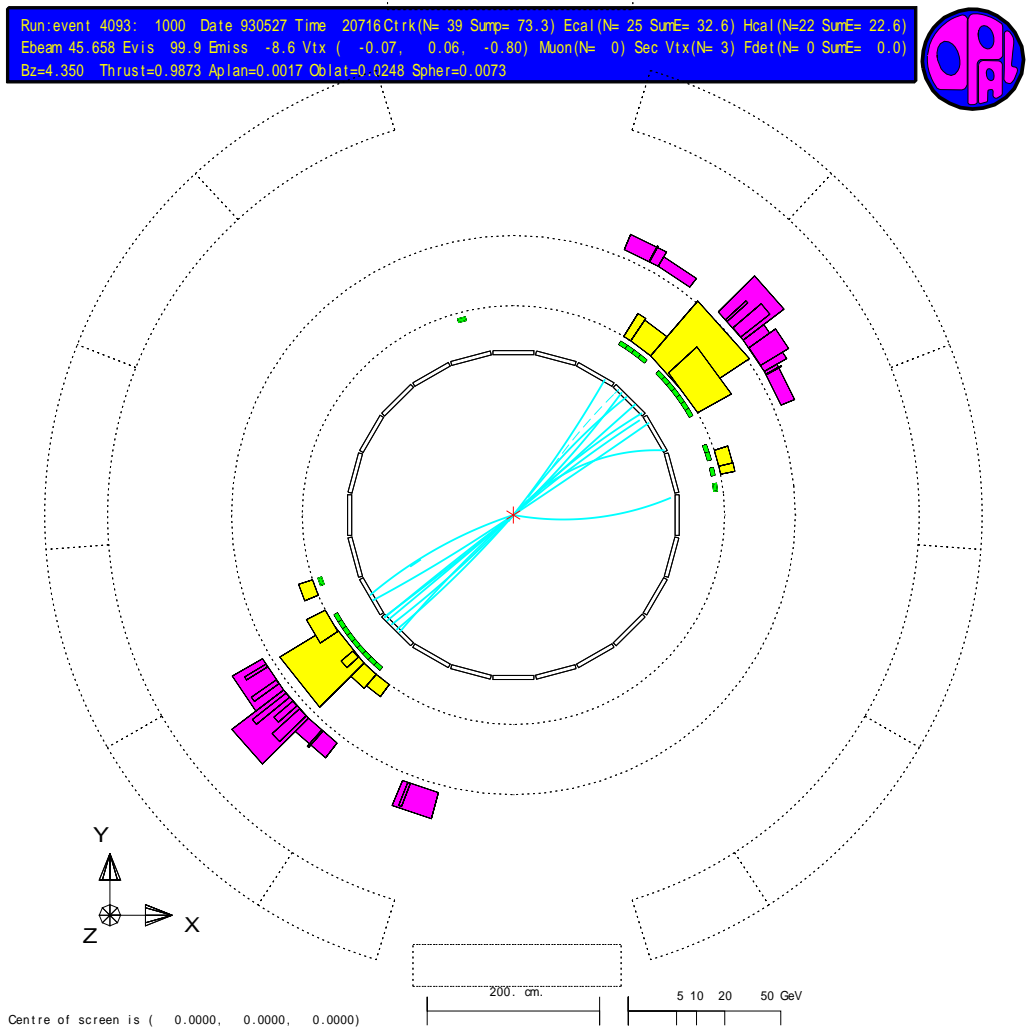}
    \end{center}
\caption{A re-enactment of a single $e^{+}e^{-}$ collision detected at CERN}\label{coll}
\end{figure}

The essence of hadronization is shown in figure \ref{hadronpic}, with the black arrow in the middle symbolising the little understood hadronization process. A computer re-enactment of a single $e^+e^-$ collision is shown in figure \ref{coll}.

It shows that for any one collision, many hadrons are produced. The precise number of hadrons produced in each collision varies, reflecting the probabilistic nature of quantum field theory. But over many collisions, we can compile an average number per collision for each hadron. A sample of these results is shown in table \ref{hadtab} (we tabulate the complete results later). The heaviest hadron detected, the $\Omega$ baryon, has a multiplicity of 0.0014, meaning that for every 10,000 collisions, it is seen fourteen times. Over the same number of collisions, we would have seen 918,000 $\pi^0$s.

\begin{table}[t]
    \begin{center}  
       \begin{tabular}{c|c}
          Hadron & Number measured \\
          \hline
          $\pi^+$ & 8.53 \\
          $\pi^0$ & 9.18 \\
          $ K+$   & 1.18 \\
          $K^0$   & 1.015 \\
          $\eta$  & 0.934 \\
          $\rho^0$& 1.21 \\
          $K^{*+}$ & 0.357 \\
          $\eta '$ & 0.13 \\
          $p$      & 0.488 \\
          $\Lambda$& 0.185 \\
          $\Omega$ &0.0014 \\
         \end{tabular} 
    \end{center}
\caption{Mean multiplicities for a typical $e^+e^-$ collision at $\sqrt{s}=91.2$ GeV \cite{thermo}}\label{hadtab}
  \end{table}

The creation of the $\bar{q}q$ pair in figure \ref{hadronpic} immediately after the $e^+e^-$ collision is a hard process, and is easily modelled by Feynman diagrams such as that shown. What as yet is unknown is how to model the process represented by the black horizontal arrow of figure \ref{hadronpic}. Although there may still be some hard processes occurring, the vast majority will be at low energies, especially as the quarks form more and more hadrons. These soft processes represent a difficult conceptual challenge.

Conservation of momentum means that the $\bar{q}q$ pair will be travelling back to back. Conservation of energy means that they will be travelling very fast. Colour confinement forbids the quarks from existing individually, and how they form colour singlet composites is the focus of several theories.\\

In the \textbf{independent model} the $\bar{q}q$ act independently and each combines with quarks and antiquarks spontaneously created from the vacuum to form hadrons.\\

Alternatively, the \textbf{Lund string theory} treats all but the highest-energy gluons as field lines, which are attracted to each other due to the gluon self-interaction and so form a narrow tube (or string) of strong colour field. This colour tube forms between the $\bar{q}q$ pair, and fragmentation of the strings in the tubes causes the formation of hadrons (see \cite{Andersson:1983ia} for more details). The Lund theory is part of a more general class of parton fragmentation models used by event generators such as PYTHIA and JETSET. However these are complex multi-parameter Monte Carlo event generators that do not reflect the simplicity of QCD, and have little predictive power.\\

A recent model which has had striking success is that of \cite{thermo}. It assumes that each quark/antiquark from the $\bar{q}q$ pair forms a hadronic gas in thermodynamical equilibrium. The gas then cools and expands, until it is no longer in equilibrium, at which point the hadrons are frozen out, and these are then the particles that travel to the detector. So this model only requires three parameters: the temperature and volume of the gas at the point of freezeout, and a strangeness suppression factor to account for the superior mass of the strange quark. The strangeness supression factor should, in principle, be derivable from the known mass of the strange quark. However, such an attempt is not made in \cite{thermo}, and $\gamma_s$ is left as a free-fit parameter. (Factors for the charmed and bottom quarks can also be introduced, but these quarks are so massive, their production rate is negligible). This model can reproduce the results of LEP and PEP-PETRA to 34\% and 38\% respectively.\\

All models of hadronization have two parts: predicting the initial yield of hadrons directly after annihilation, and then allowing for decays of those particles in transit towards the detector. Both of these processes are QCD processes, and so should be fully predictable from the theory. However just predicting one of these two processes is a major challenge: here we concentrate our efforts on the first stage. We will then use experimentally measured values for the branching ratios of all the hadrons \cite{PDG06} to predict what would actually be seen in a detector.

\section{The premise}

All AdS-QCD models provide a weakly coupled five dimensional gravitational theory that describes the hadrons of QCD. Each hadron and its excited states (e.g. the `stack' of $\rho$ meson masses: $\rho(770),\rho(1450),\rho(1700) \ldots$) is described by a five dimensional field that shares the Lorentz and global symmetries of the hadron. In the gravitational theory one seeks solutions for those fields that separate the 3+1 dimensional dependence from the extra radial dependence. More succinctly, we seek solutions of the form $g_n(r)e^{ik_nx}$ with $k_n^2=M_n^2$. There are only regular solutions on the space for discrete masses: these correspond to the meson masses. The functions $g_n(r)$ form an orthogonal basis with the appropriate weighting function.

We will assume that at the point where the quarks form hadrons,
the energy of the event is democratically available to all
hadronic channels. We describe the initial condition as some deposition of energy into the 5D model's stress-energy tensor. The radial dependence of the stress-energy
dependence can be expanded in terms of the functions $g_n$ and
will determine the relative multiplicities of each particle in a
hadron stack. The $x$ dependence will determine the energy and
momentum of the hadrons. In this paper we will simply concentrate
on the multiplicities. 

The result for the multiplicities depends on the choice of the
function expanded in terms of the $g_n$ and this represents the
matching to the underlying asymptotically free QCD dynamics. The
simple guess we will employ is that we should treat all hadronic
channels equally and pick a Gaussian for the initial condition. The height of
the Gaussian determines the absolute value of each particle's
multiplicity and hence is a free parameter which we fit (this parameter can be re-expressed as the average energy per hadron, $\kappa$, which we detail later). The width is also a free parameter which we fit. Finally
the data displays a suppression factor on the production of
strange quarks - this is not surprising since they are more
massive. We reflect this factor in the underlying dynamics by a
strangeness suppression factor $\gamma_s$ which multiplies the
Gaussian for each strange quark in the hadron (note that if mixing
occurs the strange quark content need not be an integer).
$\gamma_s$, which is also used in the thermal hadronization
models \cite{thermo}, is another fit parameter in our analysis.

For any collision with an initial energy of $\sqrt{s}$, no particle with a rest mass greater than $\sqrt{s}$ will be produced, so in principle we would need a cutoff which would depend on the centre of mass energy of the initial collision. In practice we shall simply include all known states with mass below 1.7 GeV. Above this value the experimental
data on the full spectrum becomes patchy. In addition, the high
mass states in this range are only produced with very small
multiplicities and have a minimal effect on the lighter particle
results. Since the $a_0(980)$ triplet is widely thought to be a bound kaon state \cite{PDG06} we include this as a decay channel, but not as a state that can be formed immediately after the annihilation.

The final ingredient we require is a specific AdS/QCD model of the
QCD hadrons. We will adapt a string theory derived model of chiral
symmetry breaking, based on the Constable-Myers geometry of chapter \ref{ch:CM} that includes the vector mesons and pions
\cite{Evans:2004ia}. That both the vector mesons and the pions are
included is an important feature: the mass spectra of the
pseudo-Goldstone bosons are significantly different to all other
hadrons in QCD. We will then assume that the $g_n$ functions
associated with each hadron stack are not that different from the
$\rho$-stack functions - we will simply reproduce them but with
the mass of the lightest stack member tuned to the experimental
value. Similarly the pion stack can be used to reproduce the
towers of states associated with each pseudo-Goldstone of the
chiral symmetry breaking (i.e. the pions, kaons and eta meson). The relative weighting is parameterised by $R$, a measure of $g_{\mathrm{YM}}^2N$, which we fit.

\section{Holographic Hadron Basis Functions}
Our holographic model of the hadron spectrum is based on the Constable-Myers geometry discussed in chapter \ref{ch:CM}. We then compactify it to five dimensions, in exactly the same way we did in chapter \ref{ch:IRimp} when trying to regulate the infrared of holographic descriptions of QCD. We remind the reader that this model spontaneously forms a chiral condensate, and has a complete orthogonal set of functions for both the vector meson sector and pionic-like sector. Henceforth we denote these two sets as $g_n(r)$ and $f_n(r)$ respectively, matching the notation used in chapters \ref{ch:CM} and \ref{ch:IRimp}.\\

The mass of the
lowest lying $\rho$ meson can be dialled by choosing the conformal
symmetry breaking scale in the model, and the mass of the lowest
lying pion can be dialled by choosing the asymptotic quark mass.
The excited states in both stacks are then predicted:
$m_{\rho^*}=1737$ MeV, $m_{\pi^*}=1701$ MeV (c.f. experimental
values of 1459 and 1300 MeV respectively). So whilst they don't
precisely reproduce the experimental values the pattern is at
least roughly right.

As stated above, we assume that the $g_n$
functions associated with each hadron stack are not that different
from the $\rho$-stack functions, and simply rescale the $r$
coordinate such that the mass of the lowest member of each stack
is correct.

\begin{figure}[t]
    \begin{center}  
      \includegraphics[]{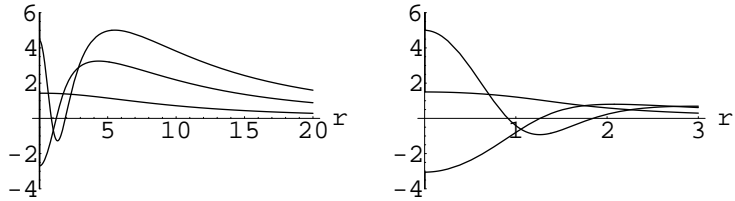}
    \end{center}
\caption{The normalised holographic hadron basis functions for the $\pi^0$ stack (left) and the $\rho$ meson stack (right)}\label{functions}
\end{figure}

We class the pions, kaons and $\eta(548)$ mesons as pionic, and use the $f_n$ functions to model them. For each pionic meson, we dial the asymptotic quark mass such that the
mass of the lowest member of each stack is correct. To give an idea of the functions that result, we plot the
first three functions for $f_n,g_n$ in figure \ref{functions}.

It must be clear to the reader that this situation is not ideal: our ad hoc model predicts over massive excited states, and we have crudely broadened the scope of a model which originally contained only the vector meson sector and a single pionic sector. We would prefer a model which contained all the fields of QCD, and predicted their masses correctly. However, this would be a major breakthrough in progress towards a holographic dual of QCD. This breakthrough has yet to be achieved, and so in the meantime we proceed with the model as presented here.

\section{Overlap Computation of Multiplicities}
With our holographic functions $f_n,g_n$ in place for each hadron
stack we can now proceed with computing the expected initial yield
in a hadronization event.

We assume that all the five dimensional fields $\Psi(r)$ have
a common initial condition of a Gaussian centred at $r=0$ and of
width $\Lambda$. To find the multiplicities of each
stack member we compute 

{\be c_n = \int_0^\infty \Psi(r) \:w(r)g_n(r) dr \ee}where $w(r)$ is the weighting function associated with the basis functions
$g_n$.

The multiplicity is simply given by $c_n^2$ multiplied by (2J+1)
where $J$ is the spin of the hadron.

There are a number of special cases, which we address in turn.

\subsubsection{Pseudo-Goldstone Bosons}
The pseudo-Goldstones are described by a separate holographic
field,  $\theta$, in our model and are represented by functions
$f_n(r)$. The relative contributions the fields make to the
stress-energy tensor are {\be T_{rr} \sim \Delta_1(r) (\partial_r \theta)^2+ \Delta_2(r)\eta^{\mu\nu}(\partial_rA_{\mu})(\partial_rA_{\nu}) + \dots \label{stress-en} \ee}$\Delta_1, \Delta_2$ are functions of $r$, completely calculable from the
holographic model, using the usual definition of the stress-energy tensor:{\be T_{\mu\nu}\equiv -2 \frac{1}{\sqrt{-g}}\frac{\delta}{\delta g^{\mu\nu}} \left(\sqrt{-g} \mathcal{L}_\mathrm{matter}\right)\ee} To ensure the fields $A_{\mu}$ and $\theta$ see the same contribution
to the stress-energy tensor we must rescale the Gaussian. If
our standard Gaussian is $\Psi(r)$ then for the energy to be equally divided between all sectors we require  $\partial_r \theta = \sqrt{\frac{\Delta_2}{\Delta_1}}\partial_r\Psi$.

\subsubsection{Strangeness Suppression Factor}%
Since the underlying asymptotically free dynamics will distinguish
the strange quark from the up and down quarks, we also multiply
the Gaussian by a factor of $(\gamma_s)^\sigma$ where $\sigma$ is
the strangeness content of the stack. $\gamma_s$ is then the
second fit parameter in our procedure. Note that this procedure is
rather crude because different members of a stack may mix to
varying degrees with other states. For example the $\eta(548)$ has
32\% strangeness content while the $\eta^{**}$ has 100\%. In these
cases we set $\gamma_s$ by the strangeness content of the lightest
member of the stack.

\subsubsection{Height of Gaussian}%
The normalisation of the Gaussian tells us the relative
multiplicities of the various hadrons in an event. An overall
multiplicative factor $\kappa$ sets the absolute number of each
species and we fit this value. $\kappa$ determines the total
number of final state particles (before allowing for decays in
transit to the detector), and hence we express it as the average
hadron energy in the collision.

\subsubsection{A Fourth Parameter}
Our choice of holographic dual also contains a free parameter, $R$, which sets the 't Hooft coupling in the gravity dual. We fit it to the data. However, $R$ is not in the same class as $\Lambda,\gamma_s,\kappa$. $R$ has a sound theoretical background, and would not be present if the holographic dual to QCD was known.

\subsubsection{Decay in transit}
Once we have calculated the initial yield of hadrons, we then have
to allow for decays of the particles in transit from the
interaction point to the detector. Branching ratios are taken from \cite{PDG06},
 and particles that can be detected at LEP (whose results
we will compare to) are set as stable. All the other particles are
allowed to decay through the decay channels until they reach one
of the stable particles. In this way we get a list of numbers
which is what our model predicts would be seen at LEP.

\newpage
\section{Predictions}
We compare our results both to $e^+e^-$  collisions performed at
LEP ($\sqrt{s}=$ 91.2 GeV), and at PEP-PETRA ($\sqrt{s}\sim$ 30 GeV). Average
multiplicities of various hadrons have been compiled in
\cite{thermo}, and we reproduce them here, along with our results,
in table \ref{resulttab}. For both sets of results we have performed a four parameter fit so as to minimise the rms error. The pion predictions are expected to be low because our holographic model predicts over-massive excited states. Thus yields of particles such as $\pi(1300)$ and $\rho(1450)$ are unnaturally suppressed which would otherwise be expected to give significant contributions to the pion multiplicities. Despite this, the fits are very good, with rms errors of 47\% and 53\% for  $\sqrt{s}=$ 91.2 GeV and $\sqrt{s}\sim$ 30 GeV respectively. The $\Omega$ and $\eta'$ yields aren't ideal, but previous models have had the same problem \cite{thermo}, especially when matching to the PEP-PETRA data. Our model is expected to have poor predictive power for these two hadrons: the $\eta'$ is technically a pseudo-Goldstone boson, but instanton effects cancel out this effect. Our model just treats it as non-Goldstone boson, which is probably over simplistic. In addition, it mixes heavily with $\eta(548)$: our model contains no good parameterisation of mixing. The $\Omega$ baryon also highlights weaknesses of our model: it is a baryon, not a meson (our functions $g_n$ originally source from a model of mesons, not baryons); it is very heavy, and so is strongly affected by our exclusion of hadrons above $1.7$ GeV; it contains three strange quarks, making it very sensitive to $\gamma_s$.

\begin{table}[t]
\begin{center}
\caption{\label{resulttab} Results of the model for hadron yields at $\sqrt{s}=91.2$ GeV (centre column) and $\sqrt{s}\sim 30$ GeV (right column). The relevant 3+1 free parameter values are shown in the final row.}\vspace{0.5cm}

\begin{tabular}{c|cc|cc}
Hadron & Model & Expt & Model & Expt\\\hline
 $\pi ^+$ & 5.95 & 8.5 &4.07& 5.35\\
 $\pi ^0 $& 6.43 & 9.2 &4.41&5.3\\
 $K^+$ & 1.09 & 1.2 &0.68&0.7\\
 $K^0$ &1.09  & 1.0 &0.68&0.69\\
 $\eta$  & 1.06 & 0.93&0.66&0.584 \\
 $\rho ^0$ & 1.33 & 1.2 &0.88&0.9\\
$ K^{*+}$ & 0.387 & 0.36 &0.28&0.31\\
 $K^{*0}$ & 0.385 & 0.37 &0.28&0.28\\
 $\eta' $ & 0.042 & 0.13 &0.03&0.26\\
 p & 0.41 & 0.406 &0.30&0.3\\
 $\phi$  & 0.03 & 0.1 &0.02&0.084\\
 $\Lambda$  & 0.172 & 0.19 &0.13&0.0983\\
 $\frac{\Sigma ^{*+}+\Sigma ^{*-}}{2}$ & 0.0120 & 0.0094 &0.0089&0.0083\\
$ \Xi^-$ & 0.012 & 0.012 &0.0088&0.0083\\
$ \Xi^{*0}$ & 0.0040 & 0.0033&n/a& n/a\\
 $\Omega$  & 0.0011 & 0.0014&0.0008&0.007\\
\hline
$\Lambda \mathrm{\:(MeV)},\kappa \mathrm{\:(GeV)}, \gamma_s, R$& \multicolumn{2}{c|}{150, 4.96, 0.97, 2.6} &\multicolumn{2}{c}{152, 2.35, 0.97, 2.4}\\
\end{tabular}

\end{center}
\end{table}

\newpage
\section{Conclusions}
We assumed that every hadron in QCD can in principal be
represented by a function in the $r$ coordinate of the 5D
holographic theory of QCD. We then proposed that hadronization can
be modelled by hypothesising that the initial yield (that is
before the particle created starts decaying) for any hadron is
given by the square of the overlap between the function which
represents the hadron, and a Gaussian, centred at the origin, with
a width of $\Lambda$. In addition we have two other
parameters in the theory; a strangeness suppression factor to
account for the heaviness of the strange quark, and $\kappa$ which
determines with what energy the particles leave the interaction
point.

With the full holographic dual to QCD currently unknown, we made
some ad hoc assumptions to achieve a full set of functions which
represented every hadron. We could then put our hypothesis to the
test, and compared the results to $e^+e^-$ collisions made at LEP
and PEP-PETRA. The results were surprisingly good.

It is also interesting to note that although we included a strangeness supression factor, $\gamma_s$, in the manner of \cite{thermo}, we found it to be almost equal to one in both fits. This suggests that in practice such a parameter is not necessary in our model. In addition, our holographic model of QCD predicts over-massive excited states, and this probably leads to a suppression of the pion yields in table \ref{resulttab}. A holographic model with a more accurate hadronic mass spectrum would almost certainly give better results.

The model of hadronization presented here is applicable to all
particle-antiparticle annihilation events, where the fireball
after the collision has no residual quantum numbers. Broadening
this model to include events such as deep inelastic proton-proton
scattering, and heavy ion collisions would clearly be desirable. We leave such an analysis
for the future.

%% file: Files/conclusion.tex
\chapter{Conclusions} \label{ch:conc}

QCD is far from being fully understood. The non-perturbative aspects of the theory are particularly intriguing: confinement, asymptotic freedom, chiral symmetry breaking, the hadron spectrum, hadronization, and the QCD phase diagram all offer exciting research opportunities. The AdS/CFT correspondence has proved to be a valuable tool in our understanding of strongly-coupled gauge theories, and the related AdS/QCD models have reproduced many aspects of QCD very successfully.

In chapter 1 we described some of the more beguiling aspects of QCD, and reviewed the very basic fundamentals of string theory. In chapter 2 we introduced the AdS/CFT correspondence, in particular explaining how every operator in the field theory can be uniquely matched with a field on the gravity side. We used chapter 3 to show how to include flavour into the correspondence, via a D7 brane probe. This gave a relationship between the quark mass and quark condensate, and we showed that in this geometry, chiral symmetry breaking could only be induced by massive quarks: the vacuum does not spontaneously break the chiral symmetry. We went on to calculate the (scalar) meson spectrum in this set-up by looking at the scalar fields on the surface of the D7 brane. We also showed that the vector meson spectrum in this geometry was identical to the scalar meson spectrum, which was to be expected by supersymmetry.

In chapter 4 we introduced a geometry, called the Constable-Myers geometry, which was singular and non-supersymmetric. It is an important geometry because it was the first known geometry to holographically demonstrate chiral symmetry breaking. We summarised its salient properties: confinement and the demonstration of a mass gap (via the glueball spectrum). We then introduced quarks with a D7 probe, and confirmed the spontaneous breaking of chiral symmetry. The pion and vector meson spectra were calculated, and we plotted $m_{\rho}$ against $m_{\pi}^2$. This was compared against equivalent lattice data for QCD at large $N$, and the results were found to be in excellent agreement.

Chapter 5 brought us into the realm of bottom-up theories, generically called AdS-QCD models. We studied in detail the most famous example of such a model because it was historically important, simple to follow, and very successful. By including just three operators in a truncated AdS space, it was possible to holographically model chiral symmetry breaking, and hence predict the masses and decay constants of the $a_1$ meson, the $\rho$ meson, and the pion. The predictions were to within about 10\% of the true QCD values. We then generalised our first AdS-QCD model to include a dilaton in the action. This had the welcome effect of making the vector mesons follow the Regge trajectory of $m_n^2 \sim n$ (c.f. $m_n \sim n$ without the dilaton), but the unwelcome effect of ruining the spontaneous chiral symmetry breaking of the model, and thus removing any meaningful predictions of the axial vector spectrum. We then briefly described the many other phenomena which have been described in AdS-QCD models.

In chapter 6 we attempted to address some of the shortcomings of the existing AdS-QCD models: the lack of asymptotic freedom, the hard IR cutoff, and no dynamical determination of the quark mass and condensate. We took the Constable-Myers geometry of chapter 4 and truncated it by removing the $S^5$ part of the spacetime. This then provided a model which was everywhere smooth, and had all the features that we analysed in chapter 4: confinement, a mass gap, and spontaneous breaking of chiral symmetry. The model, like QCD, only depended on two parameters: the light quark mass $m_q$, and the strong coupling scale $\Lambda$ (c.f. the model of chapter 5 which had three parameters). For comparison with chapter 5, we calculated the same quantities and found agreement with experiment to the 10\% level.

Whilst chapter 6 had regularized the IR of AdS-QCD models, chapter 7 dealt with regularizing the UV of such models. Without regularization in the UV, we were implicitly trying to solve a string theory in the strongly coupled regime. The simplest approach was to include a hard UV cutoff in the gravity dual. This removed the strongly coupled regime from the theory, and made predictions of the vector meson spectrum to within 2\% of accepted values. However, the UV cutoff complicated the operator-field matching process. Firstly if a field had an anomalous dimension it would have to be fitted to a different operator with the relevant dimension, and secondly irrelevant operators would become more and more dominant as we neared the non-perturbative regime. Hence they would have to be encoded in the gravity dual. Without being able to address each complication fully we instead looked at an example of each and found that, for these particular examples, the effect was not large.

The final chapter addressed hadronization. We postulated that after some annihilation event in a particle accelerator, the residual energy took on the shape of a Gaussian. The initial multiplicities of some hadron species would then be given by the overlap between this Gaussian and the radial function  which represented that particular hadron in the QCD holographic dual. Once the initial multiplicities were known they were then allowed to decay according to their published branching ratios until they reached the detector. With the full holographic dual to QCD currently unknown, we used the model of chapter 6. We compared our predictions to $e^+e^-$ collisions at centre-of-mass energies of 30 GeV and 91.2 GeV. Data which ranged over four orders of magnitude were predicted to the 50\% level.

The overall conclusion to be drawn from this thesis is that the AdS-QCD correspondence is in its infancy, but can still reproduce many aspects of QCD at very surprisingly accurate levels. Regulating the ultraviolet and infrared regimes of the dual are necessary steps that don't disrupt the good predictions. Most excitingly, it seems that AdS-QCD models can model hadronization events very well with only a few free parameters.